\documentclass[letterpaper]{mn2e}
\usepackage{psfig,subfigure,mncite}
\title[Multi-Site observations of SU~Aurigae]
      {Multi-Site observations of SU~Aurigae\thanks{Based on observations 
collected at the Canada-France Hawaii 3.6m telescope, the McDonald 2.1m
telescope, the La Palma 2.5m Isaak Newton telescope, the 1.93m 
telescope at the Observatoire de Haute-Provence, the Xinglong 2.2m telescope
and the University of Vienna's automatic photometric telescopes.
}}

\author[Y.C. Unruh, et al.]
       {Y.C. Unruh,$^1$ J.-F. Donati,$^2$ J.~M.~Oliveira,$^{3,4}$ 
	A. Collier Cameron,$^5$
        C. Catala,$^2$ 
	\newauthor
	  H.F. Henrichs,$^6$ C.M. Johns-Krull,$^7$ B. Foing,$^3$ J. Hao,$^8$ 
	  H. Cao,$^8$ J.D. Landstreet,$^{9}$
	\newauthor	
	  H.C. Stempels,$^{10}$ J.A. de Jong,$^{11}$ J. Telting,$^{12}$ N. Walton,$^{13}$ 
	  P. Ehrenfreund,$^{11}$
	\newauthor 	
	  A. Hatzes,$^{14,15}$ J.E. Neff,$^{16}$ T. B\"ohm,$^2$ 
	  T. Simon,$^{17}$  L. Kaper,$^{6}$ K.~G.~Strassmeier,$^{18}$ 
	\newauthor 
	  and Th.~Granzer$^{18}$		\\
	$^1$ Astrophysics Group, Blackett Laboratory, Imperial College of 
	Science, Technology and Medicine, London, SW7 2BW \\
        $^2$ Laboratoire d'Astrophysique de l'Observatoire Midi-Pyr\'en\'ees, Toulouse, France \\
	$^3$ Space Science Division, ESTEC/ESA, Nordwijk, The Netherlands \\
	$^4$ School of Chemistry and Physics, Keele University, Staffordshire, 
	ST5 5BG \\
        $^5$ School of Physics and Astronomy, University of St.~Andrews, Fife, KY16 9SS \\
        $^6$ Astronomical Institute ``Anton Pannekoek'', Univ.~of Amsterdam, The Netherlands \\
        $^7$ Department of Physics \& Astronomy, Rice University, Houston, Texas CA 77005, USA \\
        $^8$ Chinese Academy of Sciences, Beijing Astronomical Observatory, China \\
	$^{9}$ Department of Physics and Astronomy, University of Western Ontario, London, Ontario N6A 3K7, Canada \\
	$^{10}$ Department of Astronomy and Space Physcis, Box 515, SE-751 20 Uppsala, Sweden \\
	$^{11}$ Leiden Observatory, PO Box 9513, 2300 RA Leiden, The Netherlands \\
	$^{12}$ Nordic Optical Telescope, Apartado 474, 38700 Santa Cruz de 
La Palma, Spain \\
	$^{13}$ Institute of Astronomy, University of Cambridge, Madingley Road, Cambridge, CB3 0HA \\
        $^{14}$ Department of Astronomy, The University of Texas at Austin, 
		Austin, TX 78712, USA \\
	$^{15}$ Th\"uringer Landessternwarte Tautenburg, Karl-Schwarzschild-Observatorium, D-07778 Tautenburg \\
	$^{16}$ Department of Physics \& Astronomy, College of Charleston, 
		Charleston, SC 29424, USA \\
	$^{17}$ Institute for Astronomy, University of Hawaii, 2680 Woodlawn Drive, Honolulu, HI 96822 \\
	$^{18}$ Astrophysikalisches Institut Potsdam, An der Sternwarte 16, 14482 Potsdam, 
	Germany \\
              }

\date{Accepted 20 Nov 2003. Received 16 Sep 2003;
in original form 10 Apr 2003}

\begin{document}

\maketitle
\begin{abstract}
We present results from the 1996 MUSICOS (MUlti-SIte COntinuous Spectroscopy)
campaign on the T~Tauri star SU~Aurigae. We find a 2.7-d 
periodicity in the He~{\sc i} (587.6~nm) line and somewhat longer, less well-pronounced
periodicities in the Balmer lines and in Na~D. Our observations
support the suggestion that the wind and infall signatures are
out of phase on SU~Aur. 
We present Doppler images of SU~Aur that have been obtained from 
least-squares deconvolved profiles. Images taken about 
one rotation apart show only limited overlap, in particular at low latitudes.
This is in part due to limitations in signal-to-noise, and in part due 
to line profile deformations that arise from short-lived and/or 
non-surface features. The agreement at high latitudes is better and 
suggests that at least some longer-lived features are present. 
The analysis of Stokes V profiles yields a marginal magnetic field detection
during one of the phases. 
\end{abstract}

\section{Introduction} 
%
\subsection{T~Tauri stars}
\label{sec:intro}
One of the outstanding problems in the study of the later stages of star
formation is the nature of the mechanism by which proto-stars
lose angular momentum as they contract. T~Tauri stars (TTS) are a good
test bed for various accretion theories, in particular classical TTS which 
show strong accretion signatures. 

TTS are usually subdivided into three groups. The first group, classical 
TTS (CTTS), are very active pre-main sequence stars of about 1~$M_\odot$ 
or less.  Their spectra are typically characterised by excess 
UV and IR emission as well as strong, low-ionization emission-line 
activity. This is superimposed onto a nearly ``normal'' late-type 
stellar absorption spectrum that is thereby diluted. This dilution can 
be so strong that virtually no absorption lines are visible
in the most extreme CTTS. Most CTTS spectra can be explained by
a model in which viscous accretion (at a rate of $10^{-8}$ to 
$10^{-6} M_{\odot}yr^{-1}$) from dusty circumstellar discs with radii of 
tens to hundreds of AU \cite{beichman86,myers87} drives the emission line 
activity \cite{kenyon87,cohen89}. The way the star and its disc interact, 
however, is still somewhat controversial. CTTS usually show very strong and 
erratic light-curve variations with no clear periodicities. 
They are typically rather slowly rotating with rotation 
periods of about 5 to 10 days or so and may possess hot spots due 
to accretion shocks as well as cool starspots. 

The second group is called weak-line TTS (WTTS). They are
not accreting and show no evidence for a circumstellar disk (in 
the visible wavebands). WTTS often display higher rotation 
rates than their more active counterparts with $P_{\rm rot}$ being 
of the order of 1 to 3 days (though 
see also \scite{stassun1999PMS_rot} who found no significant difference 
between WTTS and CTTS rotation periods for stars in the Orion Nebula). 
Most WTTS have light curves that show clear periodicities and can 
be explained by invoking cool star spots. WTTS that show no
disk signatures in the infrared are sometimes also called ``naked'' T~Tauri 
stars. 

The third group consists of early-type TTS (ETTS), sometimes 
also called UXORs (after the prototype UX Ori). These are stars of 
spectral type K0 or earlier \cite{eaton95}. Strictly speaking, 
the UXOR objects include TTS as well as Herbig Ae/Be stars. 
Their light curves are characterised  by strong and erratic dimmings 
(quite often by about 1 magnitude and more in the V band). 
The most common explanation for these dimmings is the obscuring
of the stellar disk by optically thick circumstellar gas 
(see e.g.~\pcite{grinin95}). The presence of optically thick 
circumstellar disks around ETTS, however, has not yet been well 
established.
A number of ETTS show infall signatures, 
usually in the form of transient redshifted absorption lines. This
has prompted speculation that the infalling gas is due to 
evaporating solid bodies (e.g.~\pcite{grady96,grinin94}).
ETTS do not appear to show any changes in the veiling continuum. 

To date, only very few T~Tauri stars have been Doppler imaged. The most
popular targets have been the two WTTS V410 Tau
\cite{joncour92,joncour94v410,strassmeier94,hatzes94v410,rice96} and HDE~283572
\cite{joncour94hde,strassmeier98hde283572}.
The two more active TTS that have been imaged are DF~Tau \cite{unruh98dftau} 
and SU~Aur \cite{petrov96}.

Much effort has been invested recently in measuring the magnetic 
fields on TTS, mostly using Zeeman line broadening. Typical
magnetic-field measurements yield of the order of 1 to 2~kG 
for WTTS (see e.g.~\pcite{basri92,guenther96mag}). For CTTS, 
\scite{johns-krull99bptau} determined a mean surface magnetic field 
strength of $2.6 \pm 0.3$~kG for BP~Tau and \scite{guenther99mag} 
found mean magnetic fields of $2.35 \pm 0.15$~kG and $1.1 \pm 0.2$~kG 
for the T~Tau and Lk~Ca~15 respectively.
These field strengths are within the range of values needed for 
magnetic accretion models, (though see \scite{safier98} for a 
critique of the magnetospheric infall model).
Using spectro-polarimetry, magnetic fields can be detected 
directly by their Zeeman signatures (in this case no accurate
spectral line synthesis is necessary). Using this technique, 
\scite{donati97zdi} detected complex magnetic fields on 
the surfaces of WTTS V410~Tau and HD~283572. \scite{johns-krull99bptau_pol}
found the He~{\sc i} (587.6~nm) emission line of BP~Tau to be circularly 
polarised and inferred magnetic field strengths of about 2.5~kG.

\begin{table}
\caption{Parameters of SU Aur.}

\vspace{2mm}
\begin{tabular}{lll}
	\hline 
        Parameter   &   &  Ref. \\
        \hline
        spectral type & G 2     & \scite{cohen79} \\
        brightness (V)    & 9 mag & \scite{cohen79} \\
        Mass            & 2.25 M$_\odot$ & \scite{cohen79} \\
			& 1.9 M$_\odot$ & Dewarf, Guinan \& \\
			&		& Shaughnessy (1998) \nocite{dewarf1998} \\
	$v\sin i$	& $\approx 60$~km~s$^{-1}$ & \scite{johns-krull96tts}, \\
			&			   & this paper \\
        Period (photom.) & 1.55, 2.73~d & \scite{herbst87} \\
        Period (photom.) & $\leq$3.5~d & \scite{bouvier93} \\
        Period (H$\alpha$ wing) & $\approx$ 3~d & \scite{johns95suaur} \\
        Period ($v_{\rm rad}$) & $3.03 \pm 0.03$~d & \scite{petrov96} \\
	Period 		& $2.7 \pm 0.3$~d & this paper \\
        \hline\\
\end{tabular}
\label{tab:params}
\end{table}

%
%
\subsection{SU~Aur}
SU~Aur is a relatively bright T~Tauri star that has been variously
classified as a classical T~Tauri star (see e.g.~\pcite{giampapa93}),
or early-type T~Tauri star \cite{herbst94}. It has also been
used as a prototype to define an SU~Aur class of T~Tauri stars
\cite{herbig88cat}. Some of its parameters are listed
in Tab.~\ref{tab:params}. According to \scite{akeson2002}, SU~Aur has a
rather low disk mass of $\log (M_D)/M_\odot) = -5.1^{+1.4}_{-0.8}$. This
is considerably lower than typical classical TTS disk masses.

On account of its relative brightness, SU~Aur has been the subject of several
detailed spectroscopic studies
\cite{giampapa93,johns95suaur,petrov96}.
\scite{johns95suaur} put forward a magnetospheric accretion model
derived from the currently favoured model for CTTS. In the CTTS accretion
model the disk is disrupted by the magnetic field
of the star and accreting matter is channelled onto the star along magnetic
field lines \cite{konigl91,cameron93ttdisc,shu94,armitage96tts}.
In the case of SU~Aur, \scite{johns95suaur} suggest that the magnetic dipole
is slightly inclined with respect to the rotation axis of the star.
Material from the disk accretes along magnetic field lines,
but because of the inclination of the dipole, accretion is
seen preferentially for one half of the stellar rotation, whereas a
wind is observed during the other half. This model has been dubbed the
``eggbeater''.

Even though SU~Aur has been monitored frequently, the photometric
period determinations are very uncertain. This is because the light curve of
SU~Aur does not show any clear periodic variations (see also Sec.~\ref{sec:photom}). 
To date, the most conclusive period determinations have not come from photometric 
measurements, but were derived from spectral variations in 
the emission lines (see \pcite{johns95suaur}, 
\pcite{petrov96} and Tab.~\ref{tab:params}).
In fact, most of our current knowledge stems from 
time-series investigations of the Balmer profiles, though, so far, 
observations always suffered from relatively large phase gaps.

\begin{table}
\caption{List of the observatories with the wavelength coverage and resolution
of the spectrographs used.
        }
\begin{tabular}{lccc}
        \hline
        Observatory     & detector  & $\lambda\lambda$ & Resolving \\
                        &           &   [nm]           & Power     \\
        \hline
        BAO (2.2m)  & 1024$^2$ Tektronix & 557 -- 850  & 45\,000     \\
        OHP (1.9m)  & 1024$^2$ Tektronix & 389 -- 682  & 45\,000     \\
        INT (2.5m)  & 1024$^2$ Tektronix & 486 -- 849  & 35\,000     \\
        MDO (2.1m)  & 1200\,$\times$\,400 Reticon & 547 -- 673 & 55\,000 \\
        CFHT (3.5m) & 2046$^2$ STIS2 & 410 -- 814      & 35\,000     \\
\hline
\end{tabular}
\label{tab:tele}
\end{table}

Despite the uncertainties in its period, SU~Aur is, compared to 
most other classical and early-type TTS, a promising candidate for 
surface-mapping by virtue of its small veiling and high
rotation velocity. In this paper we present the observations of SU~Aur taken 
during the 1996~MUSICOS campaign and give an overview of the emission-line
and photospheric line-profile variability. In \scite{oliveira2000suaur}
the emission and upper-atmospheric lines were analysed in 
more detail.

%
\section{Observations and data reduction}
\subsection{MUSICOS}

The MUlti-SIte COntinuous Spectroscopy (MUSICOS) network was set 
up to investigate the time-dependent behaviour of stars with rotation 
periods that necessitate coordinated observations from several 
longitudes \cite{catala93mus1}.

\begin{table}
\caption{Journal of observations of SU~Aur from BAO in Xinglong, China. The columns
list the identifier for the observatory, the day (in 1996 November) on which
the spectra were taken, the UT mid time, modified Julian date (HJD$-2450400$), 
the exposure time and the number of spectra
taken per group. The last column lists the approximate S/N ratio per pixel in
the H$\alpha$ order for a single exposure. In order to obtain an estimate
of the S/N ratio per resolution element, these numbers have to be multiplied by
about 1.3.}
\begin{tabular}{ccccccc}
        \hline
Obs & Date \hspace*{-1ex} & UT & \hspace*{-1em} MHJD \hspace*{-1em} & Exp time & \hspace*{-1em} Group \hspace*{-1em}& S/N near \\
    &                     &    &                     &  [s]     & \hspace*{-1ex} size \hspace*{-1em} & H$\alpha$ \\
        \hline
BAO & 19  &  15:20 & 7.14	& 1800 &  2   & 80  \\
BAO & 19  &  18:09 & 7.26 	& 1800 &  1   & 80  \\
BAO & 19  &  19:52 & 7.33 	& 1800 &  1   & 85  \\
BAO & 20  &  14:25 & 8.10	& 1800 &  1   & 50  \\
BAO & 21  &  13:37 & 9.07	& 1800 &  1   & 50  \\
BAO & 21  &  15:02 & 9.13	& 3600 &  1   & 75  \\
BAO & 21  &  18:27 & 9.27	& 1800 &  1   & 55  \\
BAO & 21  &  19:39 & 9.32	& 1800 &  2   & 60  \\
BAO & 24  &  13:28 & 12.06 	& 1800 &  1   & 80  \\
BAO & 24  &  14:36 & 12.11 	& 1800 &  2   & 80  \\
BAO & 24  &  17:58 & 12.25 	& 1800 &  2   & 75  \\
BAO & 24  &  21:27 & 12.39 	& 1800 &  3   & 50  \\
\hline
\end{tabular}
\label{tab:jourX}
\end{table}
 
SU~Aur was chosen as one of the targets for the {MUSICOS} 1996
campaign that took place in November of that year.
A total of 126 \'echelle spectra were taken over
10 nights at five different sites. At the Beijing Astronomical Observatory
(BAO) in Xinglong, China, the setup consisted of a 2.2-m telescope with
an \'echelle spectrograph and a 1024 by 1024 Tektronix detector. The
1.9-m telescope at the Observatoire de Haute Provence (OHP, France)
was used with the Elodie spectrograph \cite{baranne96}. On the 2.1-m
Isaac-Newton-Telescope (INT)
in La Palma, Spain, the ESA {MUSICOS} spectrograph \cite{baudrand92}
was linked to the Cassegrain focus with a fibre-optic cable. At
McDonald Observatory (MDO, Texas), we used the 2.1-m Otto Struve Telescope
with the Sandiford Cassegrain echelle spectrometer \cite{mccarthy93}.
The observations at the Canada-France-Hawaii Telescope (CFHT)
included observations in Stokes V. They
were taken through a Cassegrain polarimetric unit, fibre-linked to
the {MUSICOS} echelle spectrograph \cite{donati99muspol}.
The wavelength coverage and
resolution of each telescope/spectrograph combination are shown in
Tab.~\ref{tab:tele} (see also \pcite{catala99musicos}).

\begin{table}
\caption[]{Journal of observations for SU~Aur from the OHP, France.
In order to obtain an estimate
of the S/N ratio per resolution element in the H$\alpha$ order, the numbers 
listed in the last column have to be multiplied by 1.5.
}
\begin{tabular}{ccccccc}
\hline 
Obs & Date \hspace*{-1ex} & UT & \hspace*{-1em} MHJD \hspace*{-1em} & Exp time & \hspace*{-1em} Group \hspace*{-1em}& S/N near \\
    &                     &    &                     &  [s]     & \hspace*{-1ex} size \hspace*{-1em} & H$\alpha$ \\
        \hline
 OHP    & 18    & 22:14 & 6.43	& 1800  & 3  &  25      \\
 OHP    & 19    & 03:22 & 6.64	& 1800  & 3  &  35      \\
 OHP    & 20    & 21:55 & 8.41	& 1800  & 3  &  40      \\
 OHP    & 21    & 01:32 & 8.56	& 1800  & 3  &  55      \\
 OHP    & 21    & 04:24 & 8.68	& 1800  & 2  &  50      \\
 OHP    & 23    & 22:12 & 11.43	& 1800  & 3  &  45      \\
 OHP    & 24    & 01:32 & 11.56	& 1800  & 2  &  50      \\
 OHP    & 24    & 05:00 & 11.71	& 1800  & 2  &  30      \\
 OHP    & 24    & 21:31 & 12.40	& 1800  & 2  &  50      \\
        \hline
\end{tabular}
\label{tab:jourO}
\end{table}

\begin{table}
\caption[]{Journal of observations for SU~Aur from the INT, La Palma.
The multiplication factor to obtain the S/N for each resolution 
element is 1.4.}
\begin{tabular}{ccccccc}
\hline 
Obs & Date \hspace*{-1ex} & UT & \hspace*{-1em} MHJD \hspace*{-1em} & Exp time & \hspace*{-1em} Group \hspace*{-1em}& S/N near \\
    &                     &    &                     &  [s]     & \hspace*{-1ex} size \hspace*{-1em} & H$\alpha$ \\
        \hline
 INT     &  17  &  02:08 & 4.59	& 1200 &  1 &   60     \\
 INT     &  19  &  23:05 & 7.46	& 1800 &  2 &   55     \\
 INT     &  20  &  03:56 & 7.66	& 1800 &  2 &   60     \\
 INT     &  20  &  05:55 & 7.79	& 1800 &  2 &   70     \\
 INT     &  20  &  22:40 & 8.44	& 1800 &  2 &   60     \\
 INT     &  21  &  01:09 & 8.55	& 1800 &  2 &   55     \\
 INT     &  21  &  03:25 & 8.64	& 1800 &  2 &   75     \\
 INT     &  21  &  05:39 & 8.74	& 1800 &  1 &   85     \\
 INT     &  21  &  23:18 & 9.47	& 1800 &  2 &   65     \\
 INT     &  22  &  01:27 & 9.56	& 1800 &  2 &   65     \\
 INT     &  22  &  03:35 & 9.65	& 1800 &  2 &   60     \\
 INT     &  22  &  06:10 & 9.76	& 1800 &  2 &   60     \\
 INT     &  23  &  23:14 & 11.47 & 1800 &  2 &   40     \\
 INT     &  24  &  01:08 & 11.55 & 1800 &  1 &   60     \\
 INT     &  24  &  01:47 & 11.57 & 900 &  2 &   40     \\
 INT     &  24  &  03:49 & 11.66 & 670 &  1 &   35     \\
 INT     &  24  &  21:50 & 12.41 & 1200 &  3 &   45     \\
 INT     &  25  &  00:20 & 12.51 & 1200 &  3 &   60     \\
 INT     &  25  &  03:26 & 12.64 & 1200 &  3 &   65     \\
 INT     &  25  &  22:14 & 13.43 & 900 &  2 &   35     \\
\hline 
\end{tabular}
\label{tab:jourI}
\end{table}

\begin{table}
\caption[]{Journal of observations for SU~Aur from the McDonald Observatory, 
Texas. Columns as in Tab.~\protect{\ref{tab:jourX}}, the multiplication factor 
for the S/N is 1.4.}
\begin{tabular}{ccccccc}
\hline 
Obs & Date \hspace*{-1ex} & UT & \hspace*{-1em} MHJD \hspace*{-1em} & Exp time & \hspace*{-1em} Group \hspace*{-1em}& S/N near \\
    &                     &    &                     &  [s]     & \hspace*{-1ex} size \hspace*{-1em} & H$\alpha$ \\
        \hline
 MDO     &  19  &  05:36 & 6.73	& 1800  & 1  & 110 \\
 MDO     &  19  &  06:03 & 6.75	& 1200  & 1  &  95 \\ 
 MDO     &  19  &  08:05 & 6.84	& 1300  & 2  & 105 \\ 
 MDO     &  19  &  10:33 & 6.94	& 1500  & 2  & 115 \\  
 MDO     &  19  &  12:14 & 7.01	& 1500  & 1  & 100 \\ 
 MDO     &  20  &  06:05 & 7.75	& 1500  & 2  &  95 \\ 
 MDO     &  20  &  08:38 & 7.86	& 1600  & 1  &  80 \\ 
 MDO     &  20  &  09:10 & 7.88	& 1800  & 1  &  80 \\ 
 MDO     &  20  &  12:02 & 8.00	& 1800  & 2  &  85 \\ 
 MDO     &  21  &  06:20 & 8.76	& 1800  & 2  &  95 \\ 
 MDO     &  21  &  09:04 & 8.88	& 1600  & 2  & 115 \\ 
 MDO     &  21  &  11:15 & 8.97	& 1600  & 1  &  90 \\ 
 MDO     &  21  &  11:46 & 8.99	& 1800  & 1  &  80 \\ 
 MDO     &  22  &  05:36 & 9.73	& 1600  & 1  &  80 \\ 
 MDO     &  22  &  07:42 & 9.82	& 1500  & 2  & 100 \\ 
 MDO     &  22  &  10:17 & 9.93	& 1800  & 2  & 100 \\ 
 MDO     &  23  &  05:33 & 10.73	& 1800  & 2  &  85 \\ 
 MDO     &  23  &  08:34 & 10.86	& 1800  & 1  & 110 \\ 
 MDO     &  23  &  09:03 & 10.88	& 1600  & 1  & 110 \\ 
 MDO     &  25  &  05:24 & 12.73	& 1800  & 2  &  80 \\ 
 MDO     &  25  &  08:08 & 12.84	& 1800  & 2  &  80 \\ 
 MDO     &  25  &  10:39 & 12.94	& 1800  & 2  &  80 \\ 
 MDO     &  26  &  04:14 & 13.68	& 1800  & 1  &  85 \\ 
\hline
\end{tabular}
\label{tab:jourM}
\end{table}

\begin{table}
\caption[]{Journal of observations for SU~Aur from the Canada-France-Hawaii 
Telescope. Columns as in Tab.~\protect{\ref{tab:jourX}}, the multiplication factor 
for the S/N is 1.5.
}
\begin{tabular}{ccccccc}
Obs & Date \hspace*{-1ex} & UT & \hspace*{-1em} MHJD \hspace*{-1em} & Exp time & \hspace*{-1em} Group \hspace*{-1em}& S/N near \\
    &                     &    &                     &  [s]     & \hspace*{-1ex} size \hspace*{-1em} & H$\alpha$ \\
        \hline
 CFHT   & 19 &  09:39 & 6.90	& 600  & 4  & 55 \\
 CFHT   & 19 &  12:40 & 7.03	& 600  & 4  & 50 \\
 CFHT   & 20 &  08:14 & 7.84	& 600  & 4  & 60 \\
 CFHT   & 21 &  14:18 & 9.10	& 600  & 4  & 30 \\
 CFHT   & 26 &  08:55 & 13.87	& 600  & 4  & 30 \\
 CFHT   & 26 &  11:30 & 13.98	& 600  & 4  & 50 \\
 CFHT   & 27 &  10:35 & 14.94	& 600  & 4  & 50 \\
 CFHT   & 27 &  11:34 & 14.98	& 600  & 4  & 45 \\
 CFHT   & 27 &  14:37 & 15.11	& 600  & 4  & 35 \\
\hline
\end{tabular}
\label{tab:jourC}
\end{table}

The journals of observations
are given for each observatory individually in Tabs~\ref{tab:jourX}
to \ref{tab:jourC}. The exposure time listed is the time of each
individual exposure, rather than that of the complete group. Similarly,
the signal-to-noise ratios listed are average ratios per pixel
for a single exposure in a group. The signal-to-noise ratio per
resolution element is obtained after
multiplication with the factor given in the caption of each
table.

\begin{figure}
\centerline{\hspace{-4mm} \psfig{figure=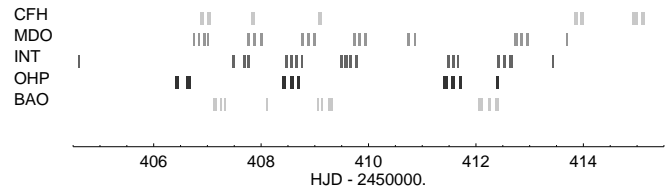,width=8.8cm}}
\caption[]{Phase coverage during the MUSICOS campaign}
\label{fig:timcov}
\end{figure}

The weather during the run was patchy, particularly at the CFHT, but
we obtained good phase coverage  between HJD 2450406.5 and
HJD~2450410.0, and reasonable phase coverage for the subsequent days up to
HJD~2450414.0 (see Fig.~\ref{fig:timcov}).

%
%
\subsection{Spectroscopic data reduction}
The data were reduced following standard procedures. After 
bias subtraction, each exposure was flat-fielded. For 
the data taken at the INT, BAO and MDO the orders were extracted
using the optimal extraction algorithm \cite{horne86extopt} as 
implemented in {\sc echomop} \cite{mills92}, an \'{e}chelle
data reduction package distributed by {\sc starlink}.
Due to internal reflections in the MDO spectrograph we had to reject
two orders in the 610~nm wavelength region.
In the case of the OHP, the data are automatically reduced
at the observatory with the {INTER-TACOS} procedure \cite{baranne96}
and were used as such.

The reduction of the polarimetric data from the CFHT was performed 
using a dedicated procedure for extracting Stokes V and I parameters
developped as part of the ``Esprit'' data reduction package by 
\scite{donati97zdi}.
Each exposure is a combination of four subexposures that have
to be taken in order to get rid of background and instrumental 
polarisation. The data reduction is described in detail in \scite{donati97zdi}.

We have not attempted to flux-calibrate our spectra as this
is notoriously difficult for \'echelle observations without 
simultaneous low-resolution spectroscopy or photometry. 
When we refer to ``flux'' we are therefore not considering absolute
flux values but the flux above (or below) the normalised continuum.
Finally, as the resolution is different for
each telescope, we have re-binned all the single-line spectra shown here
to a uniform resolution of $\lambda/\Delta\lambda =$ 30\,000, corresponding
to a velocity resolution of about 10~km~s$^{-1}$. The convolved profiles
used for the Doppler imaging in Sec.~\ref{sec:DI} are binned to a resolution
of $\lambda/\Delta\lambda =$ 40\,000.

\begin{figure*}
   \begin{center}
        \psfig{figure=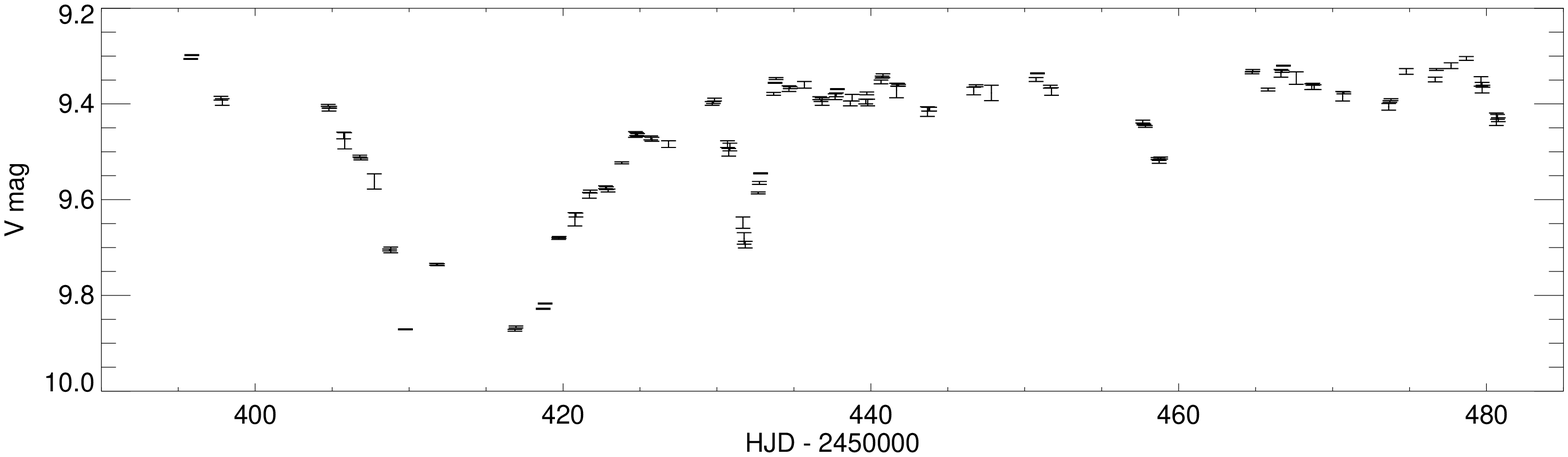,width=17.5cm} 
	\hbox{\psfig{figure=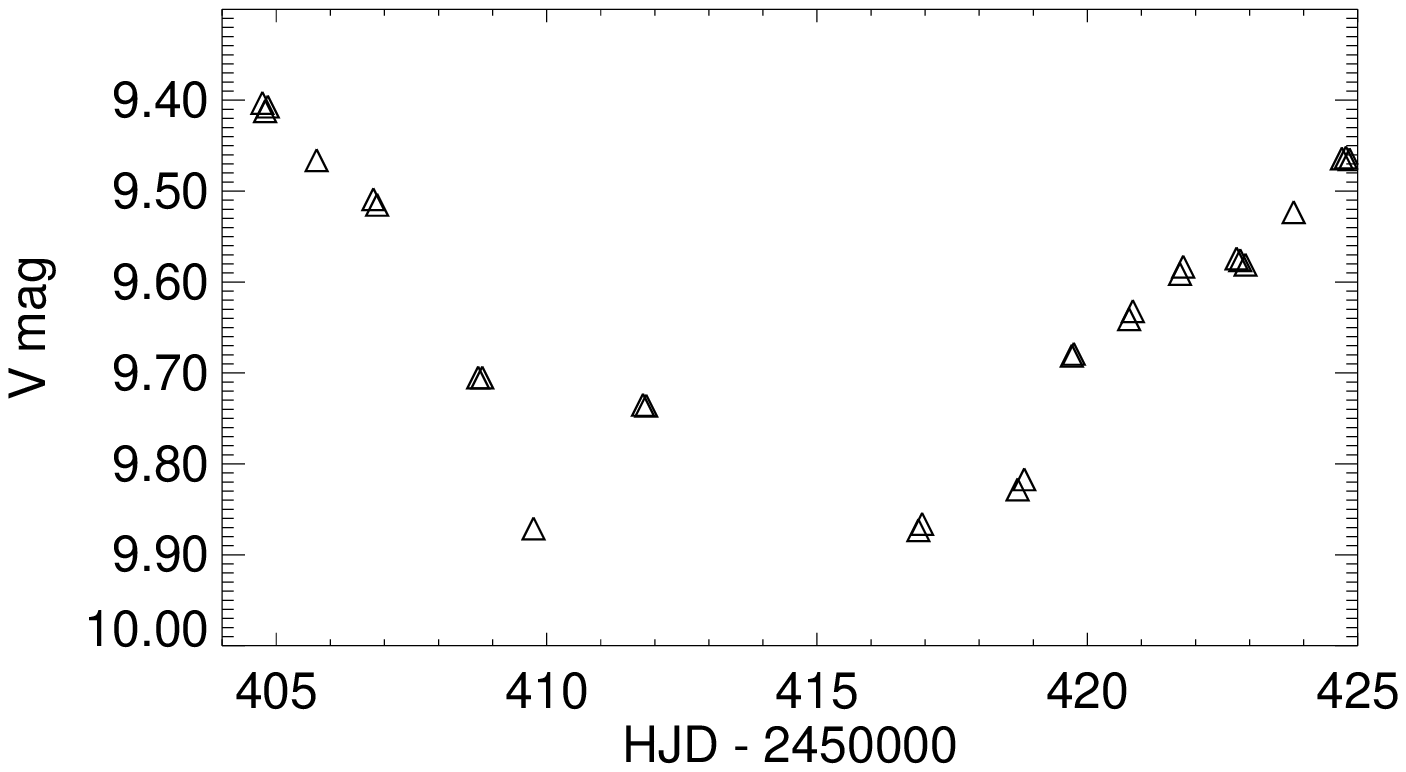,width=5.7cm}
	      \psfig{figure=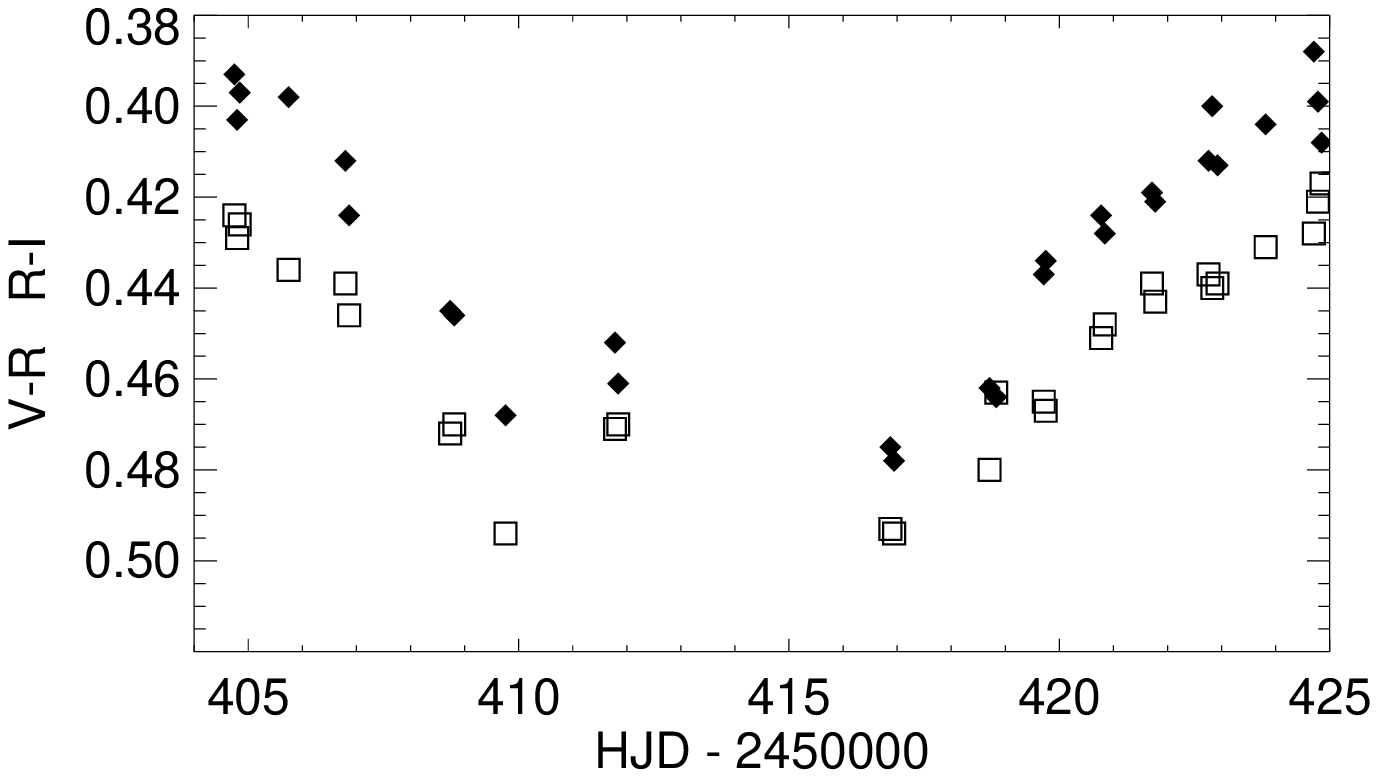,width=5.7cm}
	      \psfig{figure=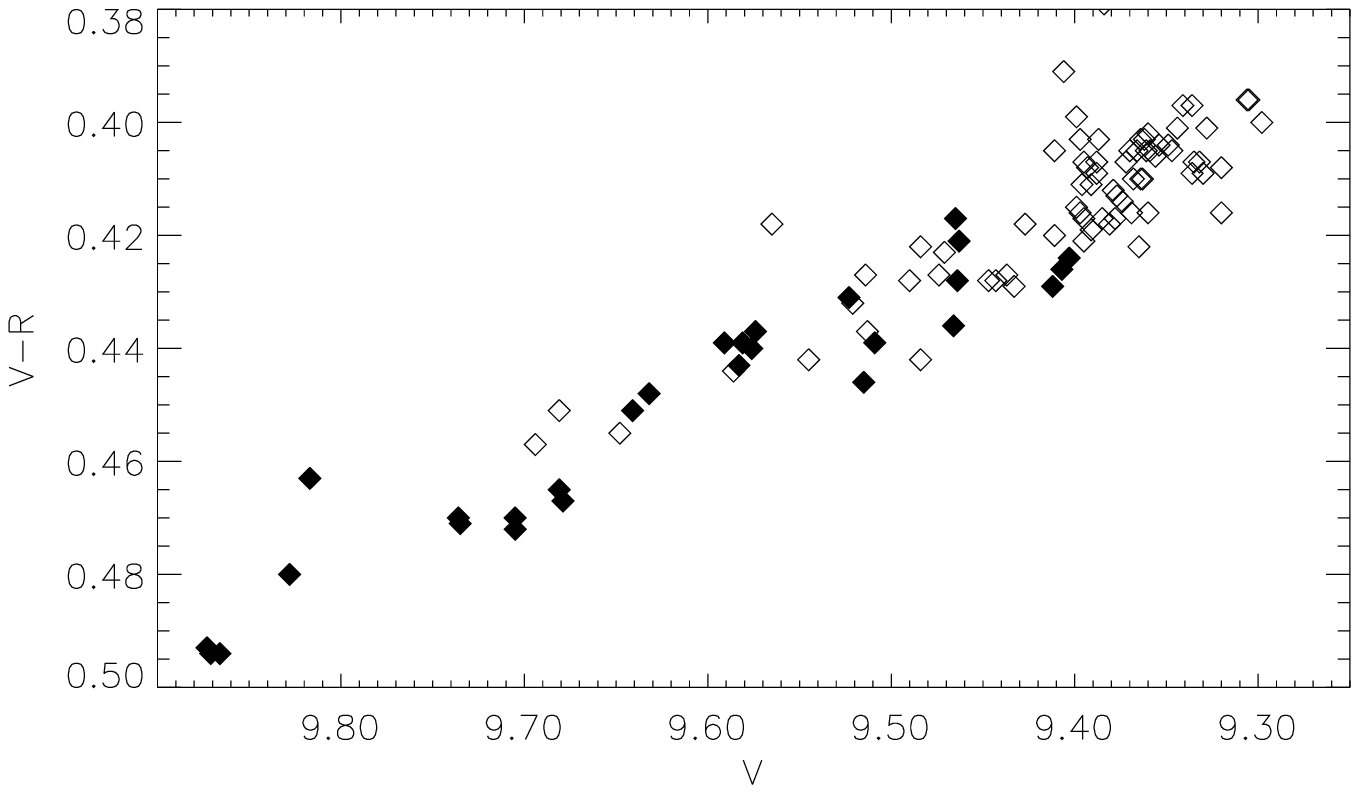,width=5.7cm}
	}
   \end{center}
\caption[]{The photometric lightcurve of SU Aur taken with the University 
of Vienna's APT in Fairbanks, Arizona. The $x$-axes show modified HJD, 
i.e.  HJD-2450000. The top graph shows the $V$-band lightcurve over the 
course of almost 90 days. 
The bottom graph on the left shows the enlarged lightcurve during the 
MUSICOS campaign, but also including the recovery phase of the obscuration 
event. The middle plot shows $V-R$ (lower curve, open squares) and 
$R-I$ (top curve, solid diamonds) during the campaign. The lower right-hand
plot shows $V-R$ vs.~$V$ during the 1996/1997 observing season. The filled
symbols are for data taken between HJD 2450405 and HJD 2450425.
        }
\label{fig:lightcurve}
\end{figure*}

%
%
\subsection{Photometry}
\label{sec:photom}
Unfortunately, there is no contemporaneous light curve with good 
phase coverage. SU~Aur was monitored during the 1996/1997 observing season
with the Automatic Photometric Telescope (APT) of the University of 
Vienna \cite{strassmeier97apt,strassmeier99apt}, but only a few data 
points could be secured during the spectroscopic campaign.
The light curve is shown in Fig.~\ref{fig:lightcurve}. Typically for SU~Aur, we
can detect no clear rotational modulation in the lightcurve. Instead, it shows
strong and erratic dimmings during which the star gets fainter by typically
about 0.5 mag in the V-band within 3 to 5 days.
The recovery is usually slower and can take 10 days or so (see 
\pcite{dewarf1998} and \pcite{nadalin2000} for a description of similar 
events). This behaviour is typical for SU Aur and
prompted \scite{herbst94} to classify SU~Aur as an UXOr variable. 
SU~Aur's dimmings are not quite as dramatic as those of other UXOrs 
that easily exceed differences of one magnitude in the $V$ band.

One such dimming event was observed starting from about HJD 2450408.5 
during our spectroscopic monitoring campaign when the star's 
V-band magnitude decreased by more than 0.5 over the
course of about 3 days. Unfortunately, the weather at the APT was such that only one
night between HJD 2450410 and 2450416 was clear. At HJD 2450417, the star was
again dimmer by 0.6 mag (in V) than ``usual'', but due to our lack of data
we can not say whether it recovered
between HJD 2450410 and 2450417. In agreement with what has been observed for
other UXOrs, SU~Aur becomes redder as it gets fainter (see lower right-hand
plot in Fig.~\ref{fig:lightcurve}). However, during the
1996/1997 observing season, it does not show the ``colour reversal'' typical 
for ETTS, where the stars become bluer as they 
enter strong brightness minima (see e.g.~\pcite{eaton95}).
This could be due to the fact that the dimmings that we observed were not very
extreme. As it is the disk that is believed to be responsible for the bluer colours 
of the systems once the star itself has been sufficiently occulted, the lack of 
colour reversal could also be due to SU~Aur's disk being unusually faint.

%

\section{Period determinations}
\label{sec:peri}
Photometric period determinations are difficult as SU~Aur's brightness
varies erratically (see Fig.~\ref{fig:lightcurve}).
Photometric period measurements range from as short as
1.7~d \cite{nadalin2000} and 1.55~d to 2.73~d
(Herbst et al.\,1987) up to 3.4~d (Bouvier et al.\,1993).
For SU~Aur, spectral line-profile variations seem to offer
a better chance of tracking the star's rotation, though
one has to bear in mind that a line such as H$\alpha$
is formed at various heights in the stellar atmosphere and under
a plethora of different conditions. The identification of
Balmer-line variations with the surface rotation of the star
therefore remains somewhat uncertain.

Several period measurements have been obtained
from analyses of the Balmer lines and of the He~D$_3$ line, all of
them indicating periods around 3~d
\cite{giampapa93,johns95suaur,petrov96}. \scite{johns95suaur}
furthermore found that the behaviour of the red wing of
H$\beta$ and the blue wings of H$\alpha$ and H$\beta$ were 
anticorrelated. This was interpreted
as the signature of a magnetic infall and
wind, in what they called the ``eggbeater'' model (see Sec.~\ref{sec:intro}
and \pcite{johns95suaur}) which has so far proven remarkably successful
at explaining a range of accretion and wind phenomena.

The measurements referred to above were taken from single-site observations. 
This introduces problems as the life times of the rotationally 
modulated features seem to be short, perhaps due to unsteady accretion.
Indeed, \scite{smith1999tts} concluded that, for SU~Aurigae, the
dominant variability time scale for H$\alpha$ is of the order of or
shorter than 1 hour.

In the following we discuss the line profile changes and associated
periods for H$\alpha$, H$\beta$, Na~D and He~{\sc i} (587.6~nm). 
Starting from 2-D periodograms 
we sum the fluxes in those velocity bins that show strong power peaks
and use three different methods to search for periods. 
Note that adding the fluxes degrades the frequency resolution somewhat, 
though we took care to only sum velocity bands with very similiar appearance 
on the original periodograms. The main results of our period search are 
presented in Tab.~\ref{tab:periods} and the period-search methods 
are outlined in the following section.

%
%
\subsection{Period-search methods}
\label{sec:period}
The three methods used to search for periods are the Lomb-Scargle
algorithm, the ``clean'' algorithm and a technique based on the
minimisation of information entropy.
The Lomb-Scargle algorithm is well suited to irregularely spaced
data and is described in \scite{horne86period}
and \scite{press92}. When applied to the summed fluxes, the false-alarm
probability (FAP) that gives an indication of the likelihood that 
a given power peak is due to random noise was calculated using a 
Monte-Carlo bootstrap method. For this we ran 1000 trials where we 
randomly selected fluxes from our original data set while leaving
the number of data points and location (i.e. observing time) unchanged.
We then calculated the periodogram for each synthetic data set, noting the
maximum power for each trial. We plotted the probability for a
given power value (in the power ranges obtained in the simulations)
and extrapolated this to predict the probability of the actual
power at the period found for the original data set. In
each case the highest values of the power found for the 1000 synthetic data
sets were less than half of the power peaks in the observed data.

\begin{figure}
\centerline{
        \psfig{figure=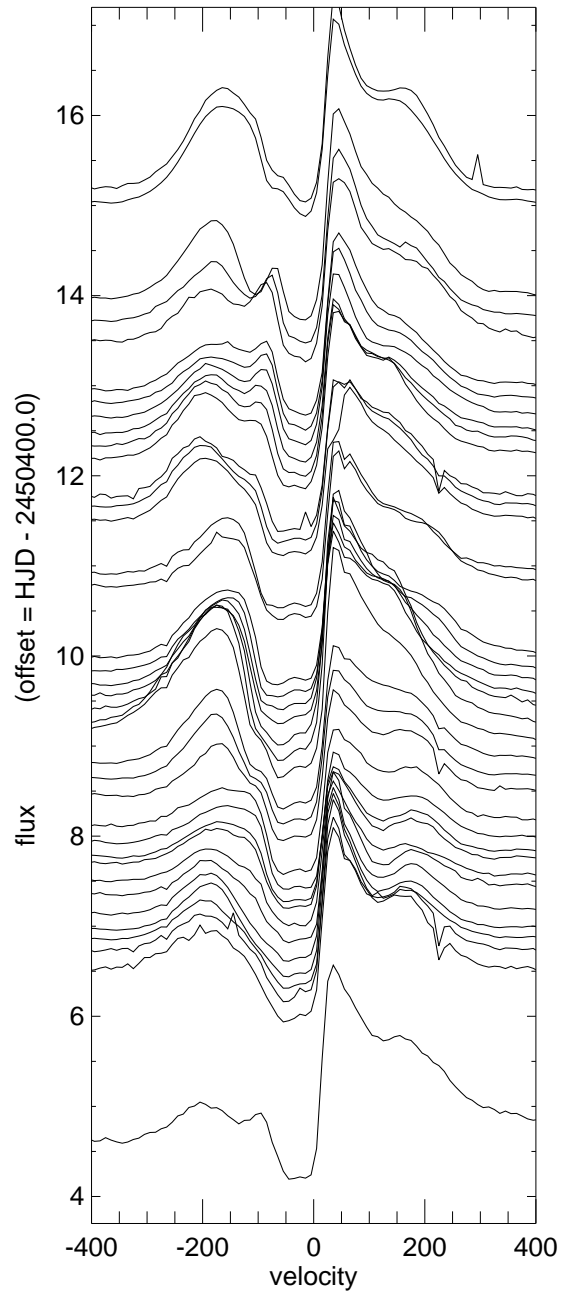,width=8.3cm}
        }
\caption[]{Stacked H$\alpha$ profiles. For clarity, we have grouped
together profiles that have been observed within 4 hours of each other.
The continuum flux has been
normalised and the y-axis offsets of the profiles
corresponds to the modified HJD of the exposure times, i.e. (HJD - 2450400).
        }
\label{fig:Ha_stack}
\end{figure}

Once a period has been determined, one obviously seeks to obtain
an error estimate for this period. Most formulae for calculating
the errors in the frequency (such as e.g.~\pcite{montgomery99,horne86period})
assume that the signal to be subtracted is sinusoidal. It turns out
that this is not a good assumption for this data set. We therefore adopt
a very conservative estimate and
list the range of the periods that lie within the FWHM of the power peaks.
The nominal error on the period is more than one order of magnitude lower
than this.

The second method that we used is the `clean''
algorithm \cite{roberts87clean} as implemented in
the programme {\sc period} \cite{dhillon97} distributed through
{\sc starlink}. The periodograms are shown in Fig.~\ref{fig:pers} and
the uncertainty
range in the period is listed in column 6 of Tab.~\ref{tab:periods}.
The implementation of the ``clean algorithm'' that we used also
calculates the FAP using the bootstrap method outlined above, however,
it only gives numerical values in excess of about $10^{-5}$, setting
all other values automatically to zero. In our case, all FAPs were nominally
``zero'', i.e. below $10^{-5}$, and are therefore not listed in Tab.~\ref{tab:periods}.

Prompted by the non-sinusoidal shape of the He~{\sc i} (587.6nm) equivalent-width
variations (see below), we decided to explore a third method for period
searches which relaxes the fundamental assumption of sinusoidal variations,
namely phase dispersion minimisation (e.g. \nocite{stellingwerf78}
Stellingwerf 1978).  For our particular purposes we use the information
entropy minimisation technique  of \scite{cincotta95}. In this scheme,
several periods are tried and for each one the data are phased up on the
given period.  The data are normalized in amplitude and phase from 0 to 1.
The idea is to measure how much of this normalized area is filled with data
points.  Random signals will fill the area with data points while periodic
signals will define a curve. The degree of space filling is quantified by the
information entropy and periodic signals produce a minimum in the entropy at
the appropriate period.  In this method, the user chooses the period range to
search, the number of trial periods to explore, and the number of squares
to divide the total normalized area (phase $\times$ amplitude) into.

Typically we searched periods ranging from 0.5 to 10
days at 2000 evenly spaced points and we found that 9 squares produced 
the best results. The period uncertainty was again taken from the 
width of the minima and the significance of the period estimated
using a Monte Carlo simulation of the data. The results from this method
are listed in the last three columns of Tab.~\ref{tab:periods}. 

\begin{figure}
  \begin{tabular}{lr}
        \psfig{figure=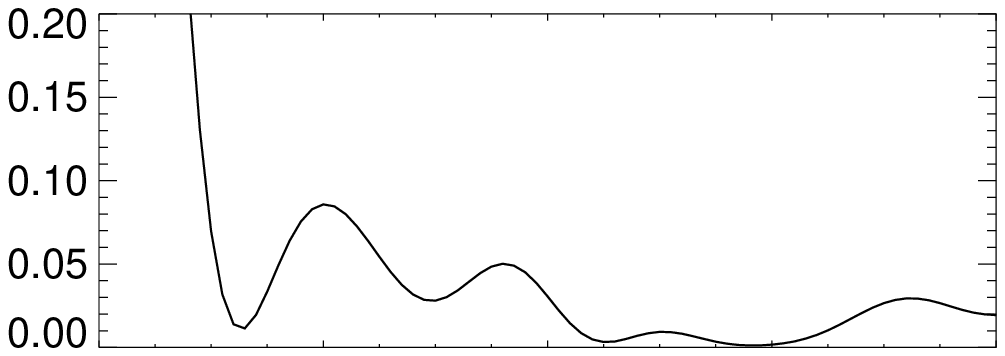,width=1.5cm,height=7.1cm,angle=90} &
        \psfig{figure=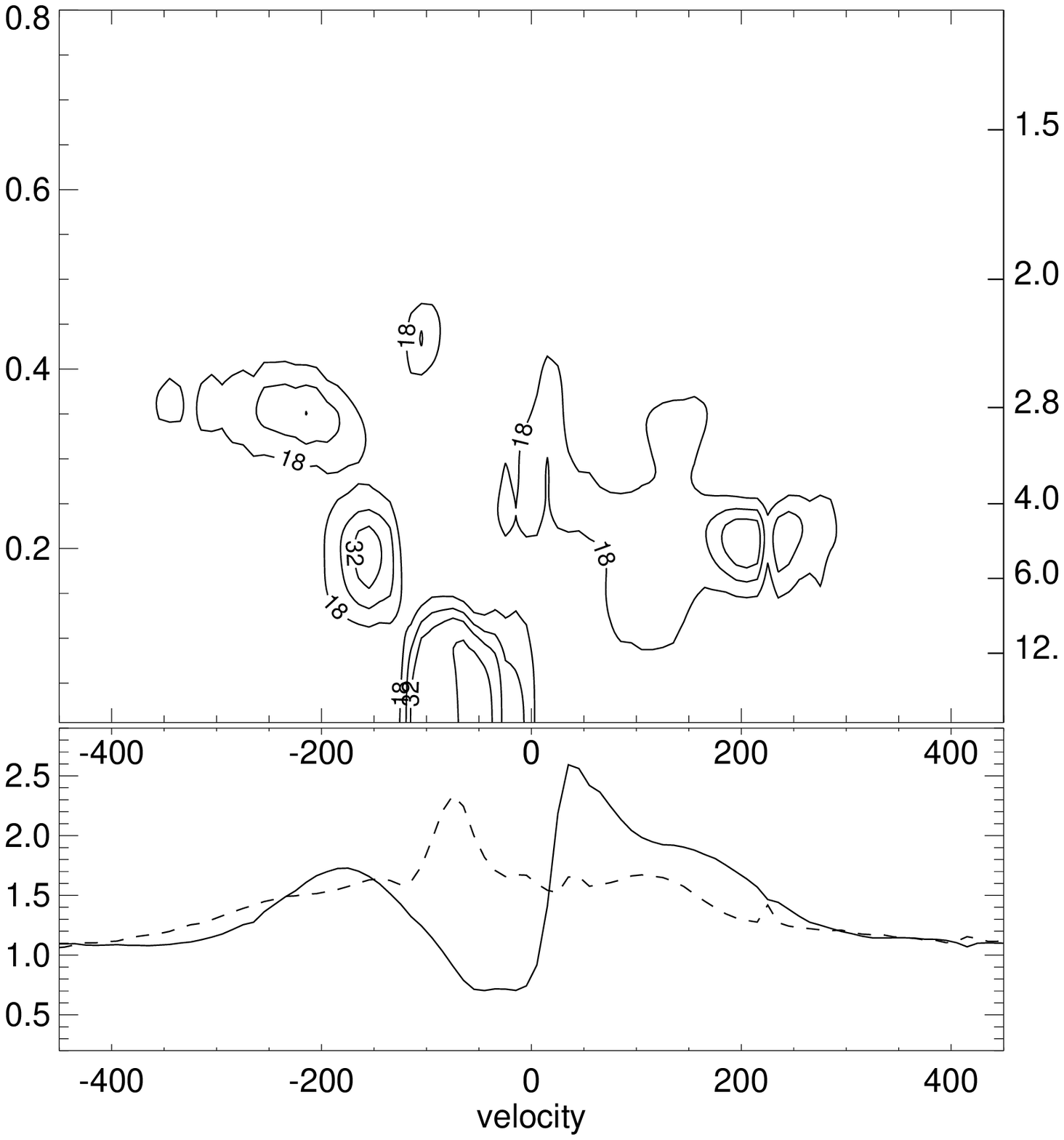,width=6.8cm}
  \end{tabular}
\caption[]{Top, right-hand side: The normalised Lomb-Scargle periodogram
of the H$\alpha$ profiles of SU~Aur. The left-hand axis shows the frequency
(in d$^{-1}$), the right-hand axis the period (in days). The contours
are normalised powers of 18, 26, 32 and 40. The lowest contours
correspond to a false-alarm probability (FAP) of about $10^{-6}$.
To the left of the periodogram, we have plotted the window function for
H$\alpha$. It has been scaled so that the power at
zero frequency is unity. Underneath the periodogram
the normalised mean (solid line) and variance profile (dashed line) are shown. The normalised
variance profile has been multiplied by a factor of 4 and offset by 1. It was calculated
according to $(\sum (f_i-\bar{f})^2 /(n-1) )^{0.5} / \bar{f}$ \cite{johns95tts},
where $\bar{f}$ is the mean profile.
}
\label{fig:Ha_period}
\end{figure}

%
%
\subsection{H\boldmath $\alpha$}
\label{sec:Halpha}
SU~Aur shows only H$\alpha$ in emission. The higher Balmer lines 
are in absorption, though they show varying degrees of filling in.
H$\alpha$ is a relatively ``typical'' wind profile with 
a marked blue-shifted absorption component that
tends to reach below the continuum. It is highly variable on 
a number of time scales as illustrated in 
Fig.~\ref{fig:Ha_stack}. Especially noteable are the redward-moving
features that appear at a velocity of about $-100$~km~s$^{-1}$
around HJD~2450408 and HJD~2450412 (see \scite{oliveira2000} 
for a discussion) and the strong enhancement of the blue emission 
wing at approximately HJD~2450409 when the star was at its faintest
during the campaign (see Sec.~\ref{sec:discussion}). 

The main plot in Fig.~\ref{fig:Ha_period} shows the Lomb-Scargle periodogram 
(see \pcite{horne86period,press92}) for H$\alpha$. The 
false-alarm probability (FAP) has been calculated for each velocity 
bin according to \scite{press92}. The FAP gives an estimate of how likely it 
is that a peak of a given height is due to random noise, a 
low value indicates a significant signal. The FAP is 
below $10^{-6}$ for all contours drawn in Fig.~\ref{fig:Ha_period}.
We note that the formula given by \scite{press92} 
can underestimate the FAP considerably, in particular when the 
data points show strong clumping. A more accurate method 
to estimate the FAP is to run a bootstrap Monte Carlo simulation, 
as we have done for selected velocity bins (see Tab.~\ref{tab:periods}). 

Along the frequency axis of each periodogram we have plotted 
the window function (as obtained from a discrete Fourier transform, 
since it is not possible to use the Lomb-Scargle analysis for windowed data). 
Underneath the periodogram we have sketched the mean profile (solid line)
and the normalised variance profile (dashed line). 
Two periodicities at around 3~d and 5~d
are apparent for H$\alpha$. The one at approximately 3~d is the one
previously identified by \scite{johns95suaur} and \scite{petrov96}.
While \scite{johns95suaur} found the 3-day period in a velocity bin
ranging from about --250 to --100~km~s$^{-1}$, we find a period of
approximately 3~days further towards the blue, in a range from
--300 to --170~km~s$^{-1}$. This 3-day period seems due to a weakening
and strengthening of the outer blue wing. At the same time, we find
a 5-day periodicity centred around a velocity of --150~km~s$^{-1}$ (where
\scite{johns95suaur} found the strongest evidence for a 3-day period)
and over for almost all of the red emission wing.

\begin{figure}
\centerline{
        \psfig{figure=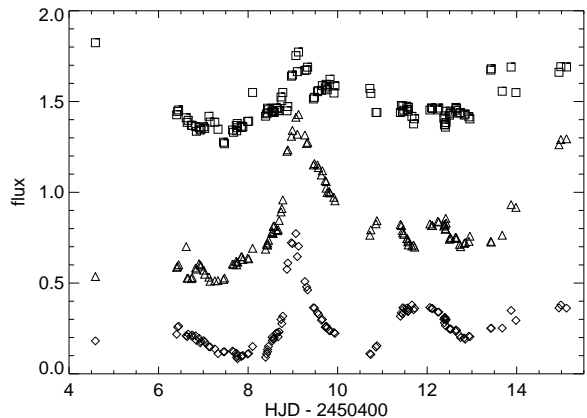,width=8.5cm}
        }	
\caption[]{The flux in H$\alpha$ above the (normalised) continuum in 
three different velocity bins. The diamonds
show the flux variations in the blue emission peak
where the 3-day period was found (--300 to --210~km~s$^{-1}$). The triangles
show the variations of a bin spanning --180 to --130~km~s$^{-1}$
where the periodogram suggests a 5-day period. The squares show the flux
variations in the red wing where the periodogram also suggests a
5-day period (170 to 250~km~s$^{-1}$). For clarity, the triangles and diamonds
have been offset by -0.7 and -1.0 respectively.
        }
\label{fig:Ha_bins}
\end{figure}

We suggest that the 5-day period is spurious, as the shape of the
window function allows for substantial leakage into the 5-day bin
(see Fig.~\ref{fig:Ha_period}). Furthermore, we do not find much
evidence for a 5-day period when the intensities of the H$\alpha$
profile are plotted against time in the regions where periodicities are suggested
in the periodogram (see Fig.~\ref{fig:Ha_bins}).
We found that exclusion of the profiles between HJD~2450409 and
HJD~2450410, that show the strong flux enhancement, leaves
the position of the power peaks largely intact, so that it is unlikely that the
flux enhancement alone is responsible for the 5-day period.

%
%
\begin{figure}
\centerline{
        \psfig{figure=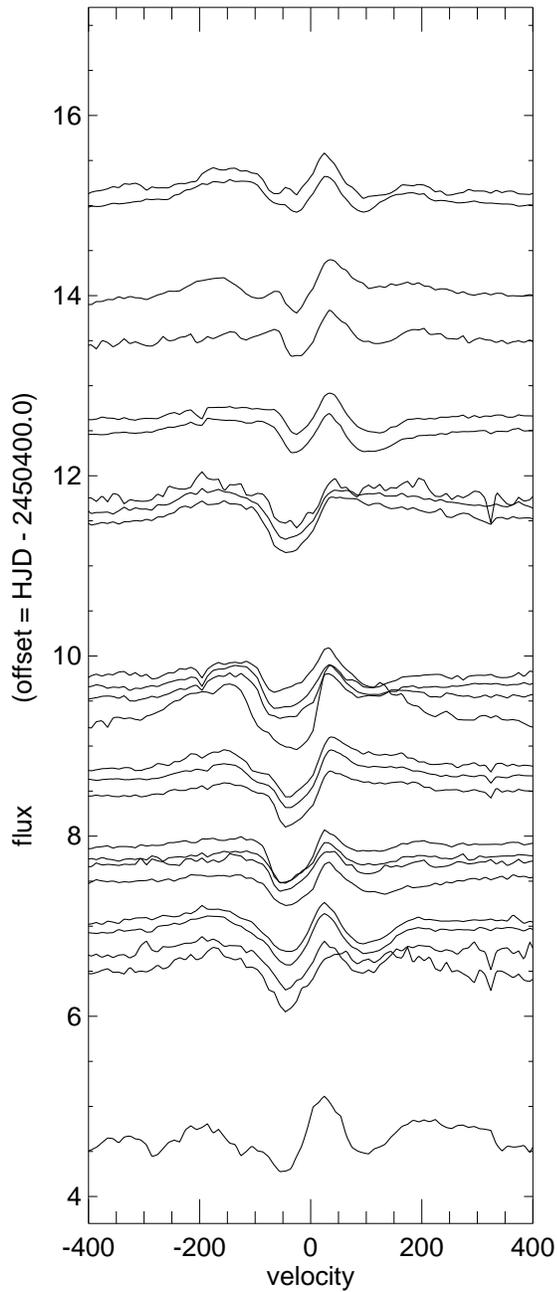,width=8.3cm}
        }
\caption[]{Stacked H$\beta$ profiles. For clarity, we have 
	 grouped together profiles in 4-hour bins.
	 The continuum flux has been
	 normalised and the y-axis offsets of the profiles
	 correspond to the observing times (HJD - 2450400).
	 The broadened profile of the Sun has been subtracted 
	 from all spectra.
        }
\label{fig:Hb_stack}
\end{figure}

\subsection{H\boldmath $\beta$}
Because of its different source function, H$\beta$ is
formed over a smaller range of temperatures and densities
than H$\alpha$. Theoretical studies
(see e.g. \pcite{hartmann94,calvet92infall})
as well as observations \cite{edwards94}, have shown that H$\beta$
reveals the signature of mass infall much more readily than H$\alpha$.
This is indeed also the case for SU~Aur.
                                                                                                                  
\begin{figure}
  \begin{tabular}{lr}
        \psfig{figure=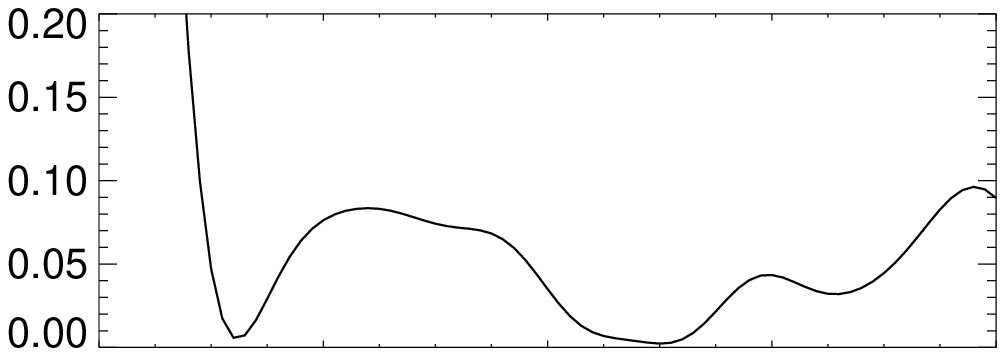,width=1.6cm,height=7.1cm,angle=90} &
        \psfig{figure=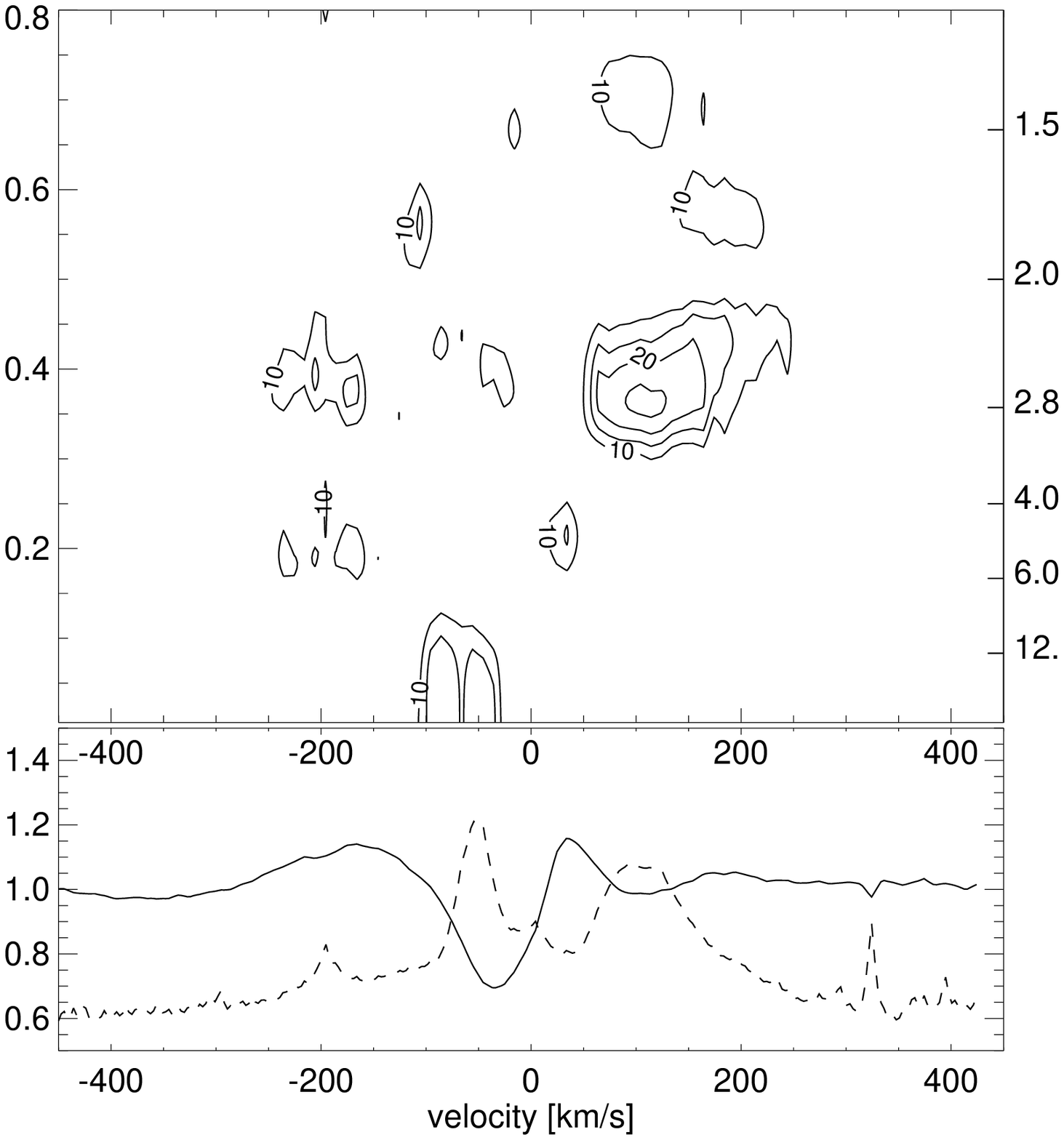,width=6.75cm}
  \end{tabular}
\caption[]{The normalised Lomb-Scargle periodogram of the H$\beta$
         profiles of SU~Aur (top figure, RHS). The y-axis show the 
	 frequency (on the left) and the period (on the right). Also shown 
	 is the window function for H$\beta$ (top left-hand plot). The FAP of
         the lowest contour is $2 \times 10^{-3}$, the FAP of the second
         contour (at a normalised power of 15) is $10^{-6}$.
	 The bottom figure shows the H$\beta$ mean profile (solid line) along with 
	 the variance profile (dashed line). In both cases the rotationally 
	 broadened solar H$\beta$ profile has been subtracted. The variance
	 profile has also been expanded by a factor of 4 and offset by 0.5. 
	 }
\label{fig:hb_period}
\end{figure}

Fig.~\ref{fig:Hb_stack} shows the stacked and grouped profiles of H$\beta$
during our run where we have subtracted the rotationally broadened
H$\beta$ profile of the Sun. We were able to observe H$\beta$
simultaneously from only three of the sites (CFHT, INT and OHP) and
therefore obtained rather poor phase coverage.
The periodogram for H$\beta$ is shown in Fig.~\ref{fig:hb_period}.
In agreement with \scite{johns95suaur}
we find that the period of approximately 3~days now occurs in the
red as well as in the blue part of the profile.
The periodicity in the red wing is not only apparent in the
periodogram, but is, in contrast to H$\alpha$, now also visible in
the stacked spectra (Fig.~\ref{fig:Hb_stack}). As already 
suggested by the lower and
less well-defined peaks in the periodogram, the periodicity in the blue
wing is harder to trace. 

\scite{johns95suaur} suggested that the red and blue wings of the H$\beta$ 
line are anti-correlated. Because of a lack of phase coverage and the 
presence of the small moving emission bumps in the blue profile wings,
it is hard to see this directly in Fig.~\ref{fig:Hb_stack}, though more detailed
analysis essentially confirms their original conjecture. We refer to 
\scite{oliveira2000suaur} for plots showing the anti-correlation between 
the red and blue wings.

%
%
\subsection{Na~D doublet}
\begin{figure}
  \begin{tabular}{lr}
	\psfig{figure=win_all.ps,width=1.5cm,height=7.1cm,angle=90} & 
        \psfig{figure=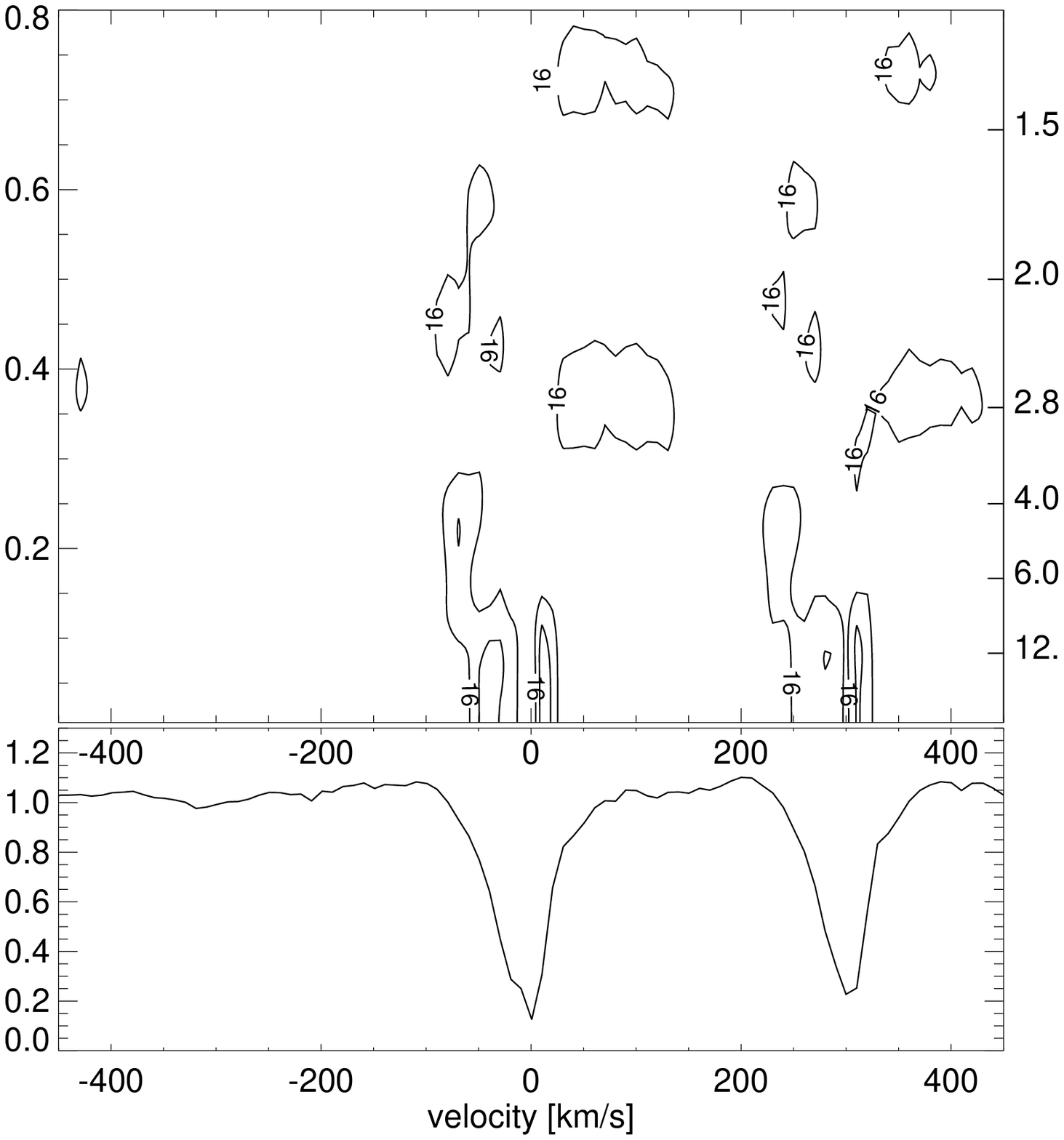,width=6.75cm}
  \end{tabular}
\caption[]{Top: The normalised Lomb-Scargle periodogram of the 
	 Na~D profiles of SU~Aur (RHS) and the window function of the 
data set (LHS). The lowest contour corresponds to a FAP of $7 \times 10^{-6}$.
Bottom: The mean and variance profiles for the Na~D lines.}
\label{fig:Na_period}
\end{figure}

Due to the interstellar absorption lines, light pollution
and lower intrinsic variability, the sodium doublet is slightly 
more difficult to analyse. However, they still show a number 
of similarities with the Balmer lines, including redward-moving 
features in their blue absorption wings (see \scite{oliveira2000} for
a detailed analysis). The periodogram for the Na~D lines is 
shown in Fig.~\ref{fig:Na_period}. We recover once again 
a period of slightly less than three days in the red parts
(from about 30 to 130~km~s$^{-1}$) of the profile. In the same velocity 
region, we can now also see a strong secondary period at about 1.4 days. 
The variability at the line centre is strong, but does not 
show any periodicities. As with H$\alpha$, there is also a 
(probably spurious) power peak close to 5 days, though now only in 
a narrow velocity range centered at --70~km~s$^{-1}$. 

%
%
\subsection{He~{\sc i} 587.6~nm}
\begin{figure}
\centerline{
        \psfig{figure=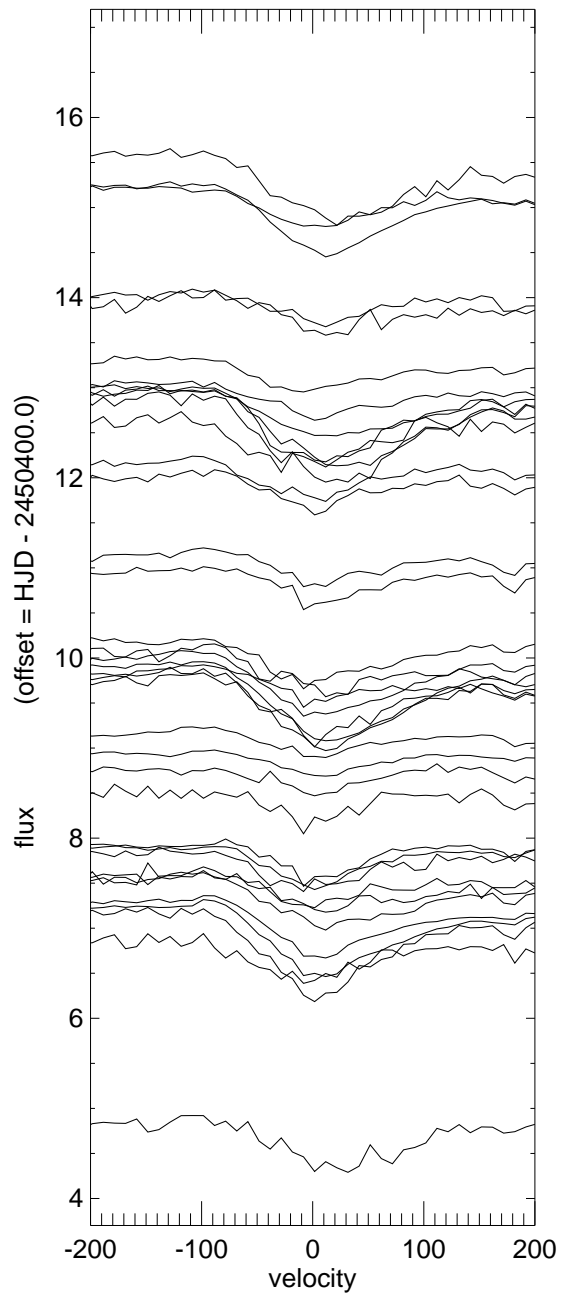,width=8.3cm}
        }
\caption[]{The stacked He~{\sc i} profiles, grouped together in 4-hour bins.
        As the line-profile changes in the helium line are not very
        large, they have been amplified by a factor of 4 for this plot.
        (Despite other appearances, the noise level for the He~{\sc i}
        profiles is in fact very similar to the noise level for Na~D).
        }
\label{fig:He_dyns}
\end{figure}

The stacked profiles of the He~{\sc i} 587.6~nm line are 
shown in Fig.~\ref{fig:He_dyns}. This plot clearly shows
the dramatic changes in line strength and the mostly asymmetric
shape of the line profile. \scite{oliveira2000suaur} argued that 
the He line can most easily be understood in terms of a superposition of two
absorption components, one at rest velocity, and one red-shifted
by about 80~km~s$^{-1}$. Note that while the minimum observed 
He equivalent width of about 50~m\AA \ is not too dissimilar to what 
is usually observed on G~dwarfs \cite{saar97helium}, the maxima
of up to 600~m\AA \ are quite exceptional. 
Solar observations show that the He~D$_3$ absorption increases in 
strength with increasing activity, but then starts to fill and, for the 
largest flares, turns into emission \cite{svestka72review}. 
\scite{bray64} reports equivalent widths of up to 400~m\AA\ in solar flares,  
but more recent observations seem to have focussed on large and 
off-limb or near-limb flares that then to show He~D$_3$ in emission. 

In their He-line-formation calculations \scite{andretta95} identify
two distinct formation regions in dwarf atmospheres, namely in the
upper chromosphere and in the much hotter plateau region.
In the density regimes considered by \scite{andretta95}
the ``turnover'' where the equivalent width starts to decrease due to
reversal is never reached and they estimate a maximum
equivalent width of 100 to 150~m\AA. This is expected to increase
for subgiants. Note that the strength of the He~D$_3$ line is
mainly electron density rather than temperature sensitive.
                                                                                                                  
\begin{figure}
  \begin{tabular}{lr}
        \psfig{figure=win_all.ps,width=1.5cm,height=7.1cm,angle=90} &
        \psfig{figure=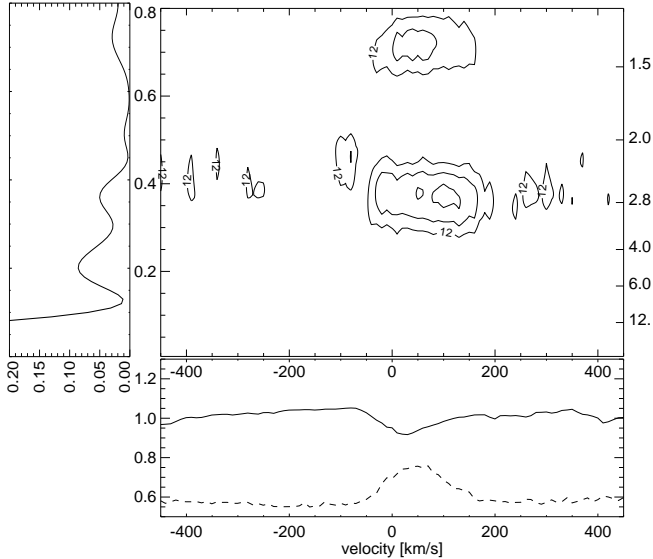,width=6.75cm}
  \end{tabular}
\caption[]{Top: The normalised Lomb-Scargle periodogram of the
        He~{\sc i} profiles of SU~Aur and the window function.
        Peaks higher than the lowest contour have a FAP of less than
        $4 \times 10^{-4}$. Bottom: The mean and variance profile of He~{\sc i} over the run.
	The variance profile (dashed line) has been amplified by a factor of 4 and offset
	by 0.5.}
\label{fig:He_period}
\end{figure}

\begin{figure}
\centerline{
        \psfig{figure=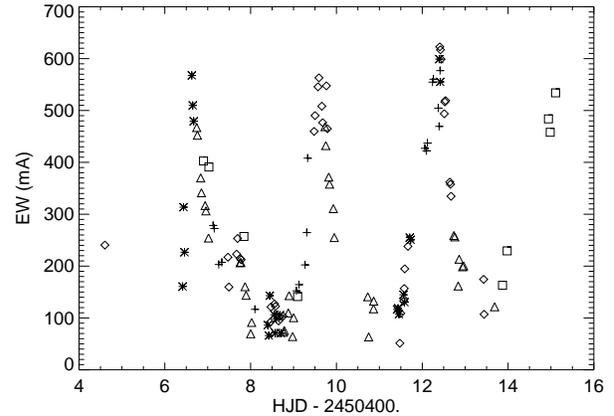,width=8.5cm}
        }
\caption[]{The He~{\sc i} equivalent width as a function of
        time. The different symbols show measurements from
        the different observatories, namely the BAO (plus signs),
        OHP (crosses), INT (diamonds), MDO (triangles) and CFHT (squares). }
\label{fig:he_ew}
\end{figure}

The 2-D Lomb-Scargle periodogram for He~{\sc D}$_3$ is shown in 
Fig.~\ref{fig:He_period}. It shows a very clear periodicity at 2.7~d with 
a strong secondary peak at 1.4~d for velocities ranging from about $-30$ to 
160~km~s$^{-1}$. \scite{petrov96} observed periodic or quasi-periodic radial-velocity
variations in the Balmer lines and in He~{\sc D}$_3$. We find that
there is indeed a trend for the radial velocity to be further redshifted
as the equivalent width of the He line increases. 
But because of the line asymmetry, the velocity shift of the line
centre is not very well determined.
The periodically varying equivalent widths of the absorption, however, 
are readily apparent, as illustrated in Fig.~\ref{fig:he_ew}. 
The variations are clearly non-sinusoidal
and display very strong and sharp peaks for the largest equivalent
widths. This is what first prompted us to use the phase-minimisation 
technique which indeed places somewhat narrower constraints on
the period of the He I variations (see Tab.~\ref{tab:periods}).

\begin{figure}
\centerline{
        \psfig{figure=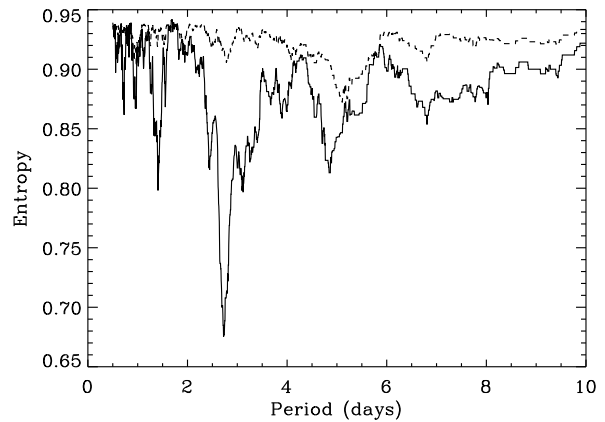,width=8.5cm,angle=90}
        }
\caption[]{
        The solid line shows the entropy as a function of period for
the He~{\sc i} equivalent width (see text). Also shown is the mean
entropy for 1000 Monte-Carlo trials where the time spacing has been
kept while the order of the equivalent width data points has been
randomized.
}
\label{fig:period_entropy}
\end{figure}

The period-dependence of the entropy is 
plotted for the He~{\sc i} line in Fig.~\ref{fig:period_entropy} and shows a 
very pronounced minimum at 2.7~d. The secondary period of about 1.4~d 
is also recovered, though it is less prominent than in the corresponding 
``clean''ed or Lomb-Scargle periodograms. The dashed line shows the mean 
entropy for 1000 independent Monte-Carlo trials. The 1000 individual
trials look very similar to the mean shown in the figure. For 
each trial, the entropy minimum was recorded. The distribution of
the 1000 trial minima peaks at an entropy of 0.8745 with a fairly
sharp cutoff above this value and a Gaussian tail extending to lower entropy
values. None of the trials produced a minimum lower than 0.843, while the
original He~{\sc i} data shows a minimum entropy of 0.675 at a period of 2.7
days. From a Gaussian fit to the peak and lower values of the trial distribution
we find that the 2.7-day period is significant at the $> 21\sigma$ level.

We find essentially the same periodicity in both components
of the He~{\sc i} line, though the shape of the periodogram 
is somewhat different in that the variability is stronger
and the secondary peak is more pronounced
in the main absorption line than in the redshifted component
(see Fig.~\ref{fig:pers}d). Note that \scite{oliveira2000suaur}
found that the variability of the redshifted absorption correlated
well with the red-shifted absorption of the NaD lines, but lagged
behind that of the helium rest-velocity component. 

%
%
\subsection{Comparison of the different lines and methods}
\begin{figure}
  \def\subfigtopskip{1pt}
  \def\subfigcapskip{1pt}
  \def\subfigbottomskip{1pt}
        \subfigure[]{\psfig{figure=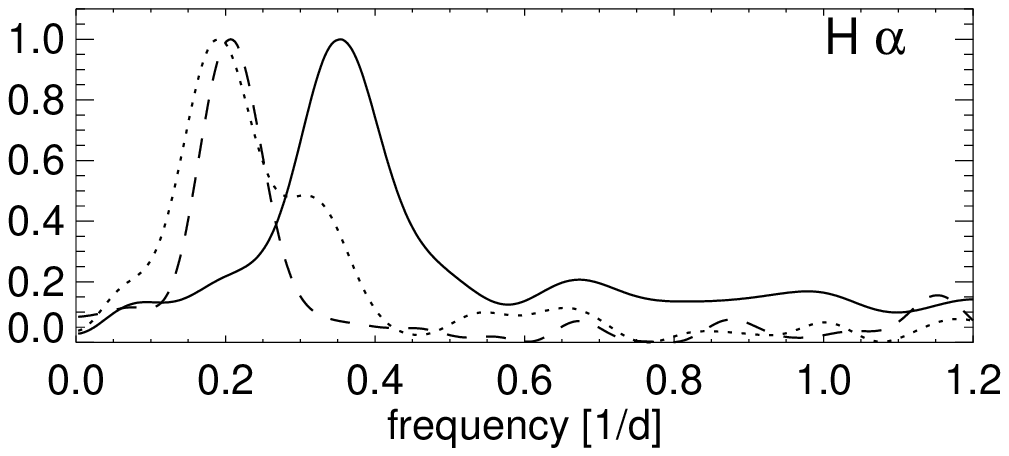,width=8.cm} }
        \subfigure[]{\psfig{figure=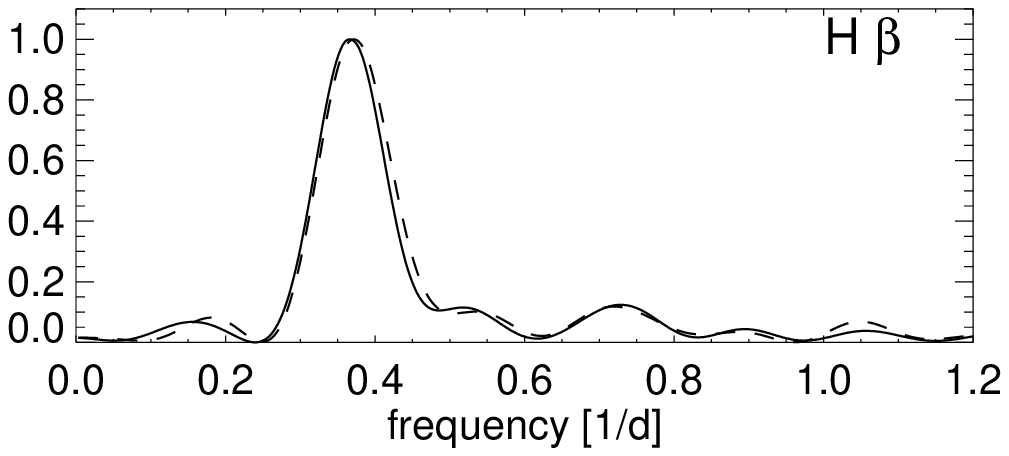,width=8.cm}}
        \subfigure[]{\psfig{figure=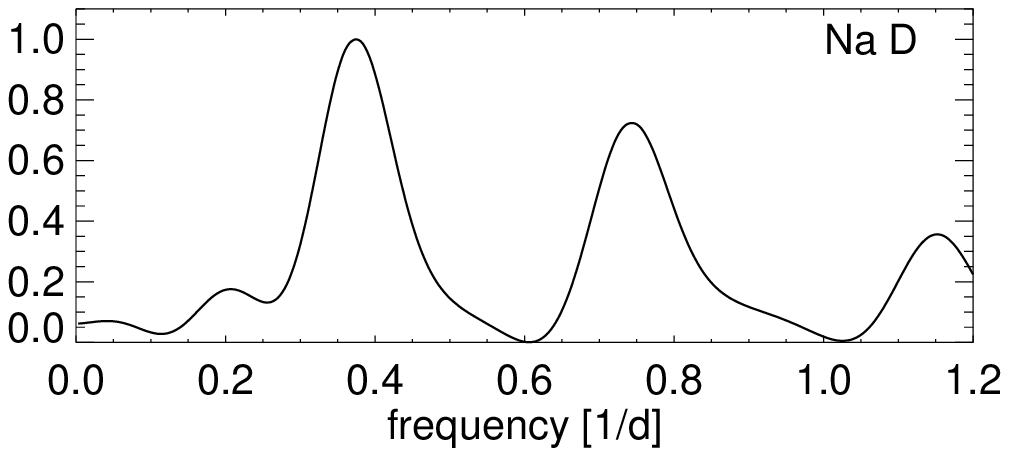,width=8.cm}}
        \subfigure[]{\psfig{figure=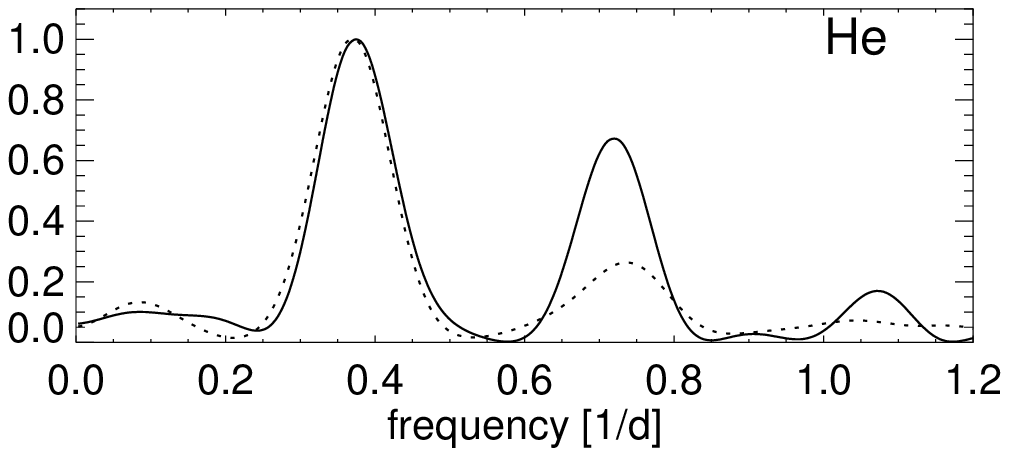,width=8.cm}}
        \caption{Top to bottom: The ``cleaned'' periodograms for H$\alpha$, H$\beta$, Na~D and He.
        The periodogram for H$\alpha$
        shows the power over several velocity ranges, namely $-300$ to $-210$~km~s$^{-1}$ (solid line), 
	$-180$ to $-130$~km~s$^{-1}$ (dotted line) and 170 to 250~km~s$^{-1}$ (dashed line). These
	bins correspond to the ones chosen for the flux plots in Fig.~\ref{fig:Ha_bins} and also 
	represent the bins with largest power in the 2-D Lomb-Scargle periodogram.
        The H$\beta$ periodogram is for the average flux between 60 and 185~km~s$^{-1}$; 
	taking a smaller bin width results in a very similar periodogram with the same main period.
	The Na~D periodogram is for flux between 50 and 70~km~s$^{-1}$. For He~{\sc i}, the solid
	line is for the average flux between 20 and 50~km~s$^{-1}$, while the dashed line is 
	for an average taken from 100 to 120~km~s$^{-1}$. 
	}
        \label{fig:pers}
\end{figure}

The results from our period searches are summarised in Tab.~\ref{tab:periods}
and, for the ``clean'' algorithm, are also plotted in Fig.~\ref{fig:pers}. 
While there are small differences in the periods found for the different
lines with the periods recovered for the H$\alpha$ line being
the largest (2.8 and 2.9~d), the widths of the power peaks are such
that a single period is compatible with all our observations. We
favour the slightly shorter periods obtained in lines other than H$\alpha$,
mainly because of the large range of regimes where H$\alpha$ is formed,
some of which may be well away from the stellar surface. There may hence
not be sufficient magnetic "locking" to force co-rotation and a single 
well-defined period. 

\begin{table*}
\caption[]{Table listing the periods found in the normalised Lomb-Scargle
periodograms (column 3), from the flux variations using the ``clean''
algorithm (column 6) and the entropy minimisation technique (column 8).
The 2nd column gives the range of velocity bins that have been co-added 
for the periodogram. 
For the periods calculated using the Lomb-Scargle algorithm, we have listed
the false-alarm probability (FAP) according to a Monte-Carlo bootstrap 
method. The FAP for the ``cleaned'' periods is below 10$^{-5}$ for all 
periods (and has not been listed, see text). The ranges listed in 
columns 4 and 7 give a very conservative estimate on the period, assuming
that the width of the power-peak at half maximum can be used as an indication 
of the possible spread in the period.
For the entropy minimisation the
error on the period as well as its significance are listed in columns 
9 and 10. The last row shows the period for the
equivalent width variations of the He~{\sc i} line.
        }
   \begin{center}
	\begin{tabular}{lccccccccc}
Line       & Velocity           & P$_{\rm LS}$ & Range  & FAP & P$_{\rm Cl}$ & Range & P$_{\rm E}$ & Error & Significance \\
           & [km/s]             & [d]         &  [d]    &     &  [d]         & [d]    & [d]     & [d]   & [$\sigma$] \\
\ \ (1)	   & (2)		& (3)	      & (4)	& (5) & (6)	     & (7)    & (8)	& (9)	& (10)	\\
\hline
H$\alpha$ & --300\ldots --210 & 2.8 & 2.4\ldots 3.5 & $2 \times 10^{-10}$ & 2.9 & 2.4\ldots 3.6 & 2.9   & 0.2   & 7 \\
H$\alpha$ & --180\ldots --130 & 5.1 & 3.6\ldots 8.9 & $5 \times 10^{-14}$ & 5.3 & 3.9\ldots 7.7 & $\sim 6$ &    & \\
H$\alpha$ & 170\ldots 250     & 4.7 & 3.9\ldots 6.8 & $2 \times 10^{-15}$ & 4.8 & 3.9\ldots 6.2 & -- & -- & --\\
H$\beta$  & 60\ldots 185      & 2.7 & 2.2\ldots 3.2 & $3 \times 10^{-12}$ & 2.6 & 2.4\ldots 3.1 & 2.84  & 0.22  & 8  \\
He~{\sc i} & 20\ldots 50      & 2.6 & 2.3\ldots 3.2 & $2 \times 10^{-14}$ & 2.7 & 2.3\ldots 3.1 & -- & -- & -- \\
He~{\sc i} & 100\ldots 120    & 2.7 & 2.3\ldots 3.3 & $4 \times 10^{-14}$ & 2.7 & 2.3\ldots 3.2 & -- & -- & -- \\
Na~D      & 50\ldots 70       & 2.6 & 2.3\ldots 3.2 & $10^{-10}$ 	  & 2.7 & 2.3\ldots 3.2 & -- & -- & -- \\
He~{\sc i} & - \ - \ -        & 2.7 & 2.3\ldots 3.3 & $3 \times 10^{-16}$ & 2.7 & 2.3\ldots 3.2 & 2.72 & 0.18 & 21 \\
\hline
	\end{tabular}
   \end{center}
\label{tab:periods}
\end{table*}

The widths of the power peaks
do not change noticeably when isolated measurements such as the
first INT exposure or the last three CFHT exposures are excluded.
We suspect that this is due to several features of
our data and the star, the main ones being that we have only covered
two to three rotation periods and that the variations are generally
non-sinusoidal. Note that the entropy minimisation technique that
does not assume sinusoidal variations, shows a much narrower peak
than either the Lomb-Scargle or the ``clean'' method.
Furthermore, we find that the window function allows for leakage
of power at a frequency
of about 0.37~d$^{-1}$, or a period of 2.6~d, which is uncomfortably
close to our main period. Inspection of the stacked
profiles also shows that while profiles that are separated by approximately
one period do show similarities in the velocity range of the power peaks,
they are certainly subject to strong variations between
one period and the next. We are hence
dealing with a star with very strong intrinsic variability.
One also has to keep in mind that the strong dimming of SU~Aur may
have introduced an additional trend in the fluxes. One of the
characteristics of ETTS is the erratic dimming with simultaneous increases
in the H$\alpha$ relative flux. Unfortunately, our photometric time
coverage is too sparse to allow cross-correlations between the photometric fluxes
and the spectroscopic data.

%
%
\section{Photospheric lines and Doppler images}
\label{sec:photlines}
\subsection{Least-squares deconvolution}
One of the original aims of the MUSICOS campaign was to obtain Doppler 
images of SU Aur and to locate possible infall sites for matter accreting 
from the stellar disk.
The presence of spots on the stellar surface should produce line-profile 
deformations in the photospheric lines. However, in our 
case, the signal-to-noise ratio is not good enough to allow us to detect 
these deformations in individual photospheric lines. One possibility
is to combine the information contained in all the photospheric
lines in the observed spectral range so as to increase 
the S/N. We have done this using the technique of least-squares
deconvolution (LSD, see e.g.~\pcite{donati97zdi,barnes98}). 
As we have gathered data at five different
observatories with different setups, it was not possible to use the
same lines in the deconvolution for each observatory. 

\begin{figure}
	\psfig{figure=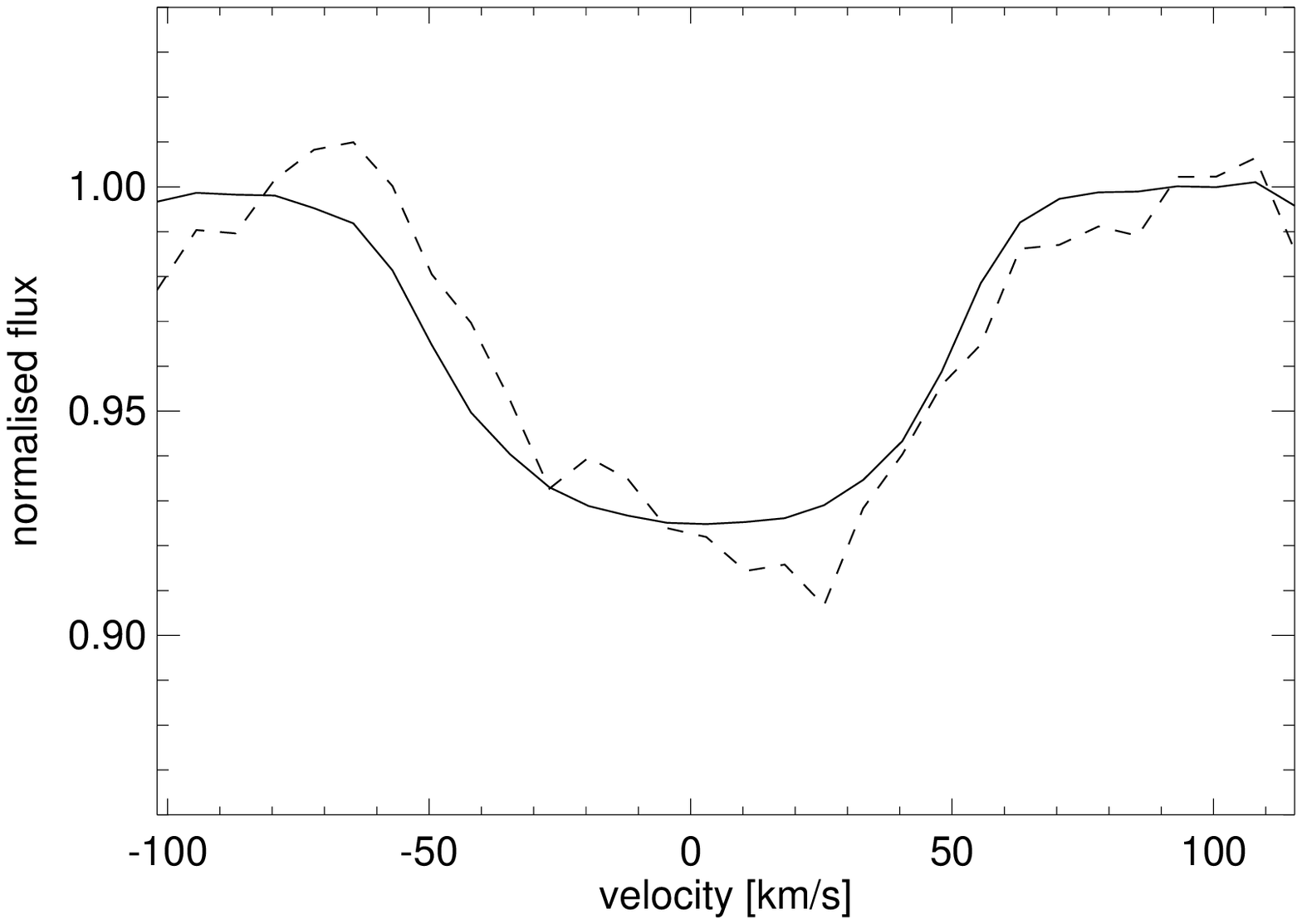,width=7.5cm}
        \psfig{figure=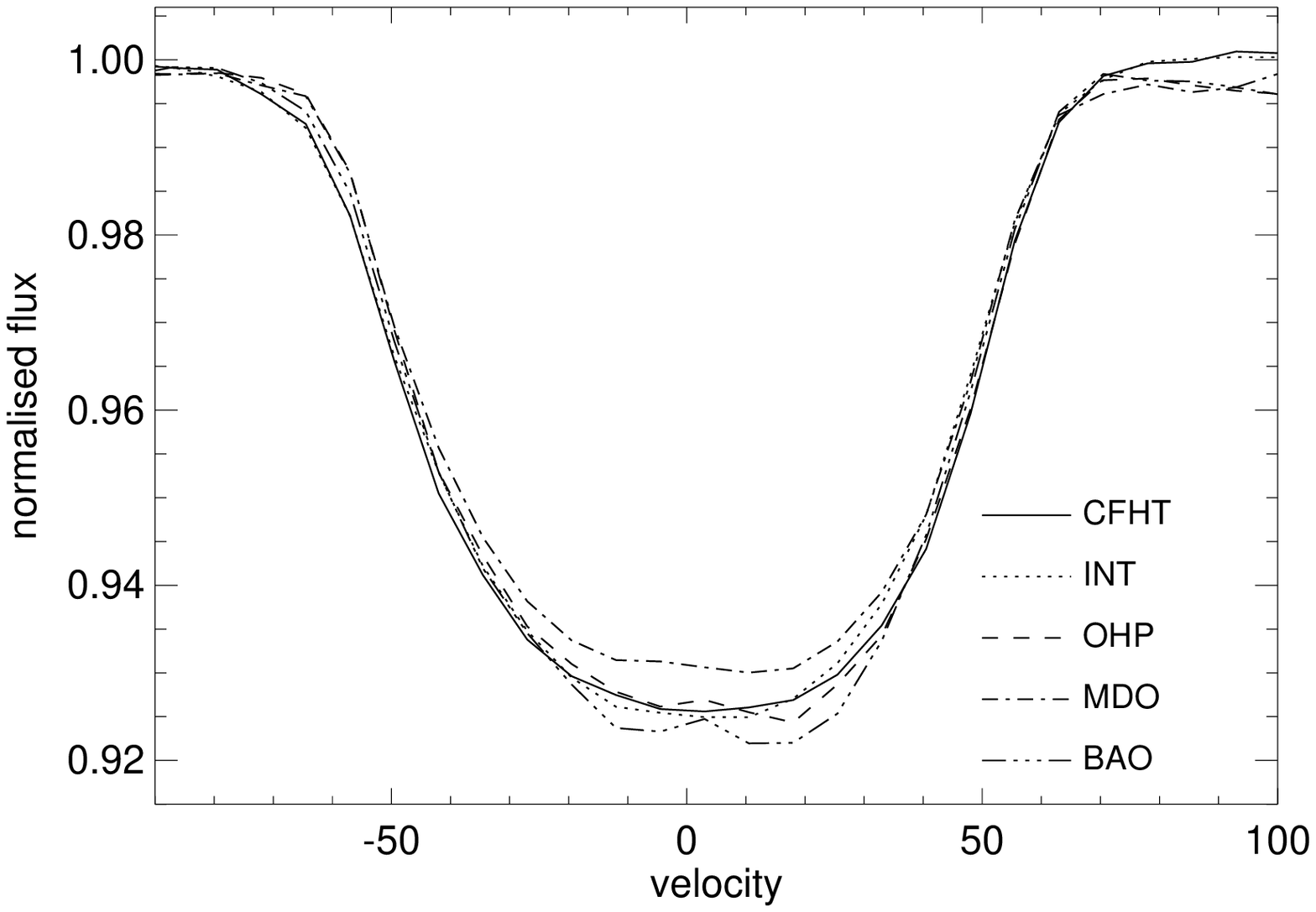,width=7.5cm}
	\psfig{figure=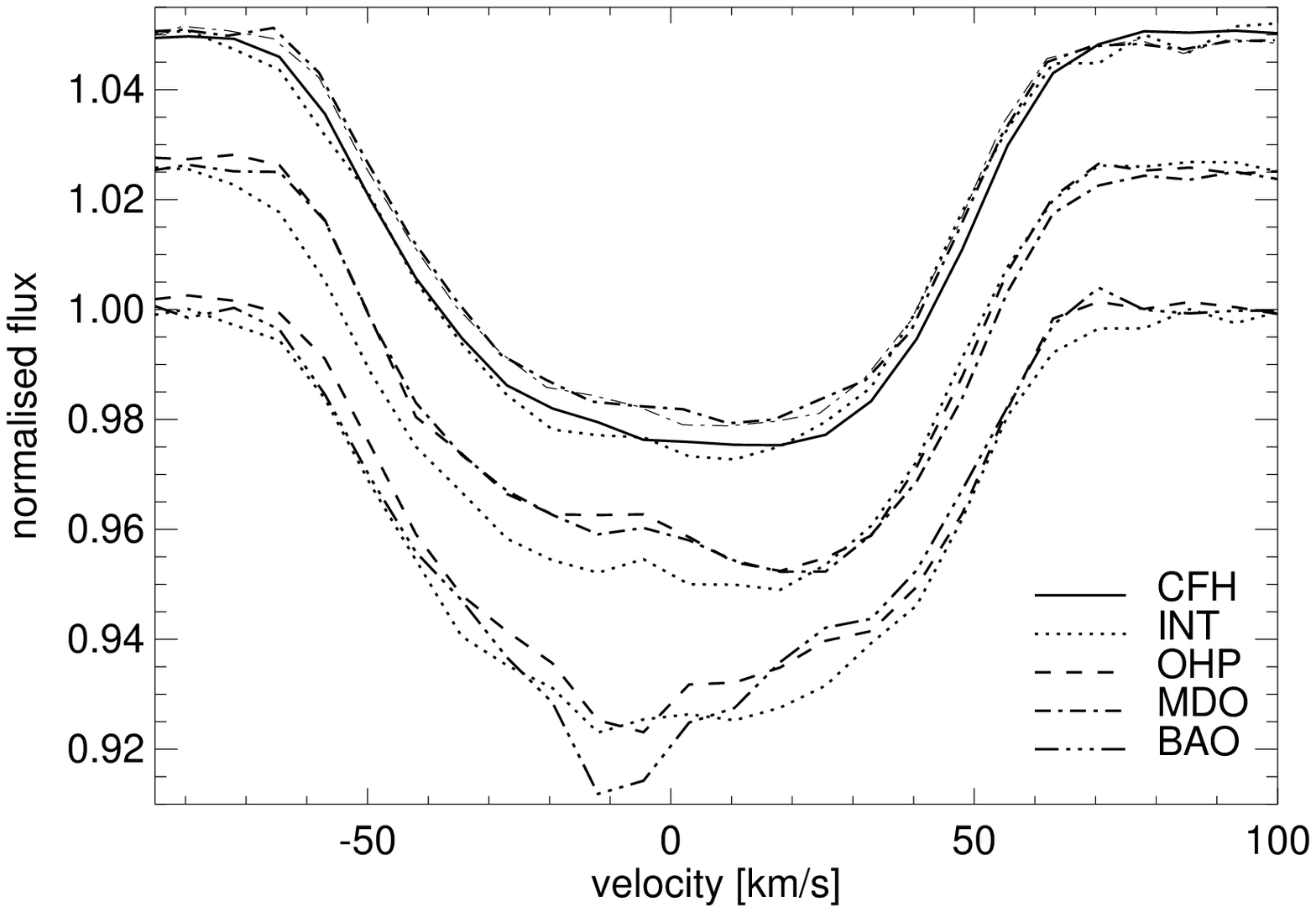,width=7.5cm}
\caption{Top figure:
Plots of different mean LSD profiles from the data taken at the 
CFHT. The solid line is the profile that has been obtained
by deconvolving all neutral lines from elements heavier than (and including)
calcium. The dashed line shows the mean profile of the four strongest 
unblended silicon lines. 
Middle figure: The rather different mean profiles at the 
5 observatories. Bottom figure: Comparison of LSD profiles taken at 
approximately the same time but at different observatories. 
The profile deformations agree reasonably well between the different
observatories, though there are some important discrepancies for some
observations. The uppermost profiles were observed around modified 
HJD (MHJD=HJD-2450400) of 12.4. The thick and thin dot-dashed profiles show 
two profiles from the MDO taken about 3 hours apart, corresponding to 
a phase difference of only 0.04. This phase difference is too small to 
show the motion of the surface features. The difference between the 
two profiles is probably due to noise. The middle profiles were observed between 
MHJD~8.7 and MHJD~8.8 and the bottom profiles between MHJD~7.7 and MHJD~7.9.
}
\label{fig:LSD_ion}
\end{figure}

One implicit assumption when using LSD profiles for Doppler images is that 
all lines used in the deconvolution should show a similar response to surface 
inhomogeneities. To test this, we generated LSD profiles with different 
line subsets, such as e.g. singly ionised lines, ``light'' elements, lines 
in the red or blue parts of the spectrum only or strong lines only. This 
was done for the CFHT data only, as it provided a large spectral range 
and the best S/N. 

Subsetting for different elements or ionisation stages is not
straightforward as neighbouring blends can easily skew the profiles.
The only element lighter than calcium that showed relatively 
unblended lines in our wavelength range was silicon and we 
were only able to use 4 lines. The dashed line in Fig.~\ref{fig:LSD_ion}a 
shows the profile for the 4 silicon lines for the CFHT. 
The solid line is the LSD profile for all elements heavier 
than and including calcium. Both profiles
are ``mean'' profiles where the six exposures with the 
highest S/N have been added. The mean silicon profile is very asymmetric 
with more absorption in the red part of the line. This is not due to 
the small number of lines that were used in the 
deconvolution or due to blending. In fact, all the strong silicon 
lines were inspected individually and showed the same asymmetric shape.  
\scite{catala99musicos} found very marked differences between 
different ionisation species and elements for AB~Aurigae, where 
an extra emission component was seen in the higher ionised species. 
Unfortunately, the singly-ionised lines in SU~Aur are too weak and 
too blended in order to carry out a similar analysis. 

As far as the line variability and the line-profile deformations 
are concerned, it appears that there is no difference between the 
various subsets or indeed the LSD profiles obtained using all available 
lines, so that using all lines in the spectrum should be a valid approach.
Unfortunately, the S/N level of the smaller subsets is too 
low to quantify the agreement (or indeed the disagreement) between them. 
We decided to err on the side of caution and exclude possible
artefacts due to the lighter elements and 
used only neutral lines of the heavier atoms for the Doppler images 
presented below\footnote{We took care to exclude a large-enough section
around the rejected lines to avoid artefacts through blended lines.}.

While least-squares deconvolution is necessary here 
to achieve a high enough signal-to-noise ratio, it introduces problems 
not usually encountered in multi-site Doppler imaging. This 
is due to the very different spectral ranges and the different
setups at the 5 sites. This is shown in the middle graph of Fig.~\ref{fig:LSD_ion} 
where the mean profiles of all 5 sites are plotted. It turns out the 
the mean profiles at the INT and CFHT agree reasonably well, but that 
the profiles at BAO and MDO look very different. Given that not all phases 
are covered equally at the different sites, small differences in the mean 
profile can be expected. Nevertheless, the differences that we observe 
are too large to be attributed to a different phase coverage. 
\begin{figure}
        \psfig{figure=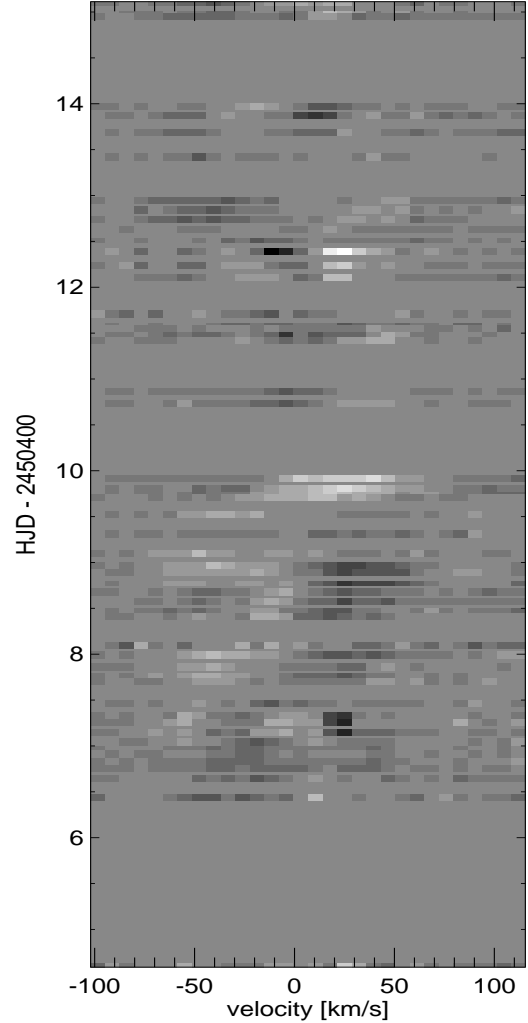,width=8.cm,height=14cm}
\caption{A stacked plot of the residuals of the deconvolved profiles.
        White is at a flux level of 1.008 and black at a flux level of 0.992.}
\label{fig:LSD_stack}
\end{figure}

\begin{table}
\caption[]{The number of lines used at the different observatories.
The first column indicates the observatory. The second column gives 
the number of different lines used for the convolution. The third 
column lists the number of line ``images'' that were actually 
used for the calculations. This number tends to be higher than 
the one given in column two as some lines were recorded twice on 
the chip. The last column gives the weighted central wavelength 
of the final profile. Note that we used the same line list for 
the CFHT and the INT.}
\centerline{\begin{tabular}{lrrr}
	\hline
        Observatory  & lines & images	& $\lambda_c$    \\ 
                     &      & 		&	[nm]    \\
        \hline
        BAO 	& 352  &  400 & 656.7  \\
        OHP 	& 1828 & 2160 & 526.7  \\
        INT 	& 950  & 1290 & 584.8  \\
        MDO 	& 311  & 433  & 601.4  \\
        CFHT 	& 1728 & 2660 & 563.3 \\	\hline
\end{tabular}}
\label{tab:lsd}
\end{table}
We suggest that the differences are mostly due to the 
different setups used at each observatory for the following
reasons.  The number of lines registered for each spectrograph setup
is different at each observatory. Therefore, even 
using the same starting list of lines can produce different mean 
profiles. This is because some lines are registered twice at some
spectrographs or fall into gaps in the spectral coverage. 
The number of lines included in the LSD line list is given in
column 2 of Tab.~\ref{tab:lsd} along with the actual number 
of line images (column 3) and the ``central'' wavelength of the
deconvolved profile (column 4).
For the deconvolution, each line is weighted according to its
line depth, giving least weight to the weakest lines.
Furthermore, lines on the edge of the chip where the continuum 
flux level (and hence the S/N) is low are also given reduced weight. This latter 
weighting differs for each observatory. We tested the influence of this 
weighting for data from the CFHT and the INT. As the overlap between 
the spectral ranges at these telescopes is very high, 
we could use the same line list to produce two different mean profiles
at each observatory, one of which takes the S/N into account. 
The mean profiles at the INT and CFHT that neglect the error 
weighting are indeed very similar so that the same template lookup 
table (see also Sec.~\ref{sec:DI}) can be used for the INT and CFHT profiles. 

More worrying than the small shape differences in the 
mean profiles of the different observatories are occasional 
differences in profile deformations observed at roughly the same
phase that can not be attributed to noisy data. This is illustrated  
in the bottom plot of Fig.~\ref{fig:LSD_ion}, where profiles that 
were taken almost simultaneously at different sites are 
compared\footnote{Note that the agreement between individual strong 
lines, such as, e.g., H$\alpha$ that were observed at different 
observatories but at similar times is very good. This suggests
that the discrepancy arises mainly because of the low S/N.}.

%
%
\subsection{Photospheric line-profile variations}
We find marked shape changes in the photospheric line-profiles as mirrored 
in the convolved profiles. This is shown in the grey-scale plot of the residual 
deconvolved line profiles in Fig.~\ref{fig:LSD_stack}.
The residuals have been produced by subtracting the mean profile of
each observatory from the individual exposures at this observatory. 
Fig.~\ref{fig:LSD_stack} illustrates the major problem for Doppler imaging 
SU~Aur. Despite the clear periodicities found in the He~{\sc i}~D$_3$ and
the Na~D lines, no obvious periodicity appears to be present in the deconvolved
photospheric lines. In fact, a periodogram analysis of the 
(mean-subtracted) deconvolved profiles does 
not show any significant peaks on the stellar rotation time scale. 
We do not think that this is due to the differences in the mean profiles 
at the different observatories, as a periodogram analysis of the CFHT and 
INT profiles that were produced with identical weighting and that agree
generally well did also not show any periodicities. This begs
the question of whether large contrasting surface features are present 
on SU~Aur or whether the line-profile changes are primarily due to 
causes other than star spots. 

One such cause could be the strong dimming (usually attributed to 
circumstellar dust obscuration) that we observed around HJD~2450410
(see Fig.~\ref{fig:lightcurve}). The time of the faintest observation
coincides reasonably well with one of the peaks in the He equivalent
width and it is preceded by a very strong flux increase in the 
H$\alpha$ line. The behaviour of the LSD profiles during the dimming 
is somewhat ambiguous, though the observations at the MDO that cover
this time span best, suggest a decrease of the equivalent
width from 5\% above average at HJD~2450409 (H$\alpha$ maximum) to 
5\% below average at HJD~2450410 (He maximum). 
Note that the scatter of the LSD equivalent widths is of the order 
of 3\% at the MDO. If the variations in the light level were 
veiling-like, e.g., if they arose from excess emission due to hot spots, 
low light levels should entail increased photospheric line widths which 
does not fit in well with our observations. 

The predictions for the He line depend on what line-formation
scenario is assumed. If the He absorption is produced 
in the hot spot, the He equivalent width should be at a maximum during large
veiling, though there might also be a dilution effect due to the excess 
emission. If, however, the spot produces He emission and line-infilling, 
we would expect the He equivalent width to anticorrelate with the 
broad-band emission. While we lack high-time resolution photospheric observations, 
a comparison between SU Aur's light curve and the variations in the 
He 587.6~nm line does suggest that their variability time scales are 
different and that their variability is not correlated. 
This favours a scenario where the periodic He equivalent 
width variations are produced in dense infall columns or hot spots while the 
more erratic photometric variability is mainly due to dust. It is likely 
that the photospheric equivalent widths are due to a combination of 
both effects thus masking any clear trends and periodicities. 

\begin{figure}
  \def\subfigtopskip{1pt}
  \def\subfigcapskip{1pt}
  \def\subfigbottomskip{1pt}
   \begin{center}
	\subfigure[]{\psfig{bbllx=100pt,bblly=65pt,bburx=380pt,bbury=622pt,figure=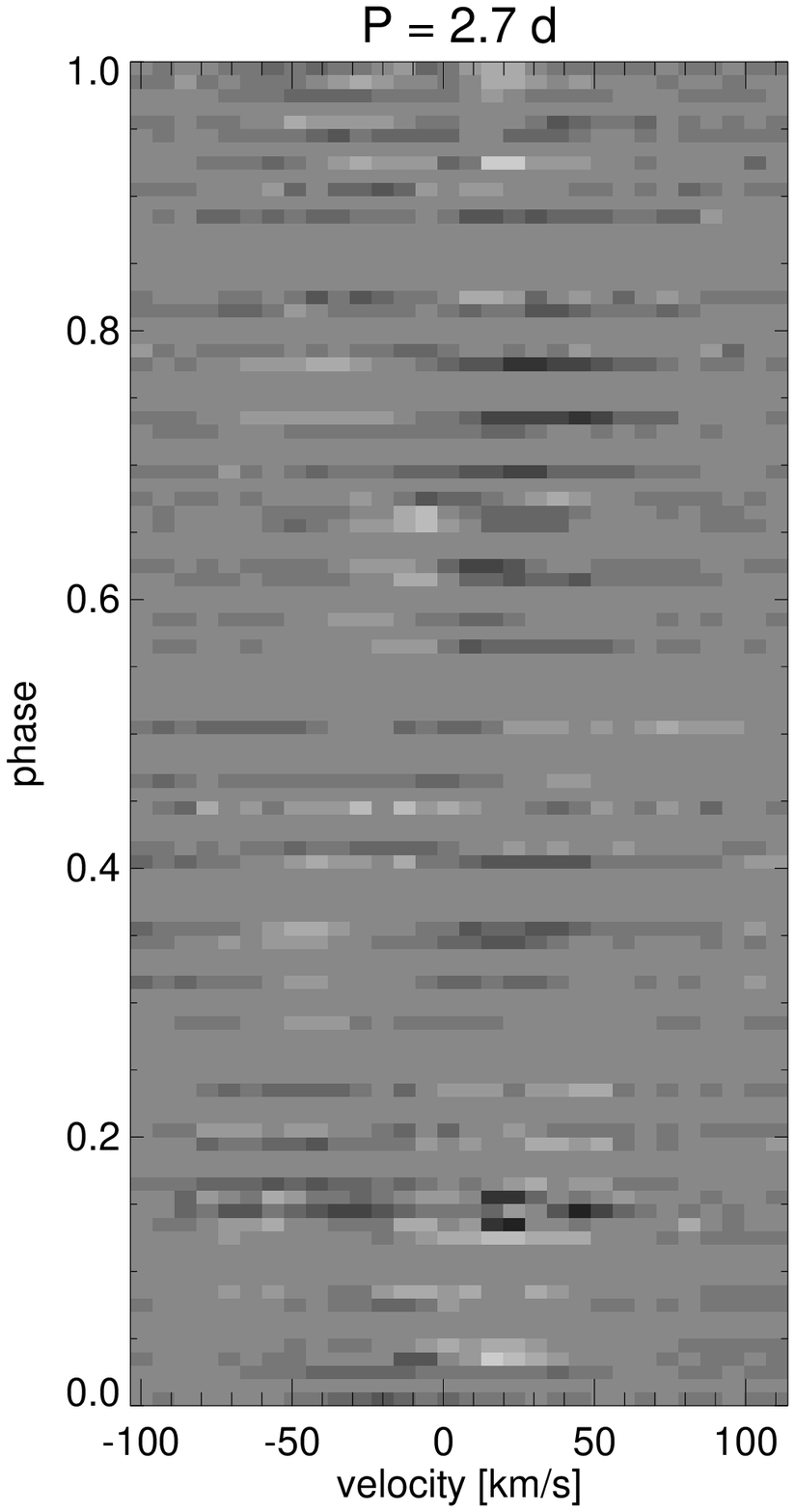,width=4cm}}
	\subfigure[]{\psfig{bbllx=100pt,bblly=65pt,bburx=380pt,bbury=622pt,figure=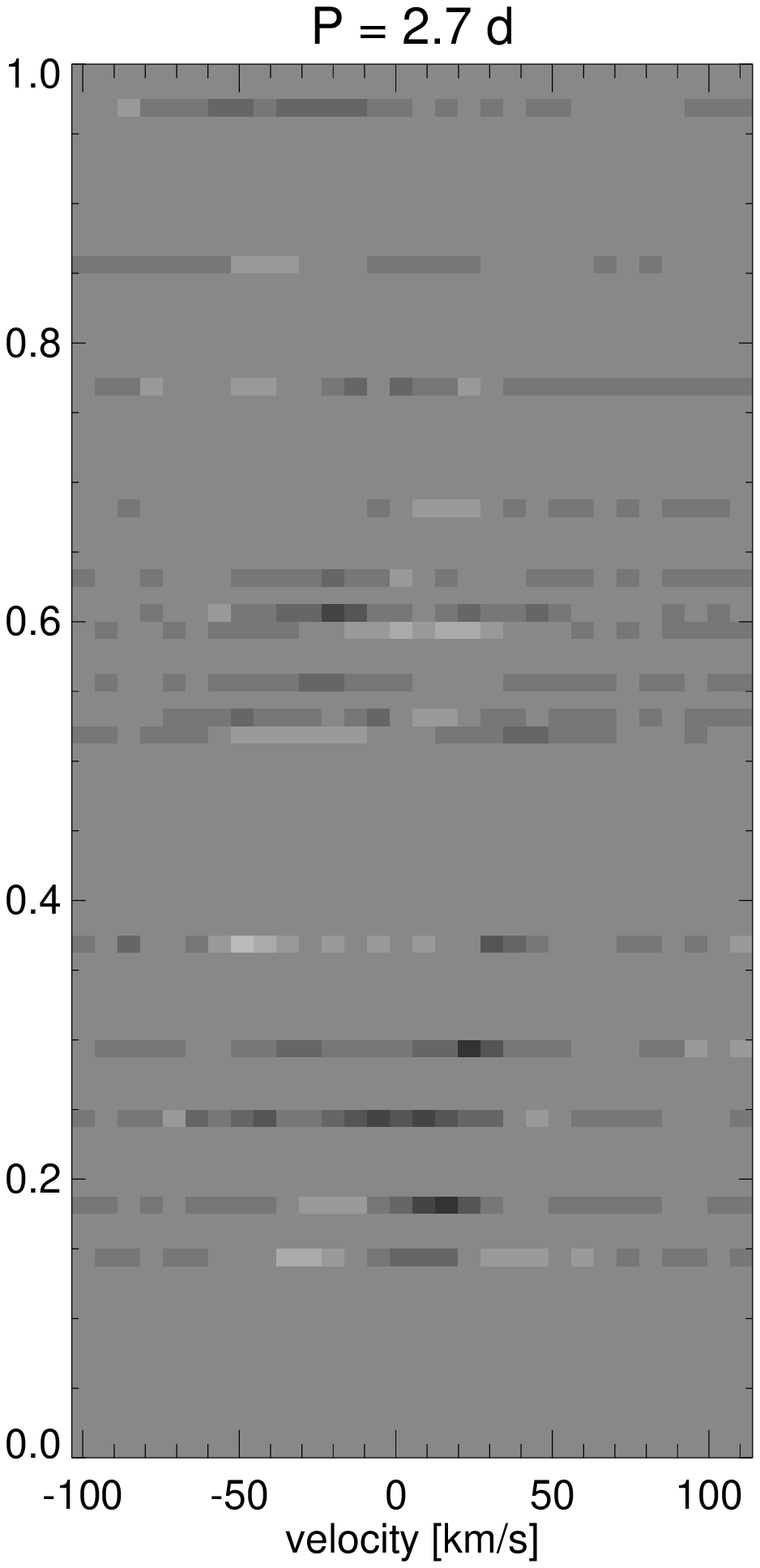,width=4cm,clip=}}
   \end{center}
\caption{The left-hand figure shows the residuals as in the previous 
figure, but binned according to a period of 2.7~d. The right-hand
figure is similar, except that only the profiles from the CFHT and INT 
have been included. We have also excluded the first profile at the INT 
(taken at HJD 2450404.6, well before the remainder of the data) 
and the profiles during the ``dimming''. The grey scale runs 
from a flux level of 0.992 (black) to 1.008 (white) for the left-hand 
figure and from 0.995 to 1.005 for the right-hand figure. It is evident 
that some profiles do not fit in well with adjacent profiles. This is in 
part due to noise (see also the previous figure), but also due
to a lack of repeated structure and the insecurity in the stellar
period. 
	}
\label{fig:LSD_phas}
\end{figure}

We checked for indications of a colour change as a consequence
of the photometric dimming. For the INT which offers a
relatively good time and wavelength coverage, there is no discernable trend
distinguishing the lines redward and blueward of 650~nm. The high-S/N data
of the CFHT did also not show any colour trend. This is in good
agreement with the findings by \scite{herbst1999}.
While we do not see any colour changes during the dimming event, we
do find that as SU~Aur became fainter, i.e.~from around HJD~2450408.0
onwards, all profiles show a strong asymmetry with the flux in the red part 
of the profile being suppressed. It appears that this effect is
strongest for the singly ionised lines while hardly visible for neutral 
lines with low atomic numbers (such as sodium, magnesium and silicon).
The profiles remain 
very similar for at least 1.2~d (see Fig.~\ref{fig:LSD_stack}). As it
takes roughly one third of a rotation period for a spot to cross the visible
hemisphere of the star, this line asymmetry is clearly too long-lived
and too stationary to be produced by a surface spot. 
Whether the asymmetry and the dimming are really linked is difficult to establish
as the photometric observations are only very sparcely sampled. 

Whatever the cause of the profile asymmetry, reconstructions based on 
the complete data set will run into difficulties. We therefore 
reconstructed images where the profiles between HJD~2450407.75 and 
2450409.0 have been excluded (see next section). Even then we 
struggle with an apparent lack of periodicity. This is visible in 
the phase-binned residuals 
shown in Fig.~\ref{fig:LSD_phas}, where some of the neighbouring spectra 
show very different features. Any selected trial period between 2.3 to 3.3~d
produced similarly disjoint plots, suggesting that SU~Aur's 
surface changes on the time scale of one stellar rotation. It is likely 
that we see a mixture of line-profile deformations that arise from non-surface 
events and that are due to dynamic short-lived events. 

%
%
\subsection{Doppler images}
\label{sec:DI}
Despite all these problems, we tried to construct several Doppler images 
over different time spans in the hope of recovering identical features 
which would allow us to exclude shorter lived disturbances. We chose
an ephemeris of HJD 2450409.6. This was picked arbitrarily to correspond
to the mid-dataset maximum He~{\sc i} absorption. 
So as to be able to differentiate between more persistent structures 
and very short-lived features as well as possible artefacts, we 
ran test reconstructions of data subsets where we excluded 
certain days or also data from individual observatories.

\begin{figure*}
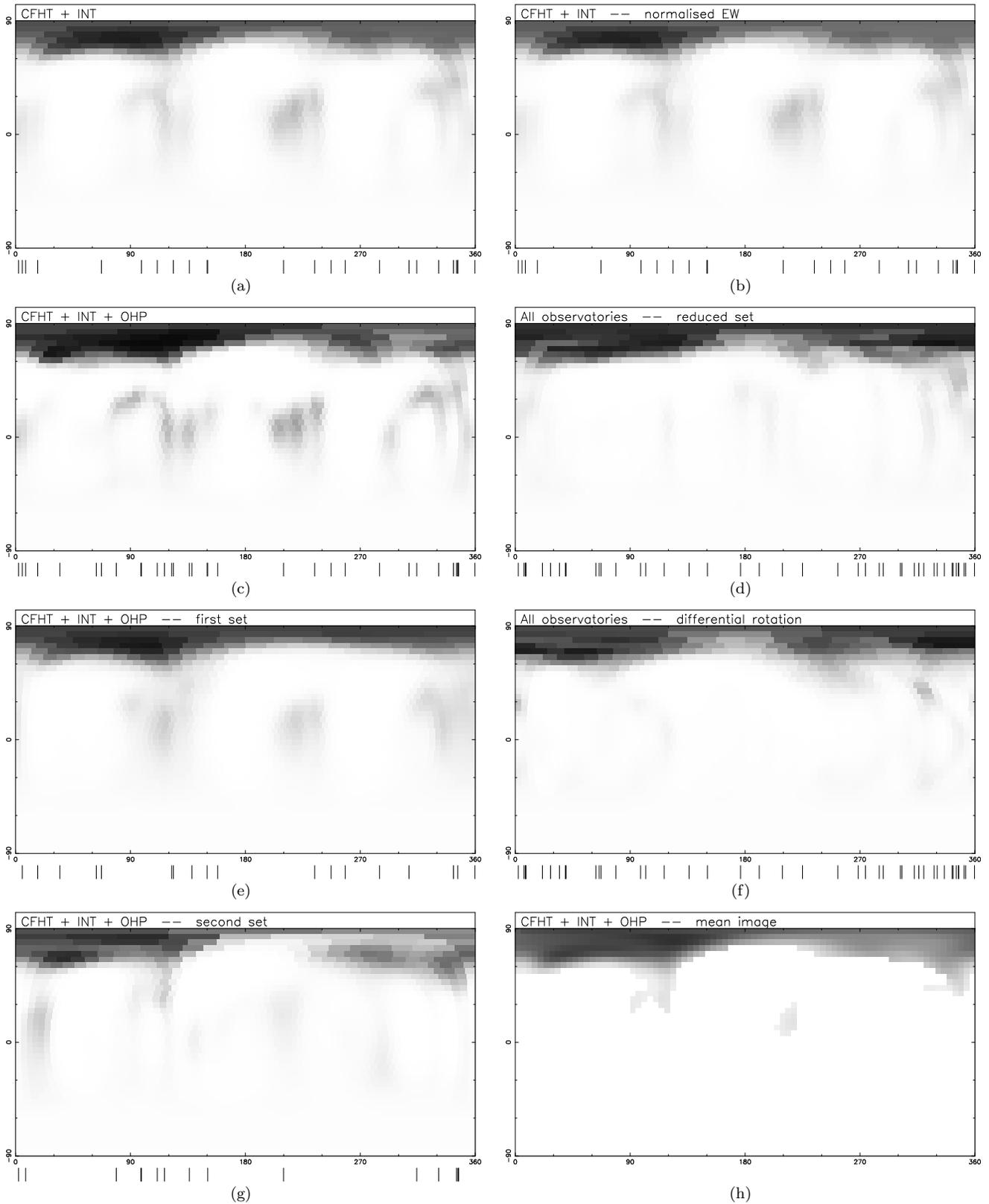

  \def\subfigtopskip{2pt}
  \def\subfigcapskip{2pt}
  \def\subfigbottomskip{4pt}
  \begin{tabular}{lr}
    \subfigure[]{\psfig{figure=fig17a.ps,width=8.5cm,angle=-90}} &
    \subfigure[]{\psfig{figure=fig17b.ps,width=8.5cm,angle=-90}} \\
    \subfigure[]{\psfig{figure=fig17c.ps,width=8.5cm,angle=-90}} &
    \subfigure[]{\psfig{figure=fig17d.ps,width=8.5cm,angle=-90}} \\
    \subfigure[]{\psfig{figure=fig17e.ps,width=8.5cm,angle=-90}} &
    \subfigure[]{\psfig{figure=fig17f.ps,width=8.5cm,angle=-90}} \\
    \subfigure[]{\psfig{figure=fig17g.ps,width=8.5cm,angle=-90}} &
    \subfigure[]{\psfig{figure=fig17h.ps,width=8.5cm,angle=-90}} \\
  \end{tabular}
\caption{Doppler images. The reconstruction parameters and the data sets
used for the different images are listed in Table~\protect{\ref{tab:di_para}},
along with the spot filling factors and the goodness of fit.
The lines underneath the images indicate the phases at which the data used 
for the reconstructions were taken.
        }
\label{fig:DI}
\end{figure*}

The reconstruction parameter of the code that we employ for the
Doppler imaging is the spot filling factor: each pixel
on the stellar surface is characterized by the amount that it is covered
in spots (see e.g.~\pcite{cameron94doppler,barnes98}). We assumed 
the photospheric temperature to be at 5500~K and used the 
LSD profile of the Sun as a lookup table for the immaculate 
photosphere\footnote{We found that using a different template 
star (e.g.~the slightly cooler HD~217014) does not change the fits as 
the equivalent width of the profile 
is adjusted to match the equivalent width of SU Aur at each observatory. 
These adjustments are usually very small, large offsets would indicate 
that the template is not suitable.}. This profile is in principle 
different for each observatory, though it turned
out that we could use the same template profile at the INT and CFHT,
provided the lines were only weighted by their depths. Whether a
depth-only weighting is chosen or whether the flux level is also taken into
account at the CFHT and INT does not have a noticeable effect on the
resulting images and fits. 

It is not entirelly clear what temperature and line profile to adopt for the 
inhomogeneities. On the one hand, the relatively constant equivalent
width suggests that any excess flux, e.g.~due to hot spots, is not 
strongly modulated. The erratic changes in the line profile, on the 
other hand, are untypical for cool spots, at least in the framework
of other active-star observations. We tried three different assumptions
for the inhomogeneities. One where the profile changes are due to cool spots, 
one where they are due to hot spots with a ``stellar'' atmospheric 
structure, and one where the hot spots only emit a continuum 
black-body spectrum, i.e.~where no absorption line is observed for the 
hot structures. 

Depending on which phases and observatories were included in our 
reconstructions, periods ranging from 2.5 to 3.1~days yielded the best 
fits with Doppler images that did not show obvious artefacts or very 
fragmented spot coverage. For the images presented here we chose a period 
of 2.7~days as this agrees best with the periods deduced from the analysis of the 
H$\beta$, Na~D and He~{\sc i} lines. We further assumed the inclination 
angle to be 60$^{\circ}$. The inclination angle can usually be estimated 
by determining the best possible fits as a function of the inclination angle.
In our case, where the stellar period is not known exactly, we determined 
the best-fit inclination angle for a range of periods. The angles  were
only weakly dependent on the period and we settled on 60$^{\circ}$ as this 
yielded good fits for periods around 2.7~d. It is also in good agreement 
with the analysis of IR and mm-observations by \scite{akeson2002} who found 
an inclination angle of 62$^{\circ \, +4^{\circ}}_{-8^{\circ}}$. For our 
values of the period, rotational velocity and inclination we calculate 
a stellar radius of approximately 3.6 solar radii (see also 
Sec.~\ref{sec:discussion}). 

\subsubsection{Cool spots}
\begin{figure*}
  \def\subfigtopskip{0pt}
  \def\subfigcapskip{0pt}
  \def\subfigbottomskip{0pt}
  \begin{tabular}{lr}
        \subfigure[BAO]{\psfig{figure=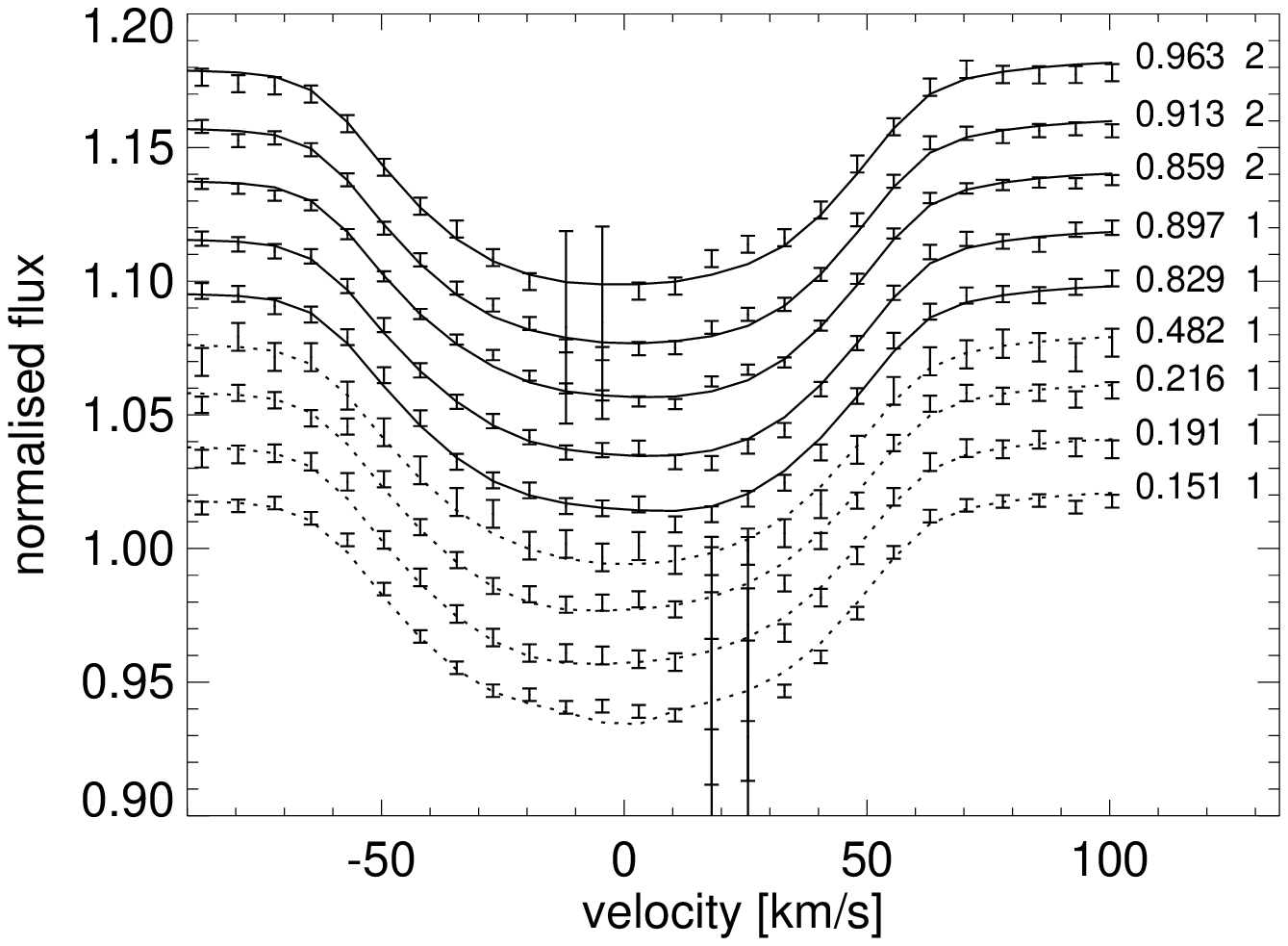,width=8.5cm}} &
        \subfigure[OHP]{\psfig{figure=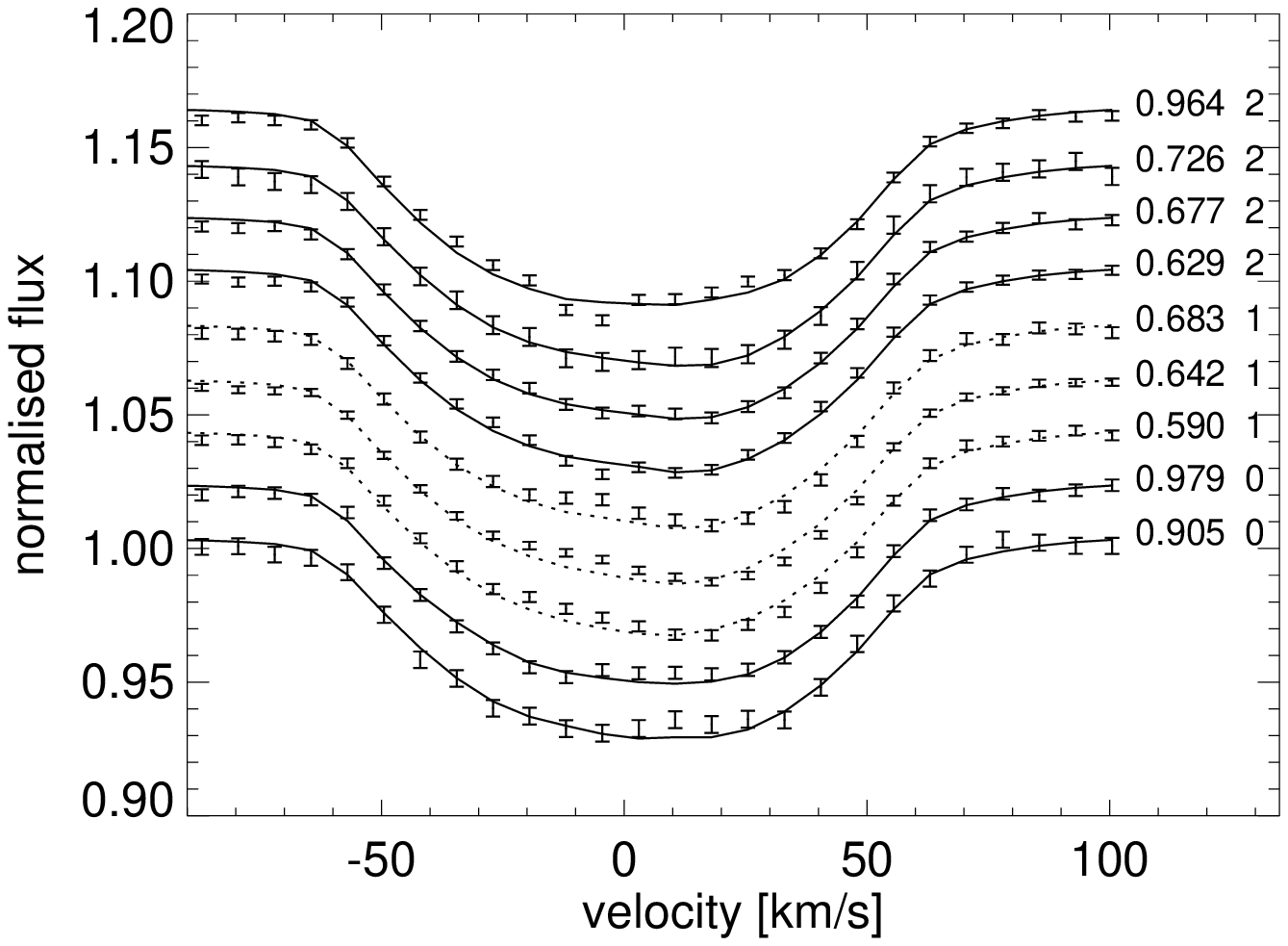,width=8.5cm}} \\
        \subfigure[INT]{\psfig{figure=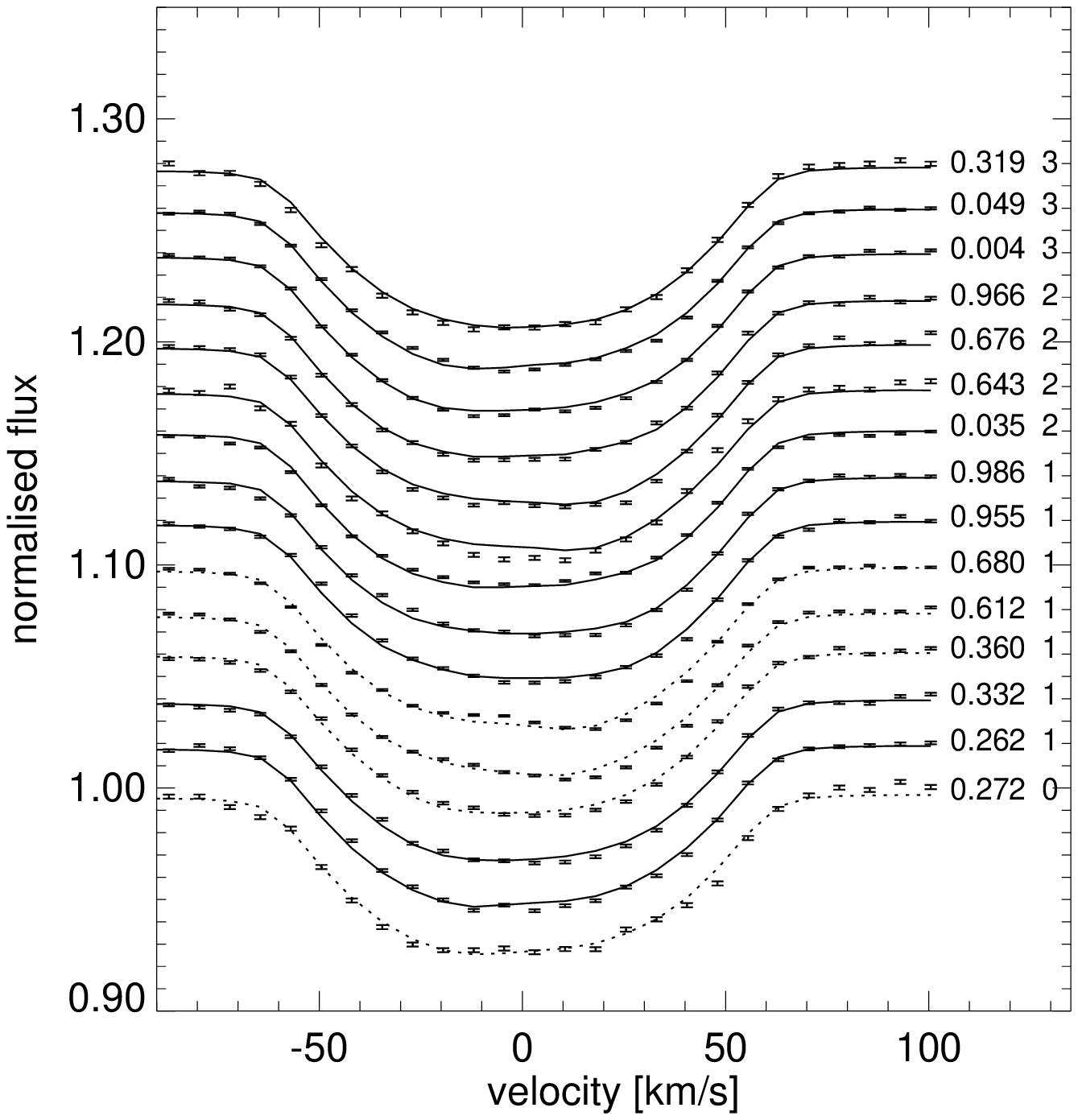,width=8.5cm}} &
        \subfigure[MDO]{\psfig{figure=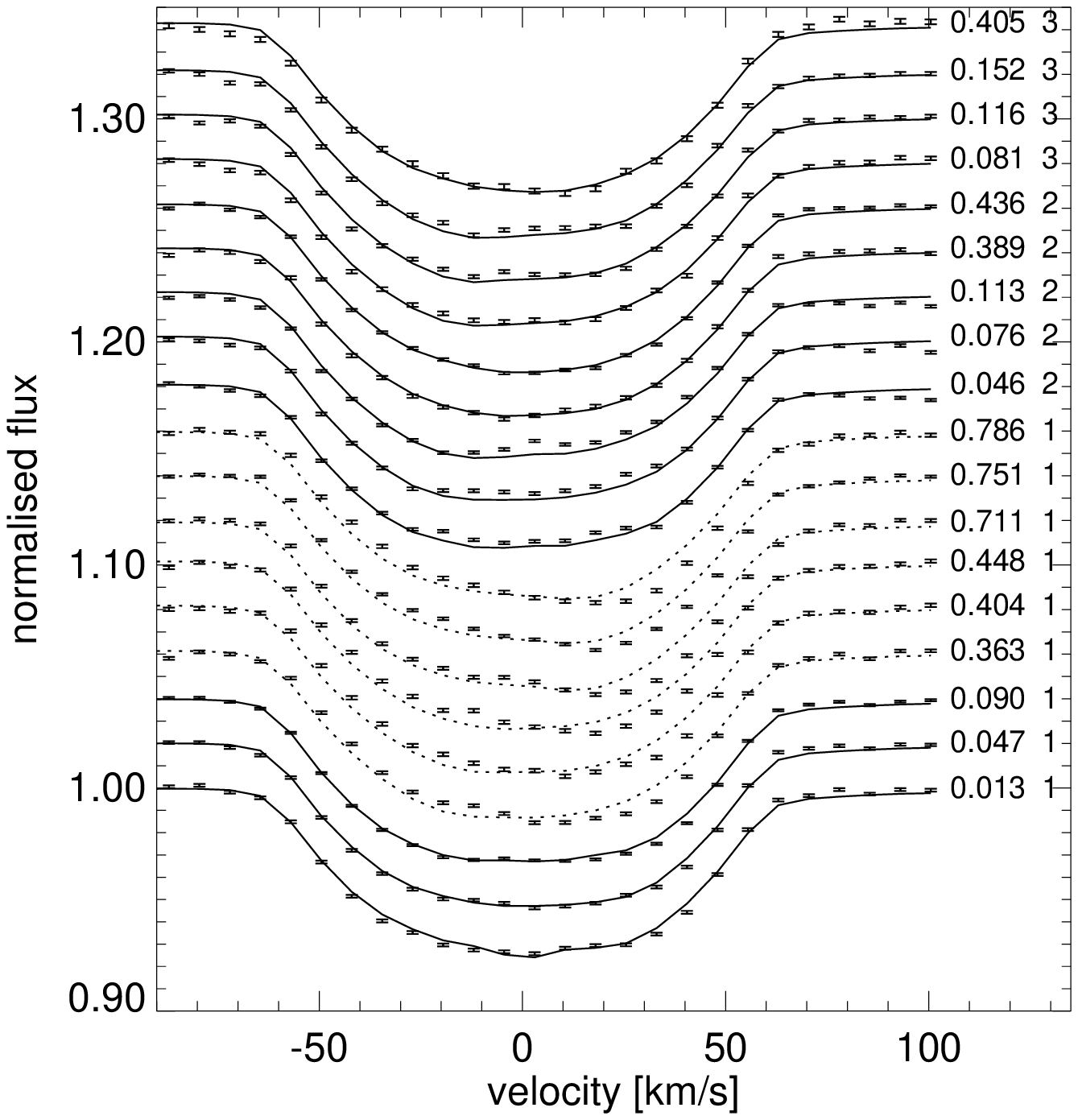,width=8.5cm}} \\
        \subfigure[CFH]{\psfig{figure=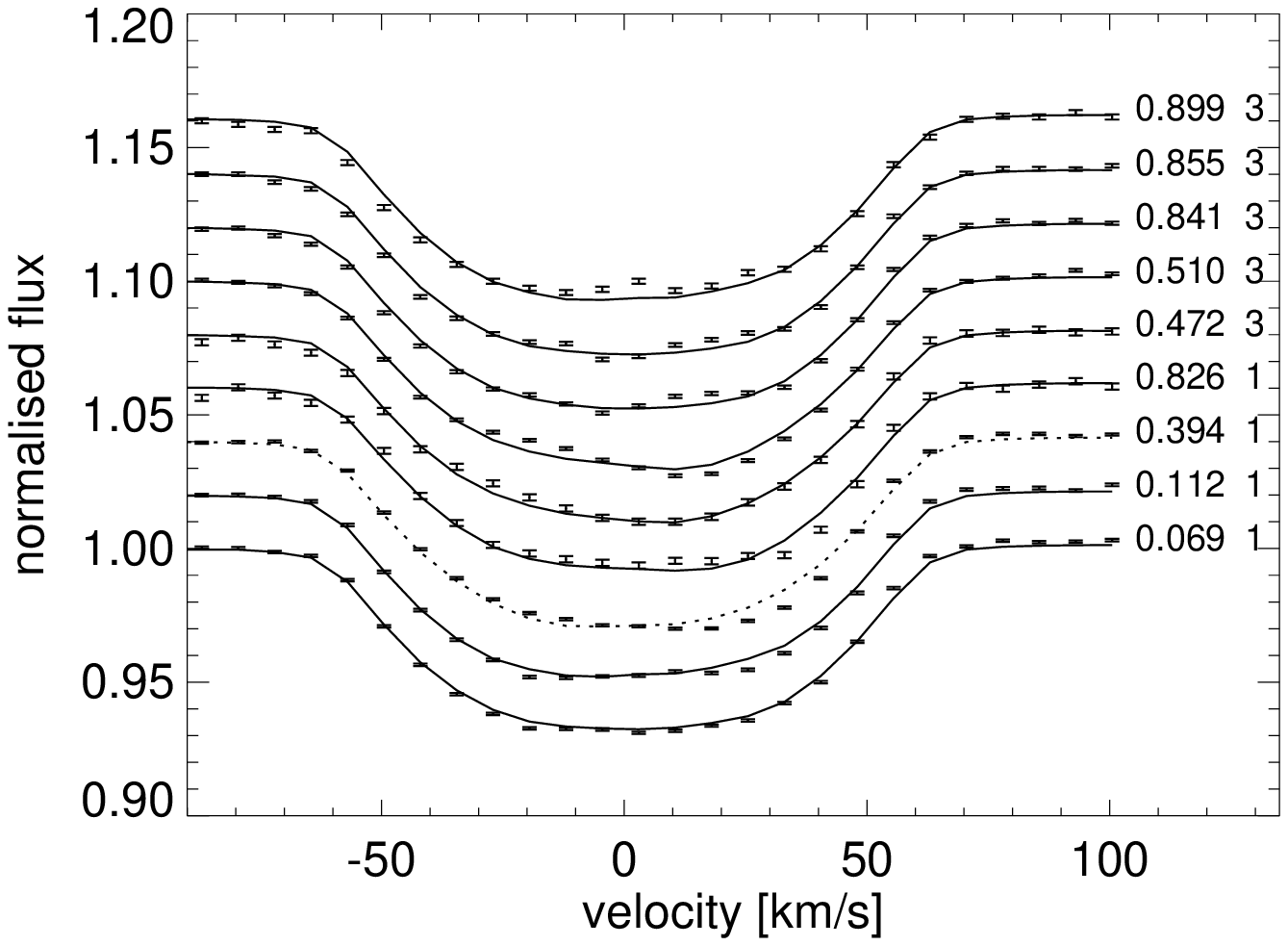,width=8.5cm}} &
  \end{tabular}
\caption[]{
The LSD profiles along with the fits (solid lines) and the
predicted theoretical profiles (dashed lines) for the different observatories.
The data are shown as dots along with their error bars. Only the
data points overlaid with solid lines were used to obtain the Doppler
image (shown in Fig.~\ref{fig:DI}d). The dashed lines are the predicted
profiles from Fig.~\ref{fig:DI}(d) for those phases that were excluded
in the fit as they might be affected by
a dimming event not related to a change in the surface structure.
At the INT, we also excluded the very first exposure as it was taken 
almost a whole rotation period before the next set of observations.
The numbers labelling the profiles are the phase of the profile 
(assuming a 2.7-d period and an ephemeris of HJD 2450409.6) and 
a rotation count number. The observatories were the data were taken are indicated 
below each subfigure.
}
\label{fig:DI_fits}
\end{figure*}

For the cool-spot solutions shown in Fig.~\ref{fig:DI} we assume a spot 
temperature of 4500~K. The top row shows two images produced from data
at the CFHT and INT. For the image on the right-hand side, the profiles 
were rescaled to the mean equivalent width. Both images are very similar, 
though the rescaled images require a smaller surface coverage to achieve an 
equally good fit. Figs~\ref{fig:DI}(c), (e) and (g) show images obtained from 
observations at the CFHT, INT and OHP. While image (c) is for all available 
data spanning more than 3 rotation periods, image (e) and (g) are for 
roughly one rotation period centred on HJD~2450407.8 and 2450413.2 respectively. 
The contributing observatories and 
the selected time spans for all the images are listed in Tab.~\ref{tab:di_para}. 
Also listed are the values for $\chi^2$ for the actual fit, the total data set, 
and for a reduced data set where the dimming event and the profiles from the 
BAO which show the lowest S/N have been excluded.
The fits that result in Fig.~\ref{fig:DI}(d) are shown in Fig.~\ref{fig:DI_fits}.
The solid lines are the fits and the dashed lines are the predicted
spectra for the phases that were not considered in the fits (due
to the dimming event).

In order to improve the fits to the line profiles, we have included differential
rotation into our Doppler imaging code, according to
$\Omega (\theta) / \Omega_{\rm eq} = 1 - D_{\rm r} \sin^2 (\theta)$.
In agreement with \scite{johns-krull96tts} we find that differential
rotation coefficients of about $D_{\rm r} = -0.1$
yield marginally better fits with smaller spot coverage (compare
rows 4 and 6 in Tab.~\ref{tab:di_para}).
Changes in the differential rotation coefficient require adjustments in the
value for $v \sin i$. For a differential rotation coefficient
of $-0.1$ we obtain the best fits for $v \sin i = 58$~km~s$^{-1}$,
while $v \sin i = 59$~km~s$^{-1}$ yields the best fits if no
differential rotation is assumed. For solid-body rotation, the spot
coverage is higher at intermediate to high latitudes than for differential rotation.
The location of the surface features, however, remains the same irrespective of
whether differential rotation is included or not.
This is illustrated in the Figs~\ref{fig:DI}(d) and (f). Image ~\ref{fig:DI}(f) 
is for a differential rotation coefficient of -0.1, image ~\ref{fig:DI}(d)
for no differential rotation.

%
%
\subsubsection{Hot spots}
The magnetospheric disk accertion model predicts hot spots or bands at
high stellar latitudes. If such hot spots are unevenly distributed and
are much hotter than the photosphere, or cover a large fraction of the
surface, their excess emission should produce a variable dilution of the
photospheric line profiles (veiling) in addition to line-shape changes.
We do not see any consistent trend above the 2.5~\%-scatter in the equivalent
widths of the LSD profiles and previous optical veiling determinations have
yielded null results for SU~Aur (see \scite{basri90,johns95suaur}, though
note also that \scite{folha99} found SU~Aur to be hightly veiled in the
infrared K and J band).
-
\begin{figure}
\psfig{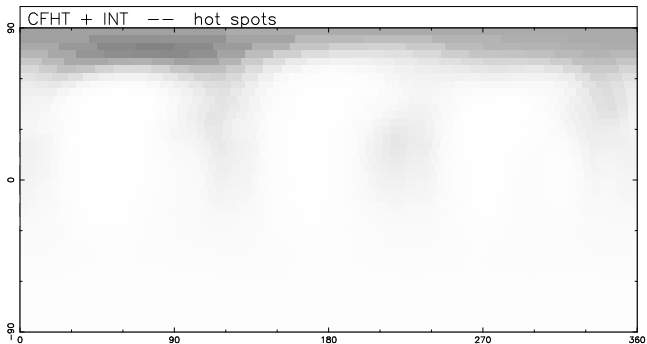}
\caption[]{
Doppler image for continuum hot (6000~K) spots. Note that as on 
previous plots, the photosphere is left white while grey 
indicates the spot filling factor. The total surface covered by the hot
spots here is 3.3~\%. 
}
\label{fig:DI_hot}
\end{figure}

\begin{table*}
   \caption[]{List of parameters used for the reconstructions of the
Doppler images shown in Fig.~\protect{\ref{fig:DI}}. The first column
lists the identifier. Columns 2 to 4  list the goodness-of-fit
for different data subsets. $\chi^2_f$ gives the
goodness-of-fit for the actual fit (using the data as listed),
while $\chi^2_t$ lists the goodness-of-fit between the complete
data set and the predicted profiles; $\chi^2_-$ lists the goodness-of-fit between
the predicted profiles and a data set that excludes profiles
from BAO and profiles taken between HJD~2450407.75 and 2450409.0
(see also dashed lines in Fig.~\protect{\ref{fig:DI_fits}}).
Column 5 shows the spot
filling factor. Columns 6 to 10 indicate which time spans of the data
have been used in the fits for each observatory. The times are listed 
in modified HJD, i.e. HJD-2450400; a dash indicates that no data from the particular
observatory was used; ``excl'' indicates that all profiles were used
for a given observatory except for profiles taken between MHJD~7.75 and
9.0. Image h) is the (thresholded) mean image obtained from images e) and g) 
and (19) refers to the hot-spot image shown in Fig.~19.
The last two rows list $\chi^2_t$ and $\chi^2_-$ for unspotted input images,
including and excluding differential rotation.
        }
   \begin{tabular}{ccccccccccl}
        \hline
	[1]   & [2]  & [3] & [4] & [5]	  & [6]	 & [7] & [8] & [9]  & [10] & [11] \\
        image & $\chi^2_f$ & $\chi^2_t$ & $\chi^2_-$ & $A_s/A_*$ & BAO & OHP & INT & MDO & CFHT & comments \\
        \hline
        (a)   &  1.2 & 2.2 & 1.7 & 5.1 \% & ---  & all & all &  --- & --- & \\
	(b)   &  1.1 & 2.2 & 1.6 & 4.8 \% & ---  & all & all &  --- & --- & normalised EW \\
	(c)   &  1.2 & 2.0 & 1.5 & 6.5 \% & ---  & all & all &  --- & all & \\
        (d)   &  1.2 & 1.9 & 1.2 & 5.1 \% & excl & excl & excl & excl & excl & \\
        (e)   &  1.3 & 2.3 & 1.8 & 5.2 \% & ---  & 6.4 - 8.7 & 7.5 - 9.7 & --- & 6.9 - 9.1 & \\
        (f)   &  1.2 & 2.1 & 1.2 & 4.8 \% & excl & excl & excl & excl & excl & diff.~rot, $D_r=-0.1$ \\
        (g)   &  1.0 & 2.4 & 1.7 & 4.4 \% & ---  & 11.4 - 12.4 & 11.5 - 13.4 & ---  & 13.9 - 15.1 & \\
        (h)   &  --- & 2.2 & 1.6 & 2.8 \% & ---  & ---  & ---  & ---  & --- &  \\
        (19)  &  1.0 & 1.8 & 1.4 & 3.3 \% & ---  & all & all &  --- & --- & \\
        blank &  --- & 4.1 & 3.3 & 0.0 \% & ---  & ---  & ---  & ---  & ---  &	\\
        blank &  --- & 4.0 & 3.0 & 0.0 \% & ---  & ---  & ---  & ---  & ---  & diff.~rot, $D_r=-0.1$ \\
        \hline
   \end{tabular}
   \label{tab:di_para}
\end{table*}

In a first attempt, we assumed that the spot spectrum would be represented
by an ordinary hot-star spectrum. It turned out, however, that no satisfactory
fits could be obtained if the convolved hot-spot spectrum showed a
significant absorption line. This led us to try fits where the spot produces
a pure black-body spectrum. Temperatures of 6000 and 6500~K yielded slightly
better fits than the cool-spot assumption while needing only very small
spot-covering fractions of around 3~\%. Fig.~\ref{fig:DI_hot} shows an
example image obtained using data from the CFHT and INT. Note that the
location of the spot features essentially agrees with that found
from the cool-spot fits (Figs~\ref{fig:DI}(a) and (b)) to the same data
set. This is not surprising as it is the decreasing total flux for the
cool spot rather than the shape of the cool-spot spectrum that acts
as the main factor in producing the line-profile deformation.
Our method does not allow us to determine an accurate hot-spot temperature,
as the size of the deformation is a function of the temperature and the
surface coverage. We find, however, that the goodness-of-fit starts to
increase and the reconstructed images start to show strongly fragmented
spots for temperatures above 7000~K. For very hot temperatures, the veiling
predicted by the spots that are invoked to fit the line-profile deformations
is no longer compatible with the relative constancy of the line equivalent width.
                                                                                                                  
%
%
\subsubsection{Comparison between different Doppler images}
Whilst the agreement between the different Doppler images
is not very convincing at first sight, there are features that can
be found on almost all of the images. To highlight
those features we created a ``thresholded mean image'' of
the Doppler images displayed in Figs~\ref{fig:DI}(e) and (g).
To create this image, we only attributed spots to those pixels where
the spot-filling factor in both contributing images was above
a certain threshold, in this case greater than 5~\%. The
features visible on the ``thresholded mean image'' (see Fig.~\ref{fig:DI}h)
are hence features that were present on both input images. As
image (e) is based on data taken between MHJD 6.4 to 9.7
whereas image (g) is based on data taken between MHJD 11.4 and 15.1,
the mean image should show the pixels where structure was present
(and possibly survived) after more than one stellar rotation had passed.
                                                                                                                  
The mean image in Fig.~\ref{fig:DI}h shows mainly high-latitude
features. In fact, most images, including Fig.~\ref{fig:DI_hot} for the
hot spot, recover two high-latitude spotted areas (between longitudes
30$^{\circ}$ and 112$^{\circ}$ and from about 270$^{\circ}$ to 350$^{\circ}$)
that are made up of a couple of stronger spots.
Also shown on most images are some low- and intermediate-latitude spots
around approximately 120$^{\circ}$ and 210$^{\circ}$ longitude.
Note that the thresholding biases against low-latitude structure.
This is because the longitude resolution is latitude dependent
and much higher at low latitudes. Errors in the period and differences in
the phase coverage produce small longitude shifts. These are particularly
noticeable for narrow elongated low-latitude features. The shifts then result
in structure being suppressed on the thresholded image.
                                                                                                                  
While we cannot exclude the presence of hot spots with stellar-like atmospheres,
the fits for cool spots or for black-body hot spots yield much better goodness-of-fit
parameters and less fragmented surface images. While we cannot decide on the
basis of our data whether the spots are hot or cold, or whether there is indeed
a mixture, the lack of repeatability in the photospheric line profiles does
favour the hot-spot scenario. The lowest high-latitude spot covering
fractions are seen around longitudes 180$^{\circ}$ to 210$^{\circ}$, i.e.
roughly out of phase with the maximum He~D$_3$ equivalent width and would
fit in very well with the ``eggbeater'' scenario.
                                                                                                                  
It is heartening that the mean image is very similar to the
image reconstructed from all the data save
those profiles that might be contaminated by the ``dimming event''
(Fig.~\ref{fig:DI}d). This event is mostly visible for the data
at the MDO, for which the discrepancy between the predicted line
profiles and the actual line profiles as shown in Fig.~\ref{fig:DI_fits}
is largest.  The goodness-of-fit values for the mean image are listed in
row (h) of Tab.~\ref{tab:di_para}.
Note that SU~Aur's light curve shows very erratic behaviour. This may mean
that the stellar surface as we observed it during the MUSICOS campaign was not
``typical'', if there is indeed a ``typical'' spot coverage for SU~Aur.
We therefore do not expect our images to show much similarity with previous
or future images.

%
%
\section{Magnetic field measurements}

\begin{figure}
        \psfig{figure=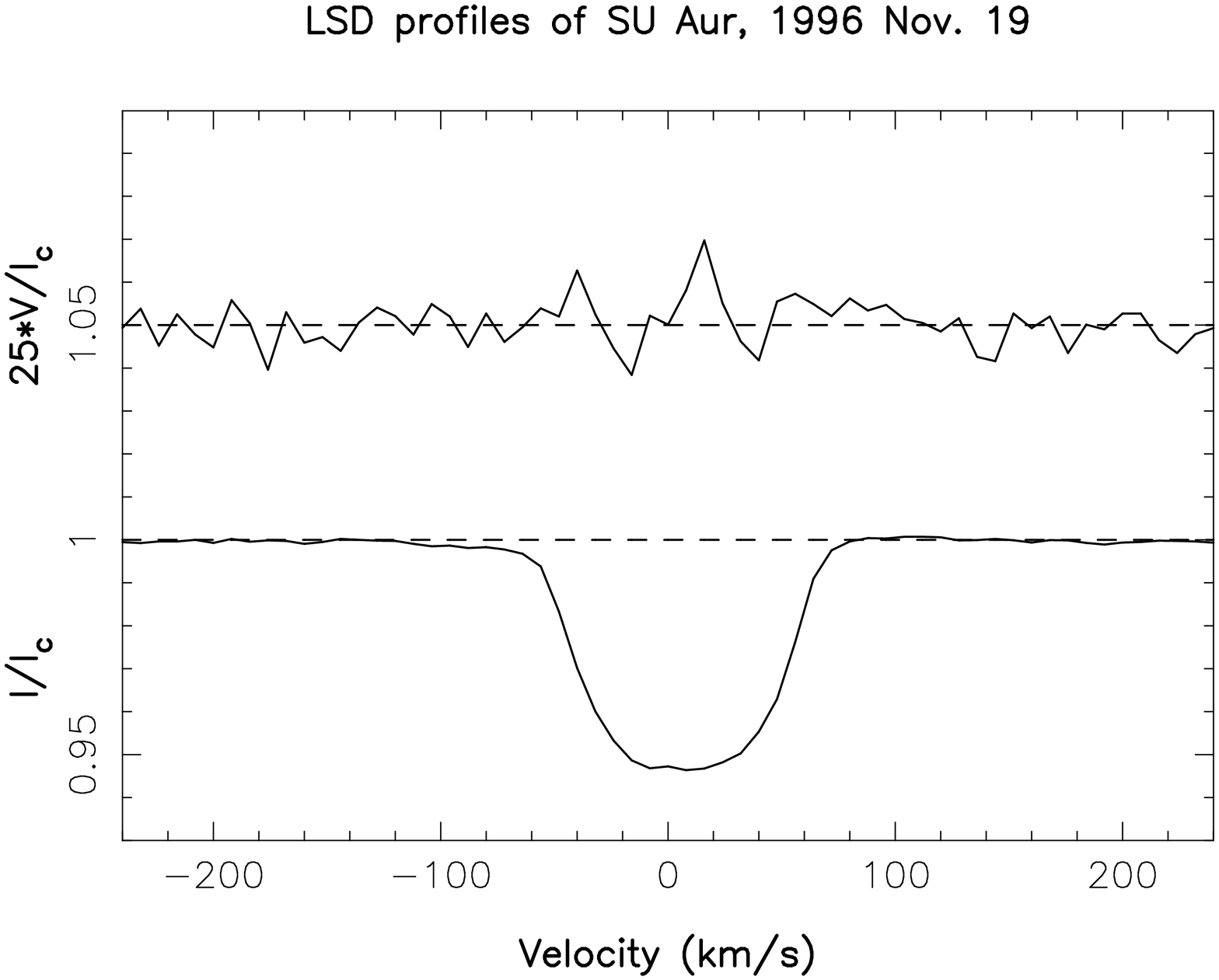,width=8.5cm,angle=-90}
\caption[]{
The Stokes I (bottom) and V (top) least-squares deconvolved profiles
of SU Aur on Nov.~19. Both profiles have been normalised. The
Stokes V profile has been shifted by 0.05 and amplified by a factor
of 25 so as to show the structure in the line more clearly.
}
\label{fig:stokes}
\end{figure}

The exposures (9 groups of 4 subexposures) obtained at the CFHT with 
the MUSICOS spectropolarimeter can be used to extract information
about the magnetic field structure of SU~Aur.  
The first step in this  process consists of extracting the LSD Stokes 
V signatures from all available photospheric lines.  
To maximise our chances of detecting the field, we convolved all spectral 
lines whose relative depth (prior to any kind of macroscopic broadening) 
exceed 10\% of the continuum level in an ATLAS9 G2 subgiant synthetic spectrum
(Kurucz 1993). As a result, about 3,500 lines were used in the analysis. 
We averaged all spectra obtained on the same nights
whose rotational phases are within a few \% of each other 
thereby decreasing the noise level in the resulting circular polarisation
profiles further.  From the original 9 profiles we thus obtain 5 LSD
Stokes V spectra that correspond to rotational phases of 0.09, 0.39,
0.83, 0.49 and 0.87, with relative noise levels of 0.031\%,
0.040\%, 0.100\%, 0.042\% and 0.037\% respectively.  The strongest
signature we see is observed at phase 0.09 on Nov.~19 (see 
Fig.~\ref{fig:stokes}).  With a peak-to-peak amplitude of 0.15\% (5$\sigma$), 
this signature is associated with a detection probability of only 94\%. 
We therefore consider that this detection is only very marginal, confirming 
that SU~Aur is an interesting target for future magnetic field
investigations with more sensitive spectropolarimeters.

One can note that the shape of the marginal Zeeman signature
discussed above, if real, is rather complex with at least three
polarisation sign switches throughout the line profile. This 
suggests that the associated surface field topology is not simple.
It is reminiscent of the 
polarisation signatures detected on the weak-line T~Tauri stars
V410~Tau and HDE~283572 with a similar instrument 
(Donati et al.\ 1997), that also argue in favour of a rather complex field
structure such as that reconstructed on the older ZAMS star AB~Dor
(Donati \& Cameron 1997).
We nevertheless investigated the suggestion of Johns-Krull \& Basri
(1995b) that the magnetic field of SU~Aur is largely dipolar with
the magnetic axis inclined with respect to the rotation axis. 
In this model accretion occurs mainly at the magnetic poles.
Although this assumption may sound
rather incompatible with the above conclusions that the field is
complex, it can nonetheless be useful to estimate the upper limit to
the dipole field strength that our data can allow.

\begin{figure}
        \psfig{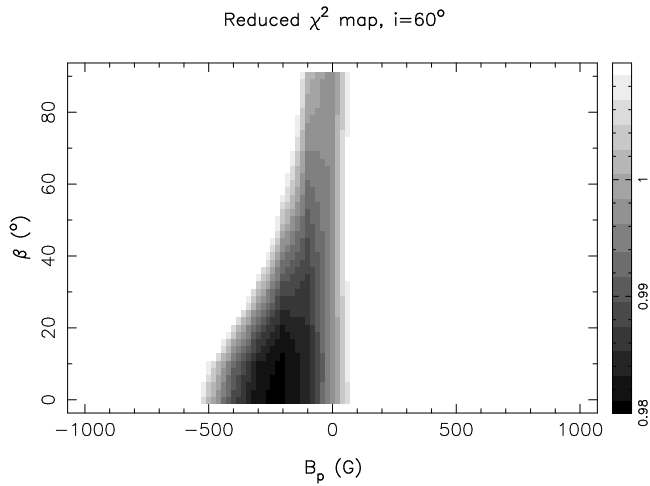}
\caption[]{
Contours of reduced $\chi^2$ between synthetic and observed Stokes V profiles.
The y-axis shows the inclination of the magnetic to the rotation axis, while
the x-axis shows the strength of the dipole field. The two
darkest gray levels correspond to models within the 68\% confidence
limit, while white corresponds to models outside the 99\% confidence
limit. Field strengths that yield
acceptable fits for an inclined magnetic axis are rather small ranging between about 
-300 and 100 Gauss.
}
\label{fig:mag_dipole}
\end{figure}

We therefore computed a number of synthetic Stokes V spectra for
various dipole models, that we compare with our 9 LSD Stokes V
profiles with the help of a simple $\chi^2$ statistics\footnote{The 
results described hold also for the co-added set of 5 LSD profiles}. 
While the phase at which the magnetic pole comes closest to the line of sight
is fixed to that at which the He~1 line at 587.6~nm exhibits its 
strongest absorption (i.e.\ at phase 0.0 in the ephemeris
mentioned above), the strength
of the dipole field and the inclination of its axis with respect to
the rotation axis are considered as free parameters, in the range of
--1 to 1~kG and 0\degr\ to 90\degr\ respectively.  For all these
models, we compute the reduced $\chi^2$ test between the synthetic
and observed Stokes V profiles, and display it as a map in 
Fig.~\ref{fig:mag_dipole}.
We can thus conclude that the strength of such a dipole field must
range between --500 and 100~G to remain compatible with our
observations (at a 68\% confidence level).  Moreover, given the
fact that the range of reasonable orientations of the magnetic axis
with respect to the rotation axis is rather 30\degr\
to 60\degr\ (to ensure that one pole is much more visible than the
other and that accretion is only observed during one half rotation
cycle), we finally obtain that the maximum dipole field strength
ranges between about --300~G and 100~G.

It is difficult to see how such a weak field will be able to 
magnetically confine material accreted at a rate of
$6\times10^{-8}$~M$_{\odot}$~yr${-1}$ at a distance of about 2 stellar radii
(where the local field strength is weaker than 20~G). This suggests 
that while the eggbeater model successfully explains many of the basic 
phenomena observed on SU~Aur and how material is accreted onto its surface, 
the details of the magnetic interation between the central star
and circumstellar disc are likely to be more complex.
We checked for circular polarisation signatures in the
He~{\sc i} line at 587.6~nm such as those reported by 
\scite{johns-krull99bptau_pol} on the classical T~Tauri star BP~Tau. 
No features above the noise level are seen in the vicinity of the 
He~{\sc i} line. 

%

\section{Discussion and Conclusions}
\label{sec:discussion}
\subsection{Stellar parameters}
Our analysis confirmed a projected rotational velocity of approximately 
$~60$~km~s$^{-1}$ for SU~Aur, while the rotation periods determined from different 
lines are all below the previously determined rotation period of 3~d. Together 
with our estimate of the inclination angle obtained
from the Doppler imaging procedure this allows us to re-evaluate the 
stellar parameters. Unfortunately, the errors on our period determinations 
and also on the inclination angle are fairly hefty so that we can only give 
a somewhat large range of probable radii. 

For a rotation period of 2.7~d, we find that the Doppler imaging process favours
inclinations between 50$^{\circ}$ and 70$^{\circ}$. The projected rotational 
velocity of 59$\pm 1$~km~s$^{-1}$ hence translates into an equatorial rotation 
velocity $v_{\rm eq} = 68^{+10}_{-6}$~km~s$^{-1}$. Using $R = P v_{\rm eq} / 2 \pi$, 
we obtain a stellar radius of $3.6^{+0.6}_{-0.3} R_{\odot}$. The smallest radius 
possible for a period of 2.7~d is 3.0~$R_{\odot}$. The uncertainty on the stellar
period is quite large, and periods of 2.5~d and 2.9~d which are also
plausible would yield smallest radii of 2.8 and 3.3~$R_{\odot}$ respectively.

\scite{akeson2002} found that an inner radius of 0.05 to 0.08~AU for the
K-band emission yielded the best model fits. This would correspond
to about 2.7 to 4.3 stellar radii. This is in very good
agreement with the usual assumption of the magnetospheric accretion
model that the inner disk starts at the corotation radius.
Depending on the exact parameters adopted for the stellar mass,
radius and rotation period, the corotation radius of SU~Aur
is at about 2.5 to 3 stellar radii. Any of the inner disk edges cited
above are large enough to prevent occultation of the star by the disk.
Indeed, for a system seen at an inclination of 60$^{\circ}$, the disk would
have to start at 0.7~stellar radii from the surface in order to occult the star.

The derived radii tie in well with the Hipparcos
parallax and SU~Aur's colour. The Hipparcos catalogue \cite{hipcat} lists a 
parallax of $(6.58 \pm 1.92) \times 10^{-3}$~arcsecs, a colour of $B-V=0.833$~mag
and a V-band brightness of $9.23 \pm 0.056$~mag. SU~Aur's 
absolute (non-extinction-corrected) V-band magnitude is hence
$3.3^{+0.6}_{-0.8}$~mag. The bolometric correction is independent of the stellar
luminosity class and is $-0.2$~mag for the colour term given above \cite{lang80}. 
The interstellar extinction has been given as $A_V=0.9$~mag by
\scite{chen95} and as $A_V=0.93 \pm 0.14$~mag by \scite{cohen79}, while the 
prescription given in \scite{hillenbrand97}
yields $A_V=0.46 \pm 0.23$~mag. Using an extinction of 0.9~mag, we obtain 
a corrected bolometric absolute magnitude of 2.2$^{+0.3}_{-0.7}$. 
For a black body, $M_{\rm bol}$ (in magnitudes) is related to 
its radius (in $R_\odot$) and temperature (see \pcite{lang80}) through
$$ M_{\rm bol} = 42.31 - 5 \log R - 10 \log T. $$
Using a temperature of $5700 \pm 100$~K, we obtain a radius of $3.2^{+1.4}_{-0.5} R_\odot$. 
Note that these radii (along with the colour of the star) suggest that 
SU~Aur is a subgiant rather than a giant. 

The bolometric magnitude of SU Aur translates into a luminosity of 
10.2$^{+9.2}_{-2.5}$~L$_{\odot}$. According to the evolutionary tracks 
by \scite{dantona94} (using the convection description by \nocite{canuto92}
Canuto \& Mazzitelli, 1992), SU~Aur's mass has to lie between about
2 and 2.5~M$_{\odot}$ while its age is around $3\times 10^6$~years.
This is in reasonable agreement with the findings of \scite{dewarf1998}, 
who determined a mass of $1.9 \pm 0.1$~M$_{\odot}$ and an
age of $4\times 10^6$~years. The large age difference is mainly due
to the different extinction corrections that were applied. 

%
%
\subsection{Variability}
SU~Aur is certainly a very complex system. While it displays 
some clear periodic features, notably in the Balmer lines, we also 
found short-lived and non-periodical features 
that make the interpretation of the system rather difficult.
We find that the model by \scite{johns95suaur} is on the whole 
successful in explaining the variability of the Balmer lines, Na~D and 
He~D$_3$, as well as the correlations between those lines. 
A more in-depth analysis of time lags and cross-correlations between the 
different lines was presented in \scite{oliveira2000suaur} where
some problems of the model along with possible extensions were also described.

During the MUSICOS~1996 campaign, we observed transient features in the 
blue wings of the H$\alpha$, H$\beta$ and Na~D lines. While they 
appear roughly contemporaneous in all lines, their velocity shifts and 
strengths differ from line to line. Whether the two events are due 
to the same mechanism is not clear as they show different strengths 
and time behaviour (see also \pcite{oliveira2000suaur}). 

From the analysis of the Balmer lines, He~{\sc i} (543.6~nm) and Na~D,
we find strong indications for a stellar rotation of the order of 2.8~days,
albeit with relatively large uncertainties, mainly due to the
non-sinusoidal nature of the variations combined with the relatively short
observation time span. The picture is less clear for the photospheric
lines. The LSD profiles do not show any clear periodicities and leaving
the period variable in the Doppler image reconstructions yields different
optimum periods (spanning 2.4 to 3.2~d) for different data subsets.

During the campaign, the brightness of SU~Aur dropped by 0.5~mag$_V$. 
This dimming seems to be coupled with a change in the line-profile
shape of the photospheric lines. The convolved profiles show a redshifted 
emission features for a duration of about 1.2~d, almost half a stellar rotation 
period. The most common explanation for the rapid dimmings of ETTS is 
that circumstellar matter obscures the stellar disk. This is not normally 
expected to influence the shapes of the photospheric line profiles. 
We suspect that not the complete stellar disk is obscured. 
There does not seem to be a change in the equivalent widths of the photospheric
lines due to the dimming. The fact that we could also not
find a wavelength-dependent behaviour of the equivalent width suggests that
the obscuring matter is grey. 

The dimming is also associated with an increase in the equivalent width of the 
(blue) H$\alpha$ emission wings, an increase in the blue emission peak and
of the red wing for velocities below about 150~km~s$^{-1}$ (see also 
Fig.~\ref{fig:Ha_bins}). 
If the equivalent width of the emission peak is plotted against magnitude,
we find that there is a clear trend of decreasing equivalent width with
increasing brightness. While the brightness drops by 0.5~mag$_V$, the 
equivalent width increases by over a factor of two. No such trend 
is observed for the blue emission peak of H$\beta$ even though there is 
otherwise a good correlation between the blue wings of H$\alpha$ and H$\beta$.
The fact that no line other than H$\alpha$ shows a correlation with the 
broadband magnitudes suggests that we are not dealing with a contrast 
effect. We would like to note, however, that the S/N is poor in the 
H$\beta$ and photospheric lines, so that available matches between 
well-determined equivalent widths and photometric measurements are few. 

While the least-squares-deconvolved profiles contain a certain amount
of non-repeated and probably also spurious structure (see e.g.~the 
line depression at about 25~km~s$^{-1}$ for several BAO profiles), 
the images reconstructed from various data sets all show some 
consistent high-latitude features as shown in Fig.~\ref{fig:DI}h.
This is a strong indication that SU~Aur is indeed covered by a few
cool surface spots that last at least a few stellar rotations. 

We also want to caution that the surface spots 
pictured in Fig.~\ref{fig:DI}h can not be the only features
producing line deformations. This is illustrated by the 
only moderate improvement of the ``goodness-of-fit'' compared to 
what one gets from a non-spotted star (see Tab.~\ref{tab:di_para}). 
Note that even the stacked residual profiles (leaving apart any 
period problems) show very few events that look like surface 
spots (see Fig.~\ref{fig:LSD_stack}). There are a couple of such 
events, e.g.\ between about HJD 2450407 and 2450408 and then
shortly before 2450410. But nothing obvious later than that.
SU~Aur very likely presents a mix of periodic (spot or accretion 
curtain induced) variability typical for late-type TTS and 
the more erratic light-level changes typical for UXORs that 
are probably due to variations in the circumstellar dust.  
Our data suggest that some of the UXOR (or type~{\sc iii}
\pcite{herbst94}) variability also affects the shape of 
the photospheric line profiles. 

%
%
\section*{ACKNOWLEDGEMENTS}
The authors gratefully acknowledge use of the SIMBAD astronomical database. 
Extensive use was made of software provided through {\sc starlink}. 
We would like to thank an anonymous referee for very helpful comments and 
suggestions.
This paper is based on observations obtained during the 1996 MUlti-SIte
COntinuous Spectroscopic campaign, collected at the Canada-France Hawaii 3.6m
telescope, the McDonald 2.1m telescope, the La Palma 2.5m Isaak Newton 
telescope, the Observatoire de Haute-Provence 1.9m telescope and the 
Xinglong 2.2m telescope. The photometric data were obtained at the 
University of Vienna's automatic photometric telescopes. We are grateful 
for the efforts of the staff at all the telescopes involved. 

AHP acknowledges the support of NSF grant 9615571 and JDL acknowledges
support from the Natural Sciences and Engineering Research Council of
Canada. JMO acknowledges the 
financial support of Funda\c{c}\~{a}o para a Ci\^{e}ncia e a Tecnologia from 
Portugal (Praxis XXI grant BD9577/96) and of the UK Particle Physics and 
Astronomy Research Council (PPARC). 

\bibliographystyle{/home/ycu/Tex/MNRAS/mn}
\bibliography{/home/ycu/Tex/Bib/mnras_journals,/home/ycu/Tex/Bib/master,/home/ycu/Tex/Bib/myown,/home/ycu/Tex/Bib/accrefs,/home/ycu/Tex/Bib/suaur}

\begin{thebibliography}{{{Johns-Krull}, {Valenti} \& {Koresko}}{1999}}

\bibitem[\protect\citefmt{Akeson {\rm et~al.}}{2002}]{akeson2002}
Akeson~R.~L., Ciardi~D.~R., {van Belle}~G.~T., Creech-Eakman~M.~J., 2002,
  Astro\-phys.~J., 556, 1124

\bibitem[\protect\citefmt{{Andretta} \& {Giampapa}}{1995}]{andretta95}
{Andretta}~V., {Giampapa}~M.~S., 1995, Astro\-phys.~J., 439, 405

\bibitem[\protect\citefmt{Armitage \& Clarke}{1996}]{armitage96tts}
Armitage~P.~J., Clarke~C.~J., 1996, Mon.\ Not.\ R.\ astr.~Soc., 280, 458

\bibitem[\protect\citefmt{Baranne {\rm et~al.}}{1996}]{baranne96}
Baranne~A. {\rm et~al.}, 1996, Astr.\ Astro\-phys.~Suppl., 119, 373

\bibitem[\protect\citefmt{Barnes {\rm et~al.}}{1998}]{barnes98}
Barnes~J.~R., Collier~Cameron~A., Unruh~Y.~C., Donati~J.-F., Hussain~G. A.~J.,
  1998, Mon.\ Not.\ R.\ astr.~Soc., 299, 904

\bibitem[\protect\citefmt{Basri \& Batalha}{1990}]{basri90}
Basri~G., Batalha~C., 1990, Astro\-phys.~J., 363, 654

\bibitem[\protect\citefmt{Basri, Marcy \& Valenti}{1992}]{basri92}
Basri~G., Marcy~G.~W., Valenti~J.~A., 1992, Astro\-phys.~J., 390, 622

\bibitem[\protect\citefmt{Baudrand \& {B\"ohm}}{1992}]{baudrand92}
Baudrand~J., {B\"ohm}~T., 1992, Astr.\ Astro\-phys., 259, 711

\bibitem[\protect\citefmt{Beichman {\rm et~al.}}{1986}]{beichman86}
Beichman~C.~A., Myers~P.~C., Emerson~J.~P., Harris~S., Mathieu~R.,
  Benson~P.~J., Jennings~R.~E., 1986, Astro\-phys.~J., 307, 337

\bibitem[\protect\citefmt{Bouvier {\rm et~al.}}{1993}]{bouvier93}
Bouvier~J., Cabrit~S., Fernandez~M., Martin~E.~L., Matthews~J.~M., 1993, Astr.\
  Astro\-phys.~Suppl., 101(3), 485

\bibitem[\protect\citefmt{{Bray}}{1964}]{bray64}
{Bray}~R.~J., 1964, Zeitschrift f\"ur Astrophysik, 60, 207

\bibitem[\protect\citefmt{Calvet \& Hartmann}{1992}]{calvet92infall}
Calvet~N., Hartmann~L., 1992, Astro\-phys.~J., 386, 239

\bibitem[\protect\citefmt{Cameron \& Campbell}{1993}]{cameron93ttdisc}
Cameron~A.~C., Campbell~C.~G., 1993, Astr.\ Astro\-phys., 274, 309

\bibitem[\protect\citefmt{Canuto \& Mazzitelli}{1992}]{canuto92}
Canuto~V.~M., Mazzitelli~I., 1992, Astro\-phys.~J., 389, 724

\bibitem[\protect\citefmt{Catala {\rm et~al.}}{1993}]{catala93mus1}
Catala~C. {\rm et~al.}, 1993, Astr.\ Astro\-phys., 275, 245

\bibitem[\protect\citefmt{Catala {\rm et~al.}}{1999}]{catala99musicos}
Catala~C. {\rm et~al.}, 1999, Astr.\ Astro\-phys., 345, 884

\bibitem[\protect\citefmt{Chen {\rm et~al.}}{1995}]{chen95}
Chen~H., Myers~P.~C., Ladd~E.~F., Wood~D. O.~S., 1995, Astro\-phys.~J., 445,
  377

\bibitem[\protect\citefmt{Cincotta, M\'endez \& N\'u\~nez}{1995}]{cincotta95}
Cincotta~P., M\'endez~M., N\'u\~nez~J., 1995, Astro\-phys.~J., 449, 231

\bibitem[\protect\citefmt{Cohen \& Kuhi}{1979}]{cohen79}
Cohen~M., Kuhi~L.~V., 1979, Astro\-phys.~J.\ Suppl., 41, 743

\bibitem[\protect\citefmt{Cohen, Emerson \& Beichman}{1989}]{cohen89}
Cohen~M., Emerson~J.~P., Beichman~C.~A., 1989, Astro\-phys.~J., 339, 455

\bibitem[\protect\citefmt{Collier~Cameron \& Unruh}{1994}]{cameron94doppler}
Collier~Cameron~A., Unruh~Y.~C., 1994, Mon.\ Not.\ R.\ astr.~Soc., 269, 814

\bibitem[\protect\citefmt{D'Antona \& Mazzitelli}{1994}]{dantona94}
D'Antona~F., Mazzitelli~I., 1994, Astro\-phys.~J.\ Suppl., 90, 467

\bibitem[\protect\citefmt{Dewarf, Guinan \& Shaughnessy}{1998}]{dewarf1998}
Dewarf~L.~E., Guinan~E.~F., Shaughnessy~T.~M., 1998, Inf.\ Bull.\ var.\ Stars,
  4551

\bibitem[\protect\citefmt{Dhillon \& Privett}{1997}]{dhillon97}
Dhillon~V.~S., Privett~G.~J., 1997, Starlink User Note~167.5, Rutherford
  Appleton Laboratory

\bibitem[\protect\citefmt{Donati {\rm et~al.}}{1997}]{donati97zdi}
Donati~J.-F., Semel~M., Carter~B., Rees~D.~E., Collier~Cameron~A., 1997, Mon.\
  Not.\ R.\ astr.~Soc., 291, 658

\bibitem[\protect\citefmt{Donati {\rm et~al.}}{1999}]{donati99muspol}
Donati~J.-F., Catala~C., Wade~G. A.~Gallou~G., Delaigue~G., Rabou~P., 1999,
  Astr.\ Astro\-phys.~Suppl., 134, 149

\bibitem[\protect\citefmt{Eaton \& Herbst}{1995}]{eaton95}
Eaton~N.~L., Herbst~W., 1995, Astron.~J., 110, 2369

\bibitem[\protect\citefmt{Edwards {\rm et~al.}}{1994}]{edwards94}
Edwards~S., Hartigan~P., Ghandour~L., Andrulis~C., 1994, Astron.~J., 108, 1056

\bibitem[\protect\citefmt{ESA}{1997}]{hipcat}
ESA, 1997, The Hipparcos and Tycho Catalogues.
\newblock ESA SP-1200

\bibitem[\protect\citefmt{{Folha} \& {Emerson}}{1999}]{folha99}
{Folha}~D.~F.~M., {Emerson}~J.~P., 1999, Astr.\ Astro\-phys., 352, 517

\bibitem[\protect\citefmt{Giampapa {\rm et~al.}}{1993}]{giampapa93}
Giampapa~M.~S., Basri~G.~S., Johns~C.~M., Imhoff~C.~L., 1993, Astro\-phys.~J.\
  Suppl., 89, 321

\bibitem[\protect\citefmt{{Grady} {\rm et~al.}}{1996}]{grady96}
{Grady}~C.~A. {\rm et~al.}, 1996, Astr.\ Astro\-phys.~Suppl., 120, 157

\bibitem[\protect\citefmt{{Grinin} \& {Tambovtseva}}{1995}]{grinin95}
{Grinin}~V.~P., {Tambovtseva}~L.~V., 1995, Astr.\ Astro\-phys., 293, 396

\bibitem[\protect\citefmt{{Grinin} {\rm et~al.}}{1994}]{grinin94}
{Grinin}~V.~P., {The}~P.~S., {de Winter}~D., {Giampapa}~M.,
  {Rostopchina}~A.~N., {Tambovtseva}~L.~V., {van den Ancker}~M.~E., 1994,
  Astr.\ Astro\-phys., 292, 165

\bibitem[\protect\citefmt{Guenther \& Emerson}{1996}]{guenther96mag}
Guenther~E., Emerson~J.~P., 1996, AA, 309, 777

\bibitem[\protect\citefmt{Guenther {\rm et~al.}}{1999}]{guenther99mag}
Guenther~E.~W., Lehmann~H., Emerson~J.~P., Staude~J., 1999, Astr.\
  Astro\-phys., 341, 768

\bibitem[\protect\citefmt{Hartmann, Hewett \& Calvet}{1994}]{hartmann94}
Hartmann~L., Hewett~R., Calvet~N., 1994, Astro\-phys.~J., 426, 669

\bibitem[\protect\citefmt{Hatzes}{1994}]{hatzes94v410}
Hatzes~A.~P., 1994, Astro\-phys.~J., 451, 784

\bibitem[\protect\citefmt{{Herbig} \& {Bell}}{1988}]{herbig88cat}
{Herbig}~G.~H., {Bell}~K.~R., 1988, Catalog of emission line stars of the orion
  population : 3 : 1988.
\newblock Lick Observatory Bulletin, Santa Cruz: Lick Observatory

\bibitem[\protect\citefmt{{Herbst} \& {Shevchenko}}{1999}]{herbst1999}
{Herbst}~W., {Shevchenko}~V.~S., 1999, aj, 118, 1043

\bibitem[\protect\citefmt{Herbst {\rm et~al.}}{1987}]{herbst87}
Herbst~W. {\rm et~al.}, 1987, Astron.~J., 94, 137

\bibitem[\protect\citefmt{Herbst {\rm et~al.}}{1994}]{herbst94}
Herbst~W., Herbst~D.~K., Grossman~E.~J., Weinstein~D., 1994, Astron.~J., 108,
  1906

\bibitem[\protect\citefmt{Hillenbrand}{1997}]{hillenbrand97}
Hillenbrand~L.~A., 1997, Astron.~J., 113, 1733

\bibitem[\protect\citefmt{Horne \& Baliunas}{1986}]{horne86period}
Horne~J.~H., Baliunas~S.~L., 1986, Astro\-phys.~J., 302, 757

\bibitem[\protect\citefmt{Horne}{1986}]{horne86extopt}
Horne~K.~D., 1986, Publ.\ astr.\ Soc.\ Pacif., 98, 609

\bibitem[\protect\citefmt{Johns \& Basri}{1995a}]{johns95tts}
Johns~C.~M., Basri~G., 1995a, Astron.~J., 109, 2800

\bibitem[\protect\citefmt{Johns \& Basri}{1995b}]{johns95suaur}
Johns~C.~M., Basri~G., 1995b, Astro\-phys.~J., 449, 341

\bibitem[\protect\citefmt{{Johns-Krull} {\rm
  et~al.}}{1999}]{johns-krull99bptau_pol}
{Johns-Krull}~C.~M., {Valenti}~J.~A., {Hatzes}~A.~P., {Kanaan}~A., 1999, ApJ,
  510, L41

\bibitem[\protect\citefmt{{Johns-Krull}, {Valenti} \&
  {Koresko}}{1999}]{johns-krull99bptau}
{Johns-Krull}~C.~M., {Valenti}~J.~A., {Koresko}~C., 1999, Astro\-phys.~J., 516,
  900

\bibitem[\protect\citefmt{Johns-Krull}{1996}]{johns-krull96tts}
Johns-Krull~C.~M., 1996, Astr.\ Astro\-phys., 306, 803

\bibitem[\protect\citefmt{Joncour, Bertout \& Bouvier}{1994}]{joncour94hde}
Joncour~I., Bertout~C., Bouvier~J., 1994, Astr.\ Astro\-phys., 291, L19

\bibitem[\protect\citefmt{Joncour, Bertout \& M\'{e}nard}{1994}]{joncour94v410}
Joncour~I., Bertout~C., M\'{e}nard~F., 1994, Astr.\ Astro\-phys., 285, L25

\bibitem[\protect\citefmt{Joncour}{1992}]{joncour92}
Joncour~I., 1992, J. Astron. Fran., 43, 31

\bibitem[\protect\citefmt{Kenyon \& Hartmann}{1987}]{kenyon87}
Kenyon~S., Hartmann~L., 1987, Astro\-phys.~J., 323, 714

\bibitem[\protect\citefmt{{K\"onigl}}{1991}]{konigl91}
{K\"onigl}~A., 1991, Astro\-phys.~J., 370, L39

\bibitem[\protect\citefmt{Lang}{1980}]{lang80}
Lang~K.~R., 1980, {Astrophysical Formulae}.
\newblock Springer-Verlag, second edition

\bibitem[\protect\citefmt{McCarthy {\rm et~al.}}{1993}]{mccarthy93}
McCarthy~J.~K., Sandiford~B.~A., Boyd~D., Booth~J., 1993, Publ.\ astr.\ Soc.\
  Pacif., 105, 881

\bibitem[\protect\citefmt{Mills}{1994}]{mills92}
Mills~D., 1994, Starlink User Note~152, Rutherford Appleton Laboratory

\bibitem[\protect\citefmt{Montgomery \& O'Donoghue}{1999}]{montgomery99}
Montgomery~M.~H., O'Donoghue~D., 1999, in Breger~M., ed, Delta Scuti Star
  Newsletter.
\newblock p.~28, Issue 13

\bibitem[\protect\citefmt{Myers {\rm et~al.}}{1987}]{myers87}
Myers~P.~C., Fuller~G.~A., Mathieu~R.~D., Beichman~C.~A., Benson~P.~J.,
  Schild~R.~E., Emerson~J.~P., 1987, Astro\-phys.~J., 319, 340

\bibitem[\protect\citefmt{Nadalin, Dewarf \& Guinan}{2000}]{nadalin2000}
Nadalin~I., Dewarf~L.~E., Guinan~E.~F., 2000, Inf.\ Bull.\ var.\ Stars, 4987

\bibitem[\protect\citefmt{Oliveira {\rm et~al.}}{2000a}]{oliveira2000}
Oliveira~J., Foing~B.~H., {van Loon}~J.~T., Unruh~Y.~C., 2000a, Astr.\
  Astro\-phys., in press

\bibitem[\protect\citefmt{{Oliveira} {\rm et~al.}}{2000b}]{oliveira2000suaur}
{Oliveira}~J.~M., {Foing}~B.~H., {van Loon}~J.~T., {Unruh}~Y.~C., 2000b, Astr.\
  Astro\-phys., 362, 615

\bibitem[\protect\citefmt{Petrov {\rm et~al.}}{1996}]{petrov96}
Petrov~P.~P., Gullbring~E., Ilyin~I., Gahm~G.~F., Tuominen~I., Hackman~T.,
  Loden~K., 1996, Astr.\ Astro\-phys., 314, 821

\bibitem[\protect\citefmt{Press {\rm et~al.}}{1992}]{press92}
Press~W.~H., Flannery~B.~P., Teukolsky~S.~A., Vetterling~W.~T., 1992, Numerical
  Recipes: The Art of Scientific Computing.
\newblock Cambridge University Press, Cambridge, 2nd Edition

\bibitem[\protect\citefmt{Rice \& Strassmeier}{1996}]{rice96}
Rice~J.~B., Strassmeier~K.~G., 1996, Astr.\ Astro\-phys., 316, 164

\bibitem[\protect\citefmt{Roberts, Lehár \& Dreher}{1987}]{roberts87clean}
Roberts~D.~H., Lehár~J., Dreher~J.~W., 1987, Astron.~J., 93, 968

\bibitem[\protect\citefmt{{Saar} {\rm et~al.}}{1997}]{saar97helium}
{Saar}~S.~H., {Huovelin}~J., {Osten}~R.~A., {Shcherbakov}~A.~G., 1997, Astr.\
  Astro\-phys., 326, 741

\bibitem[\protect\citefmt{Safier}{1998}]{safier98}
Safier~P.~N., 1998, Astro\-phys.~J., 494, 336

\bibitem[\protect\citefmt{Shu {\rm et~al.}}{1994}]{shu94}
Shu~F., Najita~J., Ostriker~E., Wilkin~F., Ruden~S., Lizano~S., 1994,
  Astro\-phys.~J., 429, 781

\bibitem[\protect\citefmt{Smith {\rm et~al.}}{1999}]{smith1999tts}
Smith~K.~W., Lewis~G.~F., Bonnell~I.~A., Bunclark~P.~S., Emerson~J.~P., 1999,
  MNRAS, 304, 367

\bibitem[\protect\citefmt{{Stassun} {\rm et~al.}}{1999}]{stassun1999PMS_rot}
{Stassun}~K.~G., {Mathieu}~R.~D., {Mazeh}~T., {Vrba}~F.~J., 1999, Astron.~J.,
  117, 2941

\bibitem[\protect\citefmt{Stellingwerf}{1978}]{stellingwerf78}
Stellingwerf~R.~F., 1978, Astro\-phys.~J., 224, 953

\bibitem[\protect\citefmt{{Strassmeier} \&
  {Rice}}{1998}]{strassmeier98hde283572}
{Strassmeier}~K.~G., {Rice}~J.~B., 1998, Astr.\ Astro\-phys., 339, 497

\bibitem[\protect\citefmt{Strassmeier {\rm et~al.}}{1997}]{strassmeier97apt}
Strassmeier~K.~G., Boyd~L.~J., Epand~D.~H., Granzer~T., 1997, Publ.\ astr.\
  Soc.\ Pacif., 109, 697

\bibitem[\protect\citefmt{Strassmeier, Serkowitsch \&
  Granzer}{1999}]{strassmeier99apt}
Strassmeier~K.~G., Serkowitsch~E., Granzer~T., 1999, Astr.\
  Astro\-phys.~Suppl., 140, 29

\bibitem[\protect\citefmt{Strassmeier}{1994}]{strassmeier94}
Strassmeier~K.~G., 1994, Astr.\ Astro\-phys., 281, 395

\bibitem[\protect\citefmt{Unruh, Collier~Cameron \&
  Guenther}{1998}]{unruh98dftau}
Unruh~Y.~C., Collier~Cameron~A., Guenther~E.~W., 1998, Mon.\ Not.\ R.\
  astr.~Soc., 295, 781

\bibitem[\protect\citefmt{{{\v S}vestka}}{1972}]{svestka72review}
{{\v S}vestka}~Z., 1972, ARA\&A, 10, 1

\end{thebibliography}

\end{document}